\documentclass[reprint, aps, prd,amsmath,amssymb,twocolumn,showkeys, superscriptaddress, floatfix]{revtex4-2}
\usepackage{graphicx}% Include figure files
\usepackage{dcolumn}% Align table columns on decimal point
\usepackage{bm}% bold math
\usepackage[colorlinks = true, allcolors = blue]{hyperref}% add hypertext capabilities
\usepackage{verbatim}
\usepackage{soul}
\usepackage{float}
\usepackage{xcolor}
\begin{document}

\title{Duality-symmetric axion electrodynamics and haloscopes of various geometries}

\author{Dai-Nam Le}
 \thanks{Equal contribution}
 \email{ledainam@tdtu.edu.vn}
 \affiliation{Atomic~Molecular~and~Optical~Physics~Research~Group, Advanced Institute of Materials Science, Ton Duc Thang University, Ho~Chi~Minh~City 700000, Vietnam}
 \affiliation{Faculty of Applied Sciences, Ton Duc Thang University, Ho~Chi~Minh~City 700000,~Vietnam}
 
\author{Le Phuong Hoang}
 \thanks{Equal contribution}
 \email{hoanglephuong1993@gmail.com}
 \affiliation{Square Lab, Hanoi University of Science and Technology, Hanoi 100000, Viet Nam}

\author{Binh Xuan Cao}
 \thanks{Corresponding author}  
 \email{binh.caoxuan@hust.edu.vn}
 \affiliation{Square Lab, Hanoi University of Science and Technology, Hanoi 100000, Viet Nam}
 \affiliation{School of Mechanical Engineering, Hanoi University of Science and Technology, Ha Noi 100000, VietNam}

\date{\today}% It is always \today, today,

\begin{abstract}
Within the dual symmetric point of view, the theory for seeking axion dark matter via haloscope experiments is derived by exactly solving the dual symmetric axion electrodynamics equation. Notwithstanding that the conventional theory of axion electrodynamics presented in [\textit{Phys. Rev. Lett.} \textbf{51}, 1415 (1983) and \textit{J. Phys. A: Math. Gen.} \textbf{19}, L33 (1986)] is more commonly used in haloscope theory, we show that the dual symmetric axion electrodynamics has more advantages to apply into haloscope theory. First, the dual symmetric and conventional perspective of axion electrodynamics coincide under long-wavelength approximation. Moreover, dual symmetric theory can obtain an exact analytical expression of the axion-induced electromagnetic field for any states of axion. {This solution has been used in conventional theory for long-wavelength approximation}. The difference between two theories can occur in directional axion detection or electric sensing haloscopes. For illustrative purposes, we consider the various type of resonant cavities: cylindrical solenoid, spherical solenoid, two-parallel-sheet cavity, toroidal solenoid with a rectangular cross-section, and with a circular cross-section. The resonance of the axion-induced signal as well as the ratio of the energy difference over the stored energy inside the cavity are investigated in these types of cavity.
\end{abstract}


\maketitle

\section{\label{sec:intro}Introduction}

It has been long time stated that dark matter (DM) plays an important role on the gravitational evolution of our Universe, especially from $10^{4}$ to $10^{10}$ years after Big Bang \cite{Trimble1987}. While there are myriad convincing observations of DM in astrophysical level, the evidences that supports the existence of DM in particle level are still insufficient. On the effort to fulfill the Standard Model (SM) of particle physics, several theories and experimental schemes have been proposed to detect this exotic matter. Axion is one of well-known candidates employed to interpret the existence of cold DM (CDM) \cite{Dine1983, Abbott1983, Preskill1983, Blumenthal1984} which implies the DM velocity much smaller than light speed. This cold axion DM is a hypothetical pseudoscalar particle resulted from Peccei-Quinn (PQ) symmetry linked to the absence of charge-parity (CP) violationswith strong interactions \cite{DINE1981, Kim1979, Wilczek1978}. Modern experimental schemes of axion detection utilize the interaction between axion and two photons which is described as axion haloscope searches \cite{Sikivie1983, Bibber1987}. The theory that explains the coupling between axion field and electromagnetic field is so-called axion electrodynamics and has been studied for a long time \cite{Sudbery1986, VISINELLI2013, Tiwari2015}. The current axion haloscope experiments employ conventional axion electrodynamics proposed by P. Sikivie in which the minimal SM is modified to include axion coupled term \cite{Sikivie1983}. However, the set of modified Maxwell equations derived from Sikivie assumptions does not fulfill the duality preservation which is the intrinsic symmetry of Maxwell equations from the invariance under $SO(2)$ \cite{Lipkin1964, Morgan1964}. Alternatively, L. Visinelli proposed a new set of Maxwell equations considering both axions and magnetic monopoles which preserves the duality relation. Both sets of equations proposed by P. Sikivie and L. Visinelli are derived from same axion electromagnetic Lagrangian \cite{VISINELLI2013} as the following:
\begin{equation}\label{eqn:lagrange}
    \mathcal{L}_{tot}=\mathcal{L}_{Maxwell} + \mathcal{L}_{a} + \mathcal{L}_{U} ,
\end{equation}
where {$\mathcal{L}_{Maxwell} = - (4 \mu_0)^{-1} F^{\mu\nu} F_{\mu \nu} - A_{\mu} J_e^{\mu} -  B_{\mu} J_m^{\mu}$} is the classical Maxwell electromagnetic Lagrangian, $\mathcal{L}_{a}=(4\mu _0 )^{-1} g_{a\gamma \gamma} a F^{\mu \nu} \tilde{F}_{\mu \nu}=-(4\mu_0 c )^{-1} g_{a\gamma \gamma} a \mathbf{E} \cdot \mathbf{B}$ is axion-biphoton coupling term, $\mathcal{L}_U= 2^{-1}\left[ (\partial^{\mu} a)(\partial_{\mu} a)- \omega_a^2 a^2 \right]$ is axion self Lagrangian which preserves the duality relation under the substitution of axion coupled term, $\omega _a$ is the angular frequency of the axion 's oscillation and $g_{a\gamma\gamma}$ is axion-biphoton coupling constant \cite{Peccei1977}. {Here $J_e^{\mu}$ and $J_{m}^{\mu}$ are  four-current density associated with electric and magnetic charges and $A_{\mu}$, $B_{\mu}$ are their corresponding four-potentials \cite{Zwanziger1971, VISINELLI2013, Ouellet2019}}. From equation \eqref{eqn:lagrange}, the new set of modified Maxwell equations proposed by L. Visinelli is manifested as the following \cite{VISINELLI2013, Tiwari2015, Hoang2017, Tercas2018}:
\begin{equation}\label{eqn:axion-intro}
\left( \square + \dfrac{m_a^2 c^2}{\hbar ^2} \right) a = \dfrac{2 g_{a\gamma\gamma}}{\mu _0 c} \mathbf{E} \cdot \mathbf{B} ,
\end{equation}
\begin{subequations}\label{eqn:em-general-intro}
\begin{eqnarray}
\pmb{\nabla} \cdot \left( \mathbf{E} - c g_{a\gamma\gamma}a \mathbf{B} \right) = &&\dfrac{\rho _e}{\varepsilon _0} , \label{eqn:div-e-general-intro}\\
\pmb{\nabla} \cdot \left( c \mathbf{B} + g_{a\gamma\gamma} a \mathbf{E} \right) = &&c \mu _0 \rho _m , \label{eqn:div-b-general-intro}\\
\pmb{\nabla} \times \left( \mathbf{E} - c g_{a\gamma\gamma}a \mathbf{B} \right) = &&- \dfrac{\partial \left( c \mathbf{B} + g_{a\gamma\gamma} a \mathbf{E} \right)}{c \partial t} - \mu _0 \mathbf{J}_m , \label{eqn:rot-e-general-intro}\\
\pmb{\nabla} \times \left( c \mathbf{B} + g_{a\gamma\gamma} a \mathbf{E} \right) = &&\dfrac{\partial \left( \mathbf{E} - c g_{a\gamma\gamma}a \mathbf{B} \right)}{c \partial t} + c \mu _0 \mathbf{J}_e \label{eqn:rot-b-general-intro},
\end{eqnarray}
\end{subequations}
whereas $\square = c^{-2} \partial _{t} \partial_{t} - \pmb{\nabla}^2$ is d'Alembert operator and the mass of axion $m_a$ links to its frequency $\omega _a$ by $m_a c^2 = \hbar \omega$.

Noticeably, in References \cite{VISINELLI2013, Tiwari2015}, these equations \eqref{eqn:em-general-intro} had been shown to be equivalent to the conventional axion electrodynamics equations in References \cite{Sikivie1983, Sudbery1986}
\begin{subequations}\label{eqn:em-conv}
\begin{eqnarray}
\pmb{\nabla} \cdot \mathbf{E} -  c g_{a\gamma\gamma} \mathbf{B} \cdot \pmb{\nabla}  a  =&& \dfrac{\rho _e}{\varepsilon _0} \label{eqn:div-e-conv}\\
\pmb{\nabla} \cdot \mathbf{B} =&& 0 \label{eqn:div-b-conv}\\
\pmb{\nabla} \times \mathbf{E} =&& -  \dfrac{\partial \mathbf{B} }{\partial t} \label{eqn:rot-e-conv}\\
\pmb{\nabla} \times \mathbf{B} -  \frac{g_{a\gamma\gamma}}{c} \mathbf{E}\times \pmb{\nabla} a =&& \dfrac{\partial \mathbf{E}}{c^2 \partial t}  -  \frac{g_{a\gamma\gamma}\mathbf{B}}{c}  \dfrac{\partial a}{\partial t} + \mu _0 \mathbf{J}_e \label{eqn:rot-b-conv},
\end{eqnarray}
\end{subequations}
when the axion field and the magnetic monopole sources meet the following conditions
\begin{subequations}\label{eqn:cond-intro}
\begin{eqnarray}
\rho _m && = c g_{a\gamma\gamma} a \rho _e + \dfrac{g_{a\gamma\gamma}}{\mu _0 c} \left( \mathbf{E} + c g_{a \gamma \gamma} a \mathbf{B} \right) \cdot \pmb{\nabla} a \label{eqn:cond-rho-m} \\
\mathbf{J}_m && = c g_{a\gamma\gamma} a \mathbf{J}_e - \dfrac{g_{a\gamma\gamma}}{c \mu _0} \dfrac{\partial a}{\partial t} \left( \mathbf{E} + c g_{a \gamma \gamma} a \mathbf{B} \right) \nonumber\\
&& \quad \quad \quad \quad \quad - \dfrac{g_{a\gamma\gamma}}{\mu _0} \left(c \mathbf{B} - g_{a\gamma\gamma} a \mathbf{E} \right) \times \pmb{\nabla} a \label{eqn:cond-J-m}.
\end{eqnarray}
\end{subequations}
Therefore, in the viewpoint of duality conservation, the conventional axion electrodynamics equations \eqref{eqn:em-conv} are valid if and only if there is magnetic monopole sources  $\rho_m \neq 0, \mathbf{J}_m \neq 0$. Inversely, when there is no magnetic monopole sources $\rho_m = 0, \mathbf{J}_m = 0$, the axion electrodynamics equations becomes \cite{VISINELLI2013, Tiwari2015, Hoang2017, Tercas2018}
\begin{subequations}\label{eqn:em-zero}
\begin{eqnarray}
\pmb{\nabla} \cdot \left( \mathbf{E} - c g_{a\gamma\gamma}a \mathbf{B} \right) &=& \dfrac{\rho _e}{\varepsilon _0} \label{eqn:div-e}\\
\pmb{\nabla} \cdot \left( c \mathbf{B} + g_{a\gamma\gamma} a \mathbf{E} \right) &=& 0 \label{eqn:div-b}\\
\pmb{\nabla} \times \left( \mathbf{E} - c g_{a\gamma\gamma}a \mathbf{B} \right) &=& -\dfrac{\partial \left( c \mathbf{B} + g_{a\gamma\gamma} a \mathbf{E} \right)}{c \partial t}  \label{eqn:rot-e}\\
\pmb{\nabla} \times \left( c \mathbf{B} + g_{a\gamma\gamma} a \mathbf{E} \right) &=& \dfrac{\partial \left( \mathbf{E} - c g_{a\gamma\gamma}a \mathbf{B} \right)}{c \partial t}  + c \mu _0 \mathbf{J}_e . \label{eqn:rot-b}
\end{eqnarray}
\end{subequations}
{It is noticed that the set of conventional axion electrodynamics equation \eqref{eqn:em-conv} is derived from the Lagrangian principle with the Lagrangian given in \eqref{eqn:lagrange} when neglecting the magnetic four-current term i.e $B_{\mu} J^{\mu}_m$.}

In this paper, we achieve the explicit solutions of these equations \eqref{eqn:em-zero} in various geometries of haloscope cavities. We demonstrate that the duality symmetric theory is valid as it comes up with the same solutions with conventional axion electrodynamics proposed by P. Sikivie in the long-wavelength regime. {Furthermore, this analytical solution can be used to describe the electromagnetic fields inside the haloscope cavities without long-wavelength approximation (LWA) in the duality symmetric theory}. It means that the theory can handle the axion electrodynamics in different states of axions with different wavelength scales, not requiring axions to be at ground state with a long wavelength. Thus, there is no need to neglect the spatial gradient of the axion field ($\pmb{\nabla} a$ is finite) which is typically assumed in the conventional approach. Our mathematical scheme aims to providing a more precise and robust template for analyzing axion haloscopes in several geometries of resonant cavities.

The paper is organized as follows: Sec. \ref{sec:3} presents the general solutions of axion modified Maxwell equation in dual symmetric perspective and its application to haloscope experiments; subsequently, some examples for different types of cavity are presented in \ref{sec:4}, including observation of resonance of axion-to-photon conversion power and the ratio of electric-magnetic energy difference over the stored axion-induced electromagnetic energy. Sec. \ref{sec:5} presents some discussions between Sikivie and dual symmetric axion electrodynamics theories while Sec. \ref{sec:conl} devotes our conclusion.

\section{\label{sec:3}How haloscope works in dual symmetric point of view?}
\subsection{\label{sec:3a}General solution of axion-induced electromagnetic field}
To solve the $SO(2)$-invariant axion Maxwell equation \cite{VISINELLI2013, Tiwari2015, Hoang2017, Tercas2018} in case of no magnetic monopole:
\begin{eqnarray}
\left( \square + \dfrac{m_a^2 c^2}{\hbar ^2} \right) a &=& \dfrac{2 g_{a\gamma\gamma}}{\mu _0 c} \mathbf{E} \cdot \mathbf{B}, \label{eqn:axion}
\end{eqnarray}
\begin{subequations}\label{eqn:em-general}
\begin{eqnarray}
\pmb{\nabla} \cdot \left( \mathbf{E} - c g_{a\gamma\gamma}a \mathbf{B} \right) = && \dfrac{\rho _e}{\varepsilon _0}, \label{eqn:div-e-general}\\
\pmb{\nabla} \cdot \left( c \mathbf{B} + g_{a\gamma\gamma} a \mathbf{E} \right) = && 0, \label{eqn:div-b-general}\\
\pmb{\nabla} \times \left( \mathbf{E} - c g_{a\gamma\gamma}a \mathbf{B} \right) = &&  \dfrac{\partial \left( c \mathbf{B} + g_{a\gamma\gamma} a \mathbf{E} \right)}{c \partial t}, \label{eqn:rot-e-general}\\
\pmb{\nabla} \times \left( c \mathbf{B} + g_{a\gamma\gamma} a \mathbf{E} \right) = && \dfrac{\partial \left( \mathbf{E} - c g_{a\gamma\gamma}a \mathbf{B} \right)}{c \partial t} + c \mu _0 \mathbf{J}_e \label{eqn:rot-b-general},
\end{eqnarray}
\end{subequations}
we may set \cite{VISINELLI2013}
\begin{equation}\label{eqn:new-EB}
\hat{\mathbf{E}} = \mathbf{E} - c g_{a\gamma\gamma}a \mathbf{B},\quad
\hat{\mathbf{B}} = \mathbf{B} + \dfrac{g_{a\gamma\gamma} a}{c} \mathbf{E},
\end{equation}
i.e the electric and magnetic field strength $\mathbf{E}$, $\mathbf{B}$ can be written in terms of $\hat{\mathbf{E}}$, $\hat{\mathbf{B}}$ as
\begin{subequations}\label{eqn:eb-sol}
\begin{eqnarray}
\mathbf{E} &=& \dfrac{\hat{\mathbf{E}} + c g_{a \gamma \gamma} a \hat{\mathbf{B}}}{1 + g_{a \gamma \gamma}^2 a^2}, \label{eqn:e-sol}\\
\mathbf{B} &=& \dfrac{\hat{\mathbf{B}} - g_{a \gamma \gamma} a \hat{\mathbf{E}} / c}{1 + g_{a \gamma \gamma}^2 a^2}. \label{eqn:b-sol}
\end{eqnarray}
\end{subequations} 
Equations \eqref{eqn:em-general} in term of $\hat{\mathbf{E}}$, $\hat{\mathbf{B}}$ read
\begin{subequations} \begin{eqnarray}
\pmb{\nabla} \cdot \hat{\mathbf{E}} &=& \dfrac{\rho _e}{\varepsilon _0} \label{eqn:div-e-hat}\\
\pmb{\nabla} \cdot \hat{\mathbf{B}} &=& 0 \label{eqn:div-b-hat}\\
\pmb{\nabla} \times \hat{\mathbf{E}} &=& -  \dfrac{\partial \hat{\mathbf{B}} }{\partial t} \label{eqn:rot-e-hat}\\
\pmb{\nabla} \times \hat{\mathbf{B}} &=& \dfrac{1}{c^2} \dfrac{\partial \hat{\mathbf{E}}}{\partial t}  + \mu _0 \mathbf{J}_e \label{eqn:rot-b-hat},
\end{eqnarray} \end{subequations} 
which have the form of standard Maxwell equations without presence of axion fields. Therefore, one may show that $\hat{\mathbf{E}}$ and $\hat{\mathbf{B}}$ are governed by the following D'Alembert  equations \cite{Jackson}:
\begin{subequations}\label{eqn:d-hat} \begin{eqnarray}
\square \hat{\mathbf{E}} & = & - \dfrac{\pmb{\nabla} \rho_e}{\varepsilon _0} - \mu _0 \dfrac{\partial \mathbf{J}_e}{\partial t}, \label{eqn:d-e-hat}\\
\square \hat{\mathbf{B}} & = & \mu _0 \pmb{\nabla} \times \mathbf{J}_e . \label{eqn:d-b-hat}
\end{eqnarray} \end{subequations} 
Regarding the linearity of these above equations, their solutions can be expressed as the linear combinations
\begin{equation}\label{eqn:linear}
\hat{\mathbf{E}} = \mathbf{E}_0 + \mathbf{E}_{rad}, \quad \hat{\mathbf{B}} = \mathbf{B}_0 + \mathbf{B}_{rad},
\end{equation}
between special solution $\mathbf{E}_0$, $\mathbf{B}_0$ which plays a role as applied electromagnetic field and can be expressed by Jefimenko formula (here $t_r = t - \left| \mathbf{r} - \mathbf{r}^{\prime} \right|/c$ is retarded time) \cite{Jackson}
\begin{widetext}
\begin{subequations} \begin{eqnarray}
\mathbf{E}_0 \left( \mathbf{r},t \right) && = \dfrac{1}{4 \pi \varepsilon _0} \int \left[ \left( \dfrac{\rho _e (\mathbf{r}^{\prime} , t_r )}{\left| \mathbf{r} - \mathbf{r}^{\prime} \right|^3} + \dfrac{\partial _t \rho _e (\mathbf{r}^{\prime} , t_r )}{c \left| \mathbf{r} - \mathbf{r}^{\prime} \right| ^2} \right) \left( \mathbf{r} - \mathbf{r}^{\prime} \right) - \dfrac{\partial _t \mathbf{J} _e (\mathbf{r}^{\prime} , t_r )}{c^2 \left| \mathbf{r} - \mathbf{r}^{\prime} \right|} \right] d^3 \mathbf{r}^{\prime} , \label{eqn:e-0-sol}\\
\mathbf{B}_0 \left( \mathbf{r},t \right) && = \dfrac{\mu _0}{4 \pi } \int \left[\left( \dfrac{\mathbf{J} _e (\mathbf{r}^{\prime} , t_r )}{\left| \mathbf{r} - \mathbf{r}^{\prime} \right| ^3} + \dfrac{\partial _t \mathbf{J} _e (\mathbf{r}^{\prime} , t_r )}{c \left| \mathbf{r} - \mathbf{r}^{\prime} \right| ^2} \right)  \times \left( \mathbf{r} - \mathbf{r}^{\prime} \right) \right] d^3 \mathbf{r}^{\prime} \label{eqn:b-0-sol},
\end{eqnarray} \end{subequations} 
\end{widetext}
and the homogeneous solution $\mathbf{E}_{rad}$, $\mathbf{B}_{rad}$ which is known as radiation electromagnetic field and can be presented in term of the following Fourier integral \cite{Jackson}
\begin{subequations}\label{eqn:eb-f-sol}
\begin{eqnarray}
\mathbf{E}_{rad} (\mathbf{r},t) && = \int \pmb{\mathcal{E}} (\mathbf{r}, \omega) e^{- \imath \omega t} d \omega ,\\
\mathbf{B}_{rad} (\mathbf{r},t) && = - \imath \int \pmb{\nabla} \times \pmb{\mathcal{E}} (\mathbf{r}, \omega) e^{- \imath \omega t} \omega ^{-1} d \omega ,
\end{eqnarray}
\end{subequations} 
whereas Fourier image $\pmb{\mathcal{E}} (\mathbf{r}, \omega)$ of free electric component is governed by Helmholtz equation
\begin{equation}\label{eqn:eom}
\left[ \pmb{\nabla}\cdot \pmb{\nabla} + \omega ^2 / c^2 \right] \pmb{\mathcal{E}} (\mathbf{r}, \omega) = 0. 
\end{equation}
In principle, the solution of this equation can be expanded as linear combination of its eigenfunction. However, the suitable eigenfunctions as well as the coefficient of this expansion depend on the boundary condition of electromagnetic field. Although we cannot provide the general expression of free electromagnetic fields without knowing the exact boundary condition, the magnitude of $\pmb{\mathcal{E}} (\mathbf{r}, \omega)$  must be proportional to axion-biphoton coupling constant $g_{a\gamma\gamma}$ due to the fact that these radiation fields vanish when there is no axion $g_{a\gamma\gamma} = 0$.

Since the coupling $g_{a\gamma\gamma}$ is weak, the Taylor expansion of general solution of axion electromagnetic fields $\mathbf{E}$, $\mathbf{B}$ in term of  $g_{a\gamma\gamma}$ upto the first order is:
\begin{subequations} \begin{eqnarray}
\mathbf{E} && = \mathbf{E}_0 + c g_{a\gamma\gamma} a \mathbf{B_0} + \mathbf{E}_{rad} + O (g^2_{a\gamma\gamma}), \label{eqn:e-expand}\\
\mathbf{B} && = \mathbf{B}_0 - \dfrac{g_{a\gamma\gamma}a}{c} \mathbf{E}_0 + \mathbf{B}_{rad} + O(g_{a\gamma\gamma}^2) \label{eqn:b-expand}.
\end{eqnarray} \end{subequations} 
In other words, the axion-induced electromagnetic fields are approximately
\begin{subequations}\label{eqn:1-order}
\begin{eqnarray}
\mathbf{E}_{ind} && \approx  c g_{a\gamma \gamma} a \mathbf{B}_0 + \mathbf{E}_{rad},\nonumber\\
&& = c g_{a\gamma \gamma} a \mathbf{B}_0 + \int \pmb{\mathcal{E}} (\mathbf{r},\omega) e^{- \imath \omega t} d \omega,\\
\mathbf{B}_{ind} && \approx\mathbf{B}_{rad} - \dfrac{g_{a\gamma\gamma}a}{c} \mathbf{E}_0, \nonumber\\
&& = - \dfrac{g_{a\gamma\gamma}a}{c} \mathbf{E}_0 - \imath \int \pmb{\nabla} \times \pmb{\mathcal{E}} (\mathbf{r},\omega) \frac{e^{- \imath \omega t}}{\omega} d \omega .
\end{eqnarray}
\end{subequations}

\subsection{\label{sec:3b}Axion haloscope under dual symmetric axion electrodynamics theory}
For general haloscope experiments (see for examples \cite{Hoang2017,Ouellet2019, Knirck2019, Kim2019, Tobar2019} and references therein), the induced electromagnetic fields $\mathbf{E}_{ind}$ and $\mathbf{B}_{ind}$ are considered as response from the presence of axion when nonzero external magnetic field $\mathbf{B}_0 = \mathbf{B}_0 ( \mathbf{r})$ and zero external electric field i.e $\mathbf{E}_0 = 0$ are applied. The Klein-Gordon equation of axion field \eqref{eqn:axion} simply becomes the one of free axion:
\begin{equation}
\label{eqn:axion-halo}
\left( \square +\dfrac{m_a^2 c^2}{\hbar ^2} \right) a = 0.
\end{equation}
Spacetime evolution of axion field is a plane wave described as the following:
\begin{equation}
\label{eqn:axion-sol}
a ( \mathbf{r}, t) = a_0 (\omega) \exp \left[ \imath \left( \mathbf{k} \cdot \mathbf{r} - \omega t \right) \right] ,  
\end{equation}
whereas $\omega = \sqrt{ \mathbf{k}^2 c^2 + \omega _a^2}$ and $\omega_a = m_a c^2 / \hbar$.

Substituting $\mathbf{B}_0 = \mathbf{B}_0 (\mathbf{r})$, $\mathbf{E}_0 = 0$ and $a (\mathbf{r},t)$ from Equation \eqref{eqn:axion-sol} into Equation \eqref{eqn:1-order}, we obtain the axion-induced electromagnetic field
\begin{subequations}\label{eqn:1-order-a}
\begin{eqnarray}
\mathbf{E}_{ind} && = c g_{a\gamma \gamma} a (\mathbf{r},t) \mathbf{B}_0 (\mathbf{r}) + \int \pmb{\mathcal{E}} (\mathbf{r},\omega) e^{- \imath \omega t} d \omega,\\
\mathbf{B}_{ind} && = - \imath \int \pmb{\nabla} \times \pmb{\mathcal{E}} (\mathbf{r},\omega) e^{- \imath \omega t} \omega ^{-1} d \omega .
\end{eqnarray}
\end{subequations}

To determine axion-induced electromagnetic field, we need to solve the Fourier field $ \pmb{\mathcal{E}} (\mathbf{r}, \omega)$ via Helmholtz equation \eqref{eqn:eom}. Denoting this Fourier field as $\pmb{\mathcal{E}} (\mathbf{r}, \omega) = - c g_{a \gamma \gamma} a (\mathbf{r},t) \pmb{\mathfrak{B}} (\mathbf{r}, \omega)$, the axion-induced electromagnetic field is expressed in a simpler form
\begin{subequations}\label{eqn:eb-2-sol} \begin{eqnarray} 
\mathbf{E}_{ind} (\mathbf{r},t) && = c g_{a\gamma\gamma} a_0 (\omega) \left[ \mathbf{B}_0 (\mathbf{r}) -  \pmb{\mathfrak{B}} (\mathbf{r}, \omega) \right] e^{\imath \left(\mathbf{k} (\omega) \cdot \mathbf{r} -  \omega t\right)} ,\nonumber\\
\mathbf{B}_{ind} (\mathbf{r},t) && = \frac{\imath}{\omega} c g_{a \gamma \gamma} a_0 (\omega) \left[ \pmb {\nabla} \times \pmb{\mathfrak{B}} (\mathbf{r},\omega) \right. \\
&& \quad \quad \quad \left. + \imath \mathbf{k} (\omega) \times \pmb{\mathfrak{B}} (\mathbf{r},\omega)  \right] e^{\imath \left(\mathbf{k} (\omega) \cdot \mathbf{r} -  \omega t\right)}
\end{eqnarray} \end{subequations} 
whereas $\pmb{\mathfrak{B}} (\mathbf{r}, \omega)$ is governed by the following Helmholtz equation:
\begin{equation}\label{eqn:frakB-hem}
\left[ \pmb{\nabla}\cdot \pmb{\nabla} + \omega ^2 / c^2 \right] \pmb{\mathfrak{B}} (\mathbf{r}, \omega) = 0  .
\end{equation}
The appearance of the term $\imath \mathbf{k} (\omega) \times \pmb{\mathfrak{B}} (\mathbf{r},\omega) $ means that the signal of axion-induced magnetic field depends on the direction of axion motion. Suppose that axion moves in homogeneous space, it is necessary to average the signal over all of possible direction of axion motion. While $e^{\imath \mathbf{k}(\omega) \cdot \mathbf{r}}$ is nothing more than a phase term of axion spacetime evolution which can be set up to unit by wisely selected choice of initial phase as well as shifted origin of coordinate system, the average of the term $\imath \mathbf{k} (\omega) \times \pmb{\mathfrak{B}} (\mathbf{r},\omega)$  vanishes because of the homogeneity assumption. Hence, the axion-induced electromagnetic signal is simply
\begin{subequations}\label{eqn:eb-3-sol} \begin{eqnarray} 
&&\mathbf{E}_{ind} (\mathbf{r},t)  = c g_{a\gamma\gamma} a_0 \left( \omega \right) \left[ \mathbf{B}_0 (\mathbf{r}) -  \pmb{\mathfrak{B}} (\mathbf{r}, \omega) \right] e^{ - \imath \omega t} ,\\
&&\mathbf{B}_{ind} (\mathbf{r},t)  = \frac{\imath}{\omega} c g_{a \gamma \gamma} a_0 \left( \omega \right) \left[ \pmb {\nabla} \times \pmb{\mathfrak{B}} (\mathbf{r},\omega) \right] e^{ - \imath \omega t} .
\end{eqnarray} \end{subequations} 
In Reference \cite{Sikivie2009}, cold axion or axion-like particles may form a Bose-Einstein condensate, then the attitude of axion field of excited axion $\omega$ is much smaller than that of ground state axion $\omega _a$ i.e $|a_0 (\omega)| \ll |a_0 (\omega _a)|$. Meanwhile, only the resonance at ground state axion $\omega = \omega _a$ is considerable in haloscope experiments. Consequently, all the problems in axion haloscopes are just to find out the boundary condition for $\pmb{\mathfrak{B}} (\mathbf{r},\omega)$. To illustrate this point, we consider the case of perfect electric conductor resonant cavity as an example.

\subsection{\label{sec:3c}Solutions for perfect electric conductor resonant cavity}
In many works of seeking axion signal, the perfect electric conductor (PEC) is often used as resonant cavity surface to enhance the axion-induced electromagnetic field. Since this surface, namely $C$, is perfectly conductive, the total electric field must vanish in $C$ i.e $\mathbf{E} (\mathbf{r} \in C, t) = 0$. The DC-current is also prepared in this resonant cavity surface i.e $\mathbf{J}_e (\mathbf{r}) = \mathbf{I}_e (\mathbf{r}_C) \delta (\mathbf{r} \in C)$. Then the boundary condition of the electric field $\mathbf{E} (\mathbf{r} \in C, t) = 0$ leads us to the boundary value problem of $\pmb{\mathfrak{B}} (\mathbf{r},\omega)$
\begin{equation}\label{eqn:frakB}
\left\{ \begin{array}{c}
     \left[ \pmb{\nabla}\cdot \pmb{\nabla} + \omega ^2 / c^2 \right] \pmb{\mathfrak{B}} (\mathbf{r}, \omega) = 0 \\
     \pmb{\mathfrak{B}} (\mathbf{r}_C, \omega) = \mathbf{B}_0 (\mathbf{r}_C)
\end{array}
\right. .
\end{equation}
In long-wavelength regime of axion field, most of axions lie in ground state i.e $a_0 (\omega) \approx a_0(\omega _a) \delta (\omega-\omega_a)$, then the induced fields are simple harmonic oscillation at the frequency $\omega = \omega _a$
\begin{subequations}\label{eqn:eb-1-sol} \begin{eqnarray} 
\mathbf{E}_{ind}^{LWA} (\mathbf{r},t) = &&  c g_{a\gamma\gamma} a_0 (\omega _a) e^{-\imath \omega _a t} \nonumber\\
&& \times \left[ \mathbf{B}_0 (\mathbf{r}) -  \pmb{\mathfrak{B}} (\mathbf{r},\omega _a) \right] ,\\ 
\mathbf{B}_{ind}^{LWA} (\mathbf{r},t) = &&  \frac{\imath}{\omega _a} c g_{a \gamma \gamma} a_0 (\omega _a) e^{-\imath \omega _a t} \nonumber\\
&&\times \left[ \pmb {\nabla} \times \pmb{\mathfrak{B}} (\mathbf{r},\omega _a)\right].
\end{eqnarray} \end{subequations} 
In practice, suitable orthogonal coordinate system should be used to solve the boundary value problem \eqref{eqn:frakB}.

In PEC cavity, the electric and magnetic energy stored inside the cavity are both conserved since they are just harmonic oscillating versus time. Thus the summation and difference between electric and magnetic stored energy are also conserved in this circumstance
\begin{equation}\label{eqn:U}
U_{\pm} (\omega)= \dfrac{\varepsilon _0}{4} \int _{C} \left( \mathbf{E}_{ind} \cdot \mathbf{E}_{ind}^{*} \pm c^2 \mathbf{B}_{ind} \cdot \mathbf{B}_{ind}^{*} \right) d^3 \mathbf{r}.
\end{equation}
To characterize the stored energy induced from axion in PEC cavity, the relevant quantities to examine are the form factor of the total energy $C_{+}$, the energy difference $C_{-}$ and their ratio $\mathfrak{R}$:
\begin{equation}\label{eqn:ratioR}
C_{\pm} (\omega) = \dfrac{U_{\pm} (\omega)}{ \dfrac{g_{a \gamma\gamma}^2 a_0^2}{2 \mu _0} \int _C \mathbf{B}_0 \cdot \mathbf{B}_0 d^3 \mathbf{r} }, \quad \mathfrak{R}  (\omega) = \left| \dfrac{C_{-}  (\omega)}{C_{+}  (\omega)} \right|
\end{equation}
It is observable that the form factor $C^{+}$ can be extremely large at some certain frequencies $\omega _{c,n}$, namely cavity resonant frequencies. At these resonant frequencies of the PEC cavity, the ratio $\mathfrak{R}$ also vanishes.  It is noticed that above results are also acceptable for imperfect electric conductor if its thickness is such small that skin effect is negligible.

\subsection{\label{sec:3d} The axion-to-photon conversion power in haloscope cavity}
In realistic experiments, the haloscope cavity is not perfectly electric conductive and the oscillation of induced electromagnetic field is damped i.e the quality factor $Q_c$ of the cavity is finite. Additionally, the oscillation of axion field is not simply harmonic one because of the thermal distribution of its velocity under finite temperature above Bose-Einstein condensate combining with effects from astrophysical motions \cite{Turner1990, Hong2014, Kim2020}. Thus the attitude of axion field is approximately distributed as Cauchy-Lorentz distribution around resonant at ground state axion $\omega _a$ with the quality factor $Q_a$ which is around $10^6$ \cite{Kim2020}. Taking into account these factors, the axion-induced electromagnetic signal near resonant mode $\omega _{c,n}$ is then modified as follows:
\begin{widetext}
\begin{subequations} \begin{eqnarray} \label{eqn:eb-3-sol-nonPEC}
\mathbf{E}_{ind} (\mathbf{r},t) && = c g_{a\gamma\gamma} a_0 \left( \omega _a \right) \int \mathcal{A} (\omega, \omega _a) \left[ \mathbf{B}_0 (\mathbf{r}) - \mathcal{F} (\omega, \omega _{c,n}) \pmb{\mathfrak{B}} (\mathbf{r}, \omega) \right] e^{ - \imath \omega t} d \omega ,\nonumber\\
\mathbf{B}_{ind} (\mathbf{r},t) && = \imath \omega ^{-1} c g_{a \gamma \gamma} a_0 \left( \omega _a \right) \int \mathcal{A} (\omega, \omega _a) \mathcal{F} (\omega, \omega _{c,n})  \left[ \pmb {\nabla} \times \pmb{\mathfrak{B}} (\mathbf{r},\omega) \right] e^{ - \imath \omega t} d \omega, 
\end{eqnarray} \end{subequations} 
\end{widetext}
in which $\mathcal{A} (\omega, \omega _a)$ and $\mathcal{F} (\omega, \omega _{c,n})$ are two functions whose modules have form of Cauchy-Lorentz distribution
\begin{subequations} \begin{eqnarray}
\mathcal{A} (\omega, \omega _a) && = \sqrt{\dfrac{\omega _a}{ 2 \pi Q _a }} \dfrac{1}{\left( \omega - \omega _a \right) + \imath \frac{\omega_a}{2Q_a}}, \\
\mathcal{F} (\omega, \omega _{c,n}) && = \dfrac{1}{\sqrt{\pi} \left[ \left( \omega - \omega _{c,n}\right) + \imath \frac{\omega _{c,n}}{2Q_c} \right]}.
\end{eqnarray} \end{subequations} 

Under this circumstance, the stored electric and magnetic field are not conserved. The loss of stored energy corresponds to the conversion energy from axion to photon: $a \to \gamma + \gamma$. Hence, the axion-to-photon conversion power in the haloscope cavity $P_{a \to \gamma \gamma} (\omega _a)$ is determined as \cite{Kim2019, Kim2020}
\begin{equation}
    P_{a \to \gamma \gamma} (\omega _a) = \dfrac{\omega _c}{Q_c} U_{tot}.  
\end{equation}
with $U_{tot}$ is the average stored axion-induced electromagnetic energy inside the cavity
\begin{equation}
U_{tot} = \left\langle \dfrac{\varepsilon _0}{4} \int _{C} \left( \mathbf{E}_{ind} \cdot \mathbf{E}_{ind}^{*} \pm c^2 \mathbf{B}_{ind} \cdot \mathbf{B}_{ind}^{*} \right) d^3 \mathbf{r} \right\rangle.
\end{equation}
Near the resonant mode $\omega _a \approx \omega_{c,n}$, the factor $\left|\mathcal{F} (\omega, \omega _c)\right|^2$ is much large than unit while the ratio of the energy difference and total energy vanishes $\mathfrak{R} \to 0$. As a result, the axion-to-photon conversion power is approximately
\begin{eqnarray}\label{eqn:P1}
     && P_{a \to \gamma \gamma} (\omega _a) \approx \dfrac{ g_{a \gamma\gamma}^2 a_0^2}{\mu _0} \int _C \mathbf{B}_0 \cdot \mathbf{B}_0 d^3 \mathbf{r} \nonumber \\
     && \quad \times \frac{\omega _c}{2 Q_c} \int \left|\mathcal{A} (\omega, \omega _a)\right|^2 \left|\mathcal{F} (\omega, \omega _c)\right|^2 C_{+} ( \omega ) d\omega
\end{eqnarray} 
Noticeably, when resonance occurs, $\omega _a \approx \omega _{c,n}$ are closed, only the form factor $C_{+} ( \omega )$ at $\omega = \omega _{a}$ contributes to the axion-to-photon  conversion power. In addition, the following integral becomes Cauchy-Lorentz distribution:
\begin{eqnarray}\label{eqn:cauchy}
I (\omega _a, \omega _{c,n}) &&= \dfrac{\omega_c}{2 Q_c} \int \left|\mathcal{A} (\omega, \omega _a)\right|^2 \left|\mathcal{F} (\omega, \omega _c)\right|^2 d\omega \nonumber\\
&& = \dfrac{\frac{\omega_a}{2Q}}{\pi \left[ (\omega _a - \omega _{c,n})^2 + \frac{\omega_a^2}{4 Q^2} \right]}.
\end{eqnarray}
whereas $Q = Q_a Q_c /(Q_a + Q_c)$ is the effective quality factor which has been recently established in Reference \cite{Kim2020}. When both $Q_a, Q_c \to + \infty$, this integral coincides to delta Dirac distribution $\delta (\omega _a - \omega _{c,n})$. This suggests that the finite quality factors $Q_a, Q_c$ move the pole of form factor $C_{+} (\omega)$ from $\omega _{c,n}$ to $\omega _{c,n} \pm \imath \frac{\omega _{c,n}}{2 Q}$. Hence, a simple way to calculate the total stored energy $U_{tot}$ in imperfect electric conductive cavity is to replace the form factor $C_{+} (\omega)$ of PEC cavity by $\left| C_{+} \left(\omega (1 + \imath 2^{-1} Q^{-1}) \right) \right|$. Finally, we obtain a simple approximation for axion-to-photon conversion power in haloscope cavity
\begin{eqnarray}\label{eqn:P}
&& P_{a \to \gamma \gamma} (\omega _a) \approx \dfrac{g_{a \gamma\gamma}^2 a_0^2}{2\mu _0} \int _C \mathbf{B}_0 \cdot \mathbf{B}_0 d^3 \mathbf{r} \nonumber\\
&& \quad \quad \quad \quad \times \frac{\omega _a}{Q} \left| C_{+} \left( \omega _a \left( 1 + \frac{\imath}{2Q} \right) \right) \right|
\end{eqnarray} 
Hence, to detect the resonance of axion-to-photon conversion signal, we will focus on the spectrum of the modified form factor $ Q^{-1} \left| C_{+} \left( \omega _a \left( 1 + \frac{\imath}{2Q} \right) \right) \right|$ versus axion angular frequency $\omega_a$. Moreover, we will observe the zeros of the spectrum of the ratio $\mathfrak{R} (\omega _a)$ in the PEC cavity to point out the validity of the assumption $\mathfrak{R} \approx 0$ near resonance. We will consider the following cases: cylindrical solenoid, spherical solenoid, two-parallel-sheet cavity, toroidal solenoid with a rectangular cross-section and with a circular cross-section in the next section.

\section{\label{sec:4}Resonance of axion-to-photon conversion power in different shapes of cavity}

We aim to consider the solution of axion electromagnetic field $\mathbf{E}_{ind}, \mathbf{B}_{ind}$ as well as the conversion power $P_{a \to \gamma \gamma} (\omega _a)$ and the radio $\mathfrak{R} (\omega _a)$ for some different shapes of haloscope cavities such as a cylindrical solenoid, spherical solenoid, two-parallel-sheet cavity, toroidal solenoid with a rectangular cross-section and with a circular cross-section.

In our numerical estimation, we set up the size of the cavity for such the resonance may occurs on the first mode i.e $\omega _a \approx \omega _{c,1}$. According to many proposals such as ABRACADABRA \cite{Kahn2016}, ADMX Collaboration \cite{Du2018, Braine2020}, HAYSTAC \cite{brubaker2017a, brubaker2017b, Zhong2018}, ORGAN \cite{goryachev2017}, BEAST \cite{McAllister2018} or MADMAX Collaboration \cite{Brun2019, Knirck2019}, the range of axion mass $m_a$ is around $0.1 \mu eV$ to $100 \mu eV$ and hence, $1.5 \times 10^{8} rad/s < \omega _a < 1.5 \times 10^{11} rad/s$. For illustrative purpose, we will set first resonant frequency of the cavity $\omega _{c,1} = 1.5 \times 10^{10} rad/s$. Meanwhile, our result is for searching the axion like particle around $m_a \sim 10 \mu eV$. Based on References \cite{Hoang2017, Kim2019, Kim2020}, the quality factor of axion field $Q_a$ is around $1 \times 10^6$ while the typical value of quality factor of the cavity is around $10^4$ to $10^6$. Hence, we examine the effective $Q$ in the range from $10^4$ to $10^5$.

\begin{figure}[H]
   \centering
   \includegraphics[width = 0.45 \textwidth]{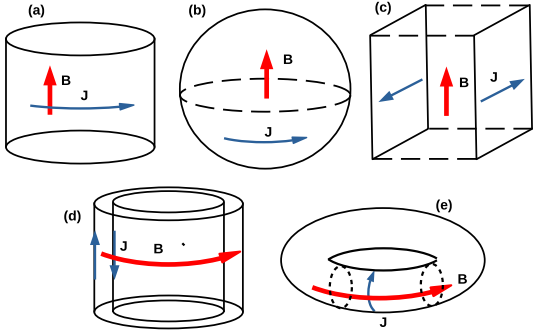}
   \caption{Different shapes of cavity considered in this works: (a) cylindrical solenoid, (b) spherical solenoid, (c) two-parallel-sheet cavity, (d) toroidal solenoid with rectangular cross section and (e) with circular cross section. The red arrow is the magnetic field and the blue arrow is the surface current.}
  \label{figcavity}
\end{figure}

\subsection{\label{sec:4a}Infinite cylindrical solenoid}
In case of an infinite cylindrical solenoid whose radius is $R$, the surface DC-current and its applied magnetic field are given in cylindrical coordinates $(\rho, z, \phi)$ as
\begin{eqnarray}
\mathbf{J}_e (\mathbf{r}) =  \dfrac{B_0}{\mu _0} \delta ( \rho - R ) \pmb{\hat{\phi}} ,\quad
\mathbf{B}_0 (\mathbf{r}) =  B_0 \theta(R - \rho) \hat{\mathbf{z}} \label{eqn:case1},
\end{eqnarray}
with $\delta$ and $\theta$ are notations of Dirac delta and Heaviside theta functions. Replacing $\mathbf{B}_0$ of Equation \eqref{eqn:case1} into Equation \eqref{eqn:frakB} and solving this boundary value problem in cylindrical coordinates $(\rho, z, \phi)$ give us 
\begin{subequations} \begin{eqnarray}\label{eqn:frakB-case1}
\pmb{\mathfrak{B}} (\rho, \omega) = B_0 \dfrac{J_0 (\rho \omega / c) }{J_0 (R \omega / c)} \hat{\mathbf{z}} ,
\end{eqnarray} \end{subequations} 
with $J_n(x)$ denotes for Bessel J function. Consequently, the axion-induced fields are explicitly determined
\begin{subequations} \begin{eqnarray}
\mathbf{E}_{ind} (\rho,t) && = c g_{a\gamma\gamma} a_0 B_0  e^{ - \imath \omega _a t} \nonumber\\
&& \times \left[ \theta ( R-\rho ) - \dfrac{J_0 (\rho \omega _a / c) }{J_0 (R \omega _a / c)} \right] \hat{\mathbf{z}} , \label{eqn:E1-res-case1}\\
\mathbf{B}_{ind} (\rho, t) && =  \imath g_{a\gamma\gamma} a_0 B_0  \dfrac{J_1 (\rho \omega _a / c) }{J_0 (R \omega _a / c)} e^{ - \imath \omega _a t} \pmb{\hat{\phi}}, \label{eqn:B1-res-case1}
\end{eqnarray} \end{subequations} 
as the same as results in Reference \cite{Kim2019}. Since the explicit expression of axion-induced electromagnetic field is known, it is easy to determine both total/difference EM form factor $C_{\pm}$ and their ratio $\mathfrak{R}$ 
\begin{subequations} \begin{eqnarray}
C_{+} && = \dfrac{J_1^2 (R \omega _a / c) }{J_0^2 (R \omega _a / c)} - \dfrac{3 J_2 (R \omega _a / c) }{2 J_0 (R \omega _a / c)}  \label{eqn:total-case1}\\
C_{-} && = - \dfrac{J_2 (R \omega _a / c) }{2 J_0 (R \omega _a / c)} \label{eqn:diff-case1}\\
\mathfrak{R} && = \dfrac{1}{\left| 3 - \dfrac{2 J_1^2 (R \omega _a / c) }{J_0 (R \omega _a / c) J_2 (R \omega _a / c)} \right|}. \label{eqn:R-case1}
\end{eqnarray} \end{subequations} 
As can be seen from Equation \eqref{eqn:total-case1}, the form factor $C_{+} (\omega_a )$ is resonant at $\omega _{c,n} =  \chi _n c / R$, whereas $\chi _n$ is $n^{th}$-zero of $J_0 (\chi)$ function e.g $\chi _n = 2.40483, \; 5.52008, \; 8.65373, \; 11.7915, \; 14.9309 \ldots$ for $n = 1,2,3,4,5\ldots$. 

\begin{figure}[H]
    \centering
    \includegraphics[width = 0.45 \textwidth]{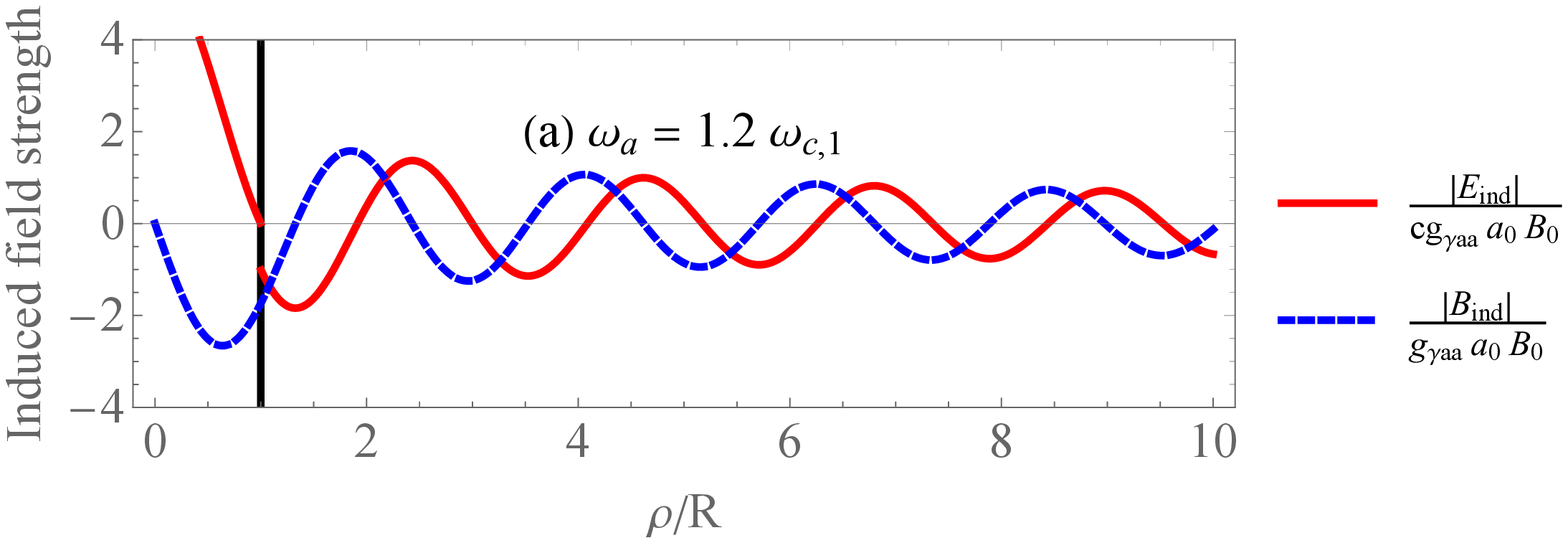}
    \includegraphics[width = 0.45 \textwidth]{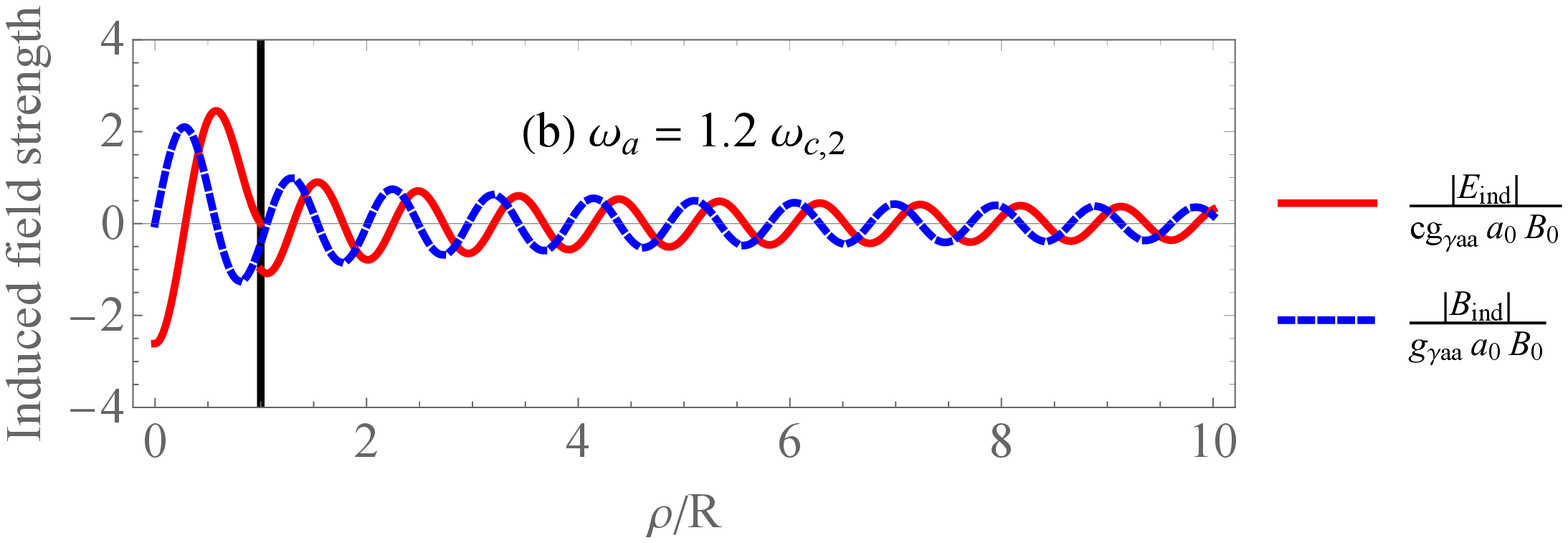}
    \includegraphics[width = 0.45 \textwidth]{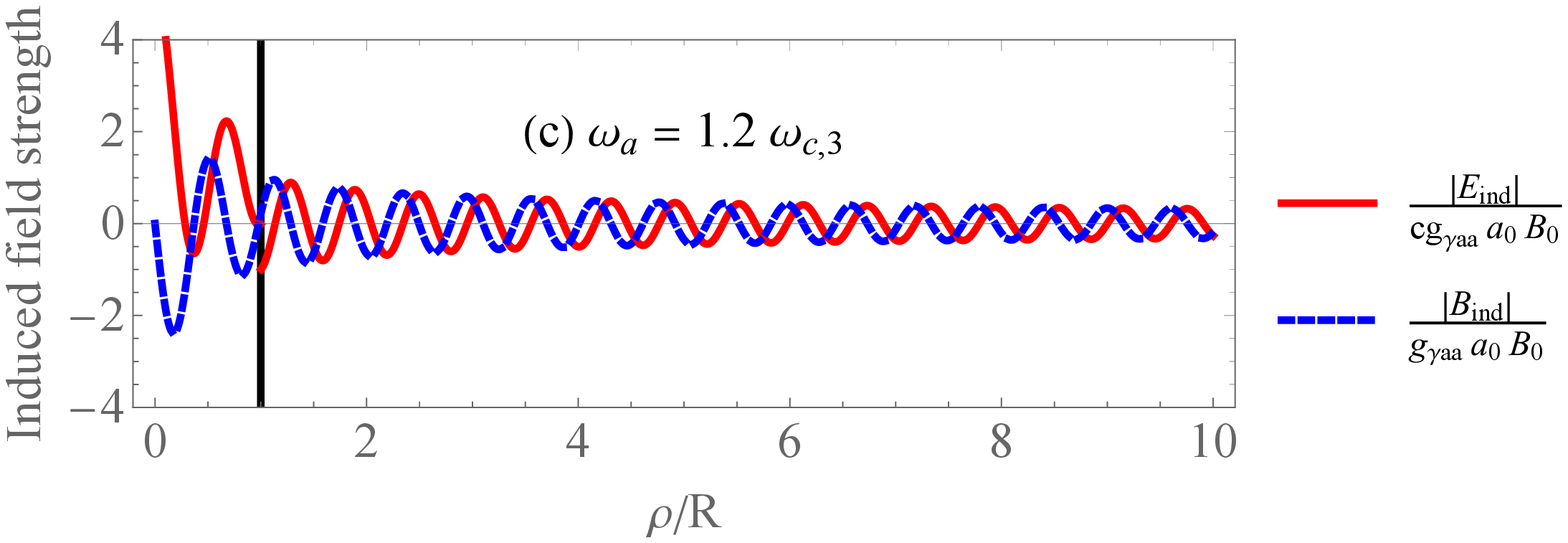}
    \caption{(a,b,c) The axion-induced electromagnetic fields strength $E_{ind}, B_{ind}$ near first three resonant modes $\omega _a = 1.2 \omega _{c,1}, 1.2 \omega _{c,2}, 1.2 \omega _{c,3}$ vary in space respectively.}
    \label{fig:case1EM}
\end{figure}

\begin{figure}[H]
    \centering
    \includegraphics[width = 0.45 \textwidth]{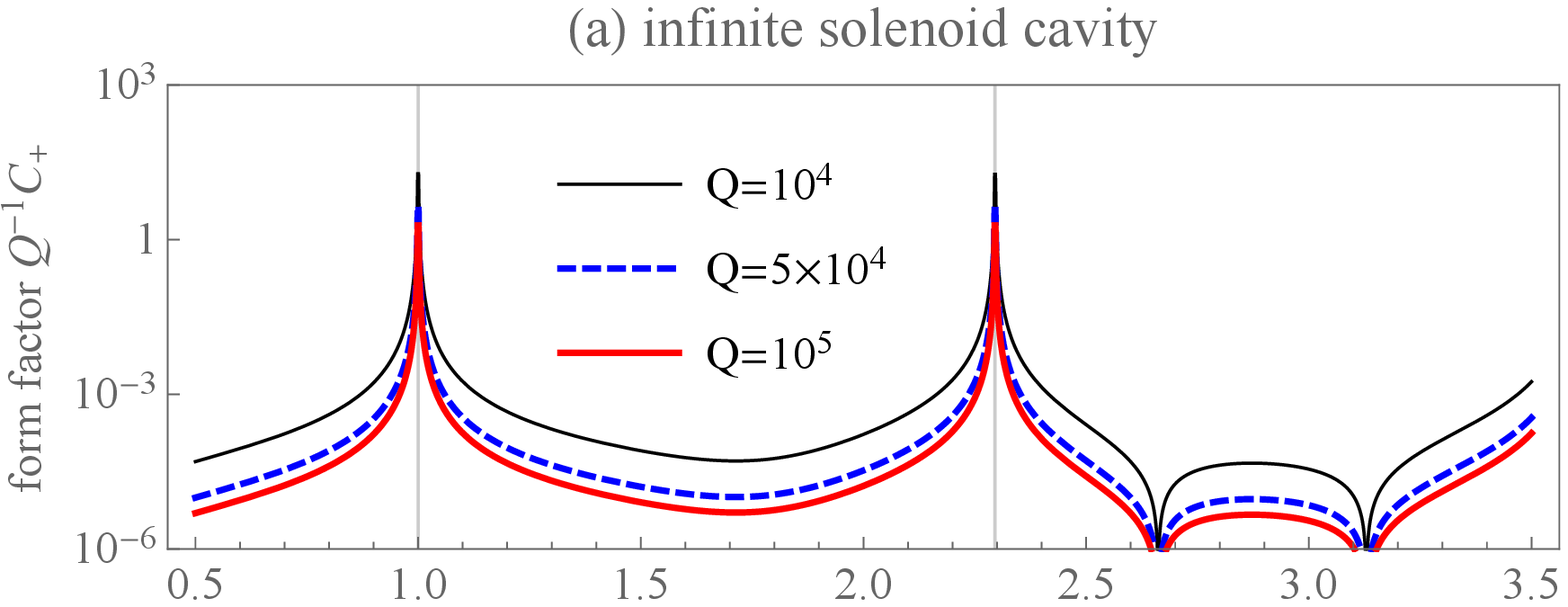}
    \includegraphics[width = 0.45 \textwidth]{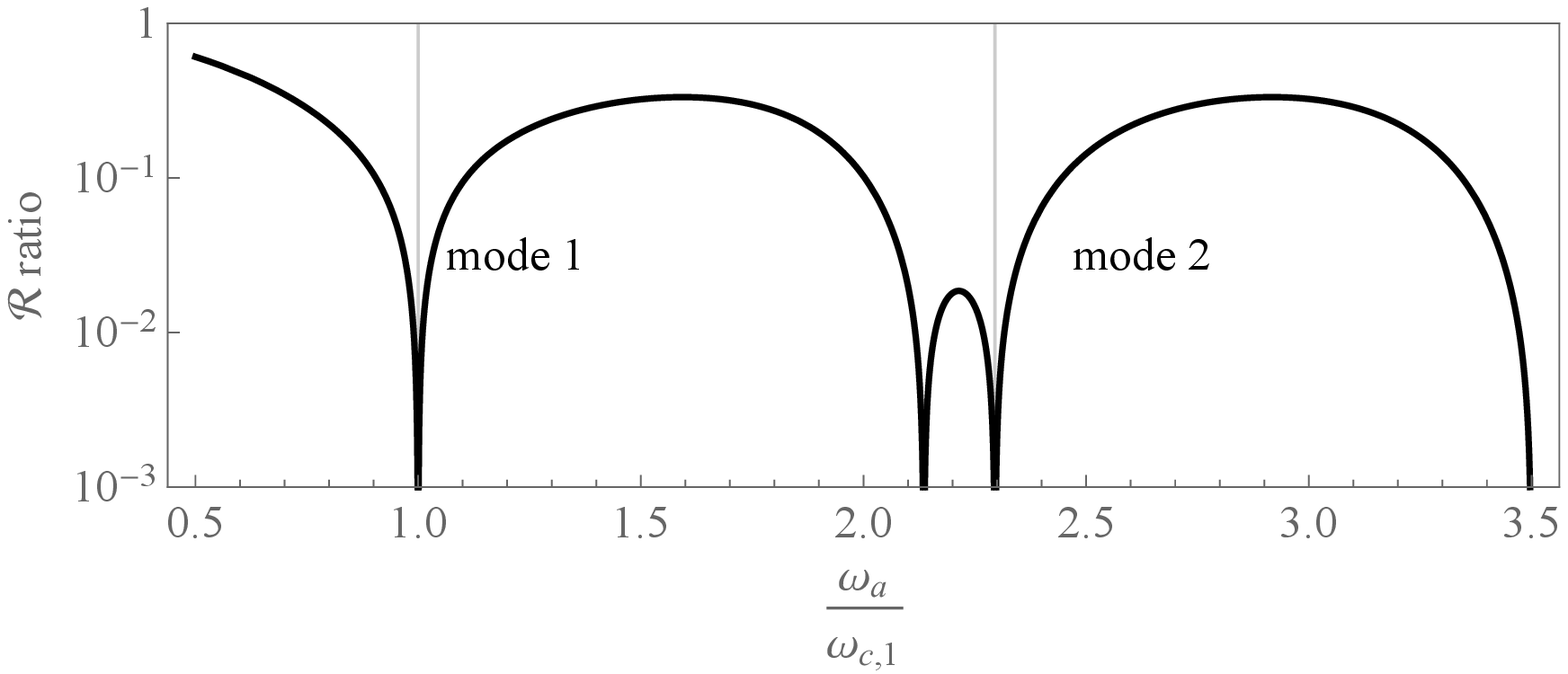}
    \caption{Profile of the modified form factor of imperfect electric conductive cavity $\left| Q^{-1} C_{+} \left(\omega_a \left(1 + \imath 2^{-1} Q^{-1} \right) \right) \right|$ as well as the ratio $\mathfrak{R} (\omega _a)$ versus frequency ratio $\omega_a / \omega _{c,1}$ for infinite cylindrical solenoid cavity. Black, blue and red curves are for effective $Q$-factor are $10^4$, $5\times 10^4$, $10^5$ respectively. The higher quality factor is, the sharper resonant peaks are.}
    \label{fig:case1}
\end{figure}

We choose the solenoid radius $R = 4.80965 \text{ cm}$ so that the axion signal is resonant at $m_a \sim 10 \mu eV$. Figure \ref{fig:case1EM} shows axion-induced electromagnetic fields strength $E_{ind}, B_{ind}$ near first three resonant modes varying in space while Figure \ref{fig:case1EMstore} indicates the distribution of the axion-induced electromagnetic stored energy density. Figure \ref{fig:case1} illustrates the modified form factor of imperfect electric conductive cavity $\left| Q^{-1} C_{+} \left(\omega_a \left(1 + \imath 2^{-1} Q^{-1} \right) \right) \right|$ as well as the ratio $\mathfrak{R} (\omega _a)$ versus $\omega_a / \omega _{c,1}$.

\subsection{\label{sec:4d}Spherical solenoid}
If the resonant cavity is a spherical superconductor whose radius is $R$, the spherical coordinates $(r, \theta, \varphi)$ should be used. The spherical surface DC-current and its corresponding magnetic field \emph{inside the cavity} are:
\begin{eqnarray}\label{eqn:case4}
\left\{
\begin{array}{l}
\mathbf{J}_e (\mathbf{r}) = \dfrac{3 B_0 \sin \theta}{2 \mu _0} \delta (r-R) \left[ \hat{\mathbf{y}} \cos \varphi - \hat{\mathbf{x}} \sin \varphi \right],\\
\mathbf{B}_0 (\mathbf{r}) = B_0 \hat{\mathbf{z}} \text{ for } r < R.
\end{array}
 \right. .
\end{eqnarray} 
According to the procedure in Equation \eqref{eqn:frakB} and Equation \eqref{eqn:eb-1-sol}, we obtain inside the cavity $r < R$ that
\begin{subequations} \begin{eqnarray}
\pmb{\mathfrak{B}} (r, \omega) && = B_0 \dfrac{R}{r} \dfrac{\sin \left( r \omega /c \right)}{\sin \left( R \omega /c \right)} \hat{\mathbf{z}}, \label{eqn:frakB-case4} \\
\mathbf{E}_{ind} (r, t) && = c g_{a\gamma\gamma} a_0 B_0 \left[ 1 - \dfrac{R}{r} \dfrac{\sin \left( r \omega /c \right)}{\sin \left( R \omega /c \right)} \right] e^{-\imath \omega_a t} , \label{eqn:E1-res-case4} \\
\mathbf{B}_{ind} (r, t) && = \imath g_{a\gamma\gamma} a_0 B_0 \dfrac{R}{r} \dfrac{\sin \left( r \omega /c \right)}{\sin \left( R \omega /c \right)} e^{-\imath \omega_a t} \times \nonumber\\
&& \times \left[ \dfrac{c}{r \omega _a} - \cot \left( \dfrac{r \omega_a}{c} \right) \right] \left[ \hat{\mathbf{y}} \cos \varphi - \hat{\mathbf{x}} \sin \varphi \right].  \label{eqn:B1-res-case4} 
\end{eqnarray} \end{subequations} 
It is noticed that in this case, the applied field as well as the axion-induced field do not vanish outside the cavity. Only the the one stored inside the cavity is significantly considered. Hence, the total/difference EM form factor and their ratio inside the cavity are
\begin{subequations} \begin{eqnarray}
C_{+} && = 2 + \dfrac{3}{2} \left[ \cot \left( \dfrac{R \omega _a}{c} \right) + \dfrac{c}{R \omega _a} \right]^2 - \dfrac{6 c^2}{R^2 \omega _a^2} \label{eqn:total-case4}\\
C_{-} && = \dfrac{1}{2} + \dfrac{3 c}{2 R \omega _a} \left( \cot \left( \dfrac{R \omega _a}{c} \right) - \dfrac{c}{R \omega _a}  \right) \label{eqn:diff-case4}\\
\mathfrak{R} && = \left| \dfrac{1}{2} + \dfrac{3 c}{2 R \omega _a} \left( \cot \left( \dfrac{R \omega _a}{c} \right) - \dfrac{c}{R \omega _a}  \right) \right| \times \nonumber\\
&& \times \left|  2 + \dfrac{3}{2} \left[ \cot \left( \dfrac{R \omega _a}{c} \right) + \dfrac{c}{R \omega _a} \right]^2 - \dfrac{6 c^2}{R^2 \omega _a^2} \right|^{-1}. \label{eqn:R-case4}
\end{eqnarray} \end{subequations} 
The resonance of the axion-photon signal occurs when $cotan$ in the form factor $C_{+}$ is infinity, it means that the axion frequency approaches $\omega_{c,n} = n \pi c /R$ for $n=1,2,3, \ldots$.

\begin{figure}[H]
    \centering
    \includegraphics[width = 0.45 \textwidth]{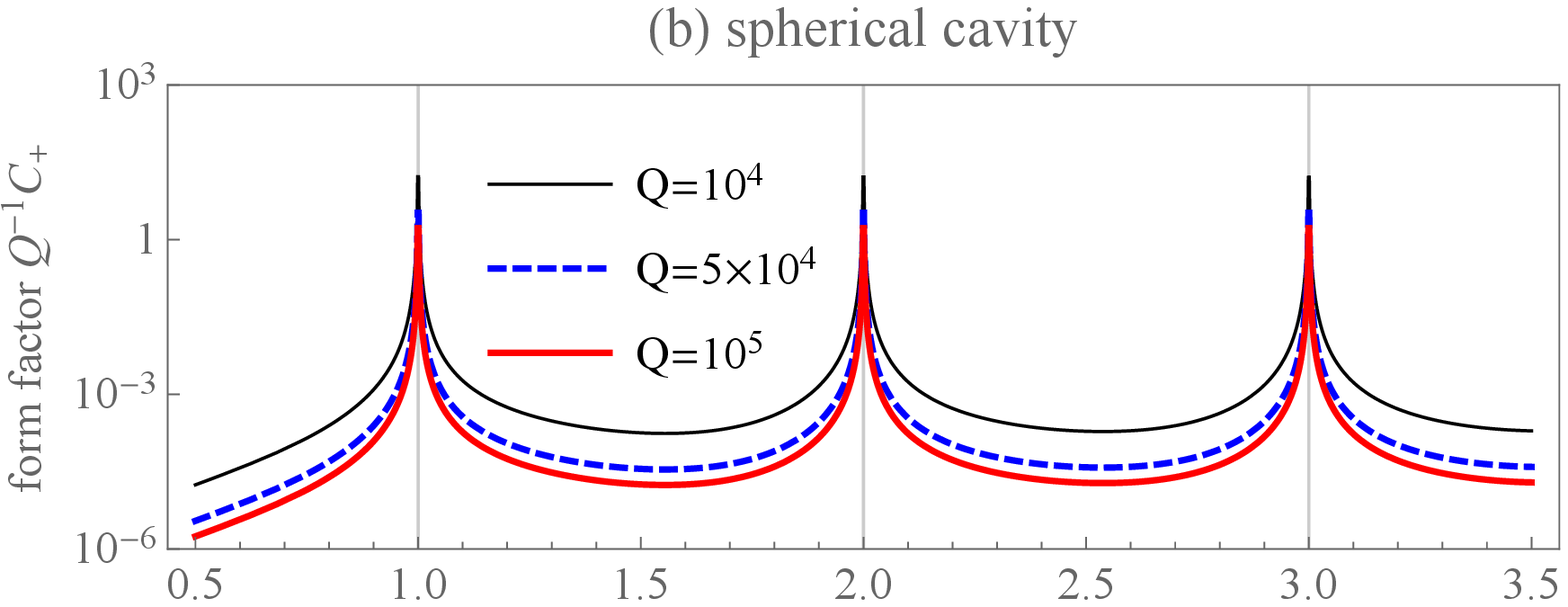}
    \includegraphics[width = 0.45 \textwidth]{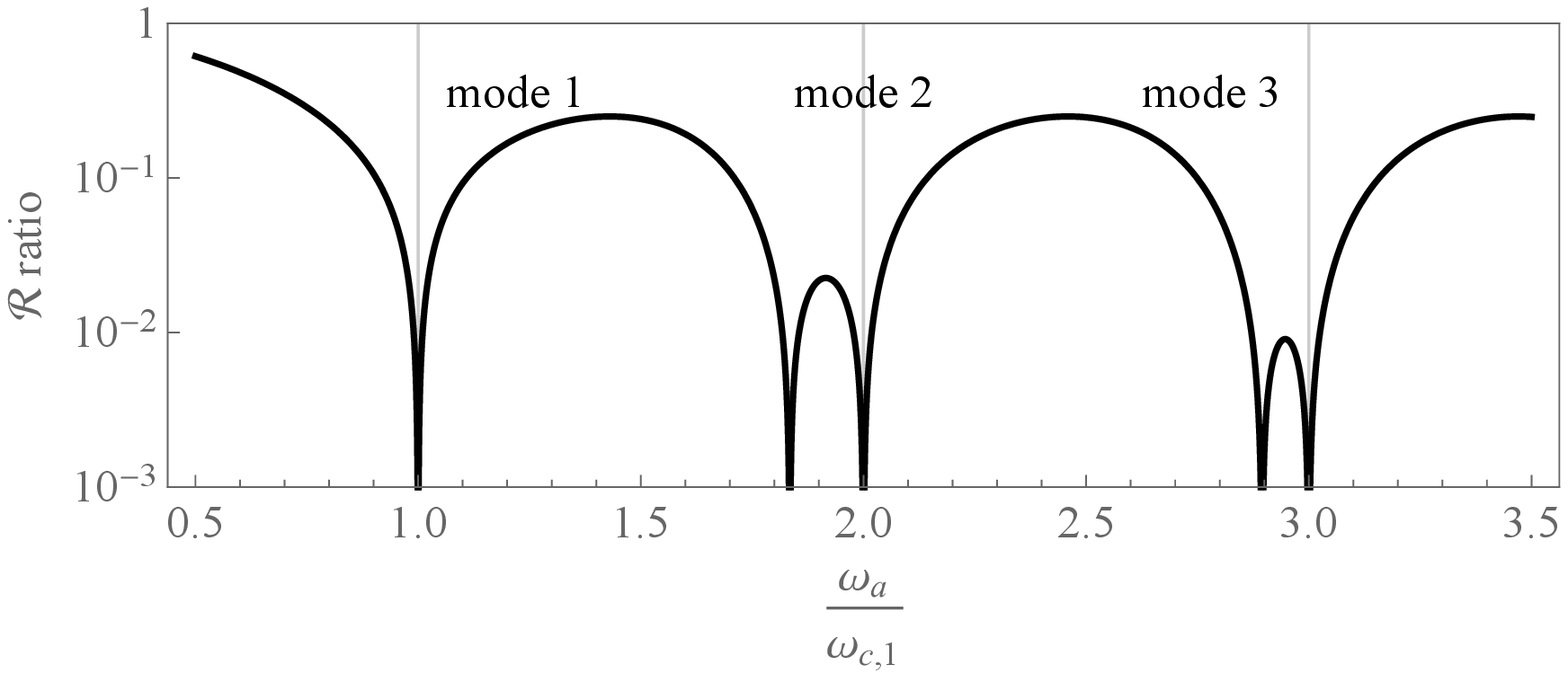}
    \caption{Profile of the modified form factor of imperfect electric conductive cavity $\left| Q^{-1} C_{+} \left(\omega_a \left(1 + \imath 2^{-1} Q^{-1} \right) \right) \right|$ as well as the ratio $\mathfrak{R} (\omega _a)$ versus frequency ratio $\omega_a / \omega _{c,1}$ for spherical cavity. Black, blue and red curves are for effective $Q$-factor are $10^4$, $5\times 10^4$, $10^5$ respectively. The higher quality factor is, the sharper resonant peaks are.}
    \label{fig:case4}
\end{figure}

To enhance the axion signal at $m_a \sim 10 \mu eV$, we set the radius $R = 6.28319 \text{ cm}$. The profile of the modified form factor of imperfect electric conductive cavity $\left| Q^{-1} C_{+} \left(\omega_a \left(1 + \imath 2^{-1} Q^{-1} \right) \right) \right|$ as well as the ratio $\mathfrak{R} (\omega _a)$ versus $\omega_a / \omega _{c,1}$ are shown in Figure \ref{fig:case4}.

\begin{figure}[H]
    \centering
    \includegraphics[width = 0.45 \textwidth]{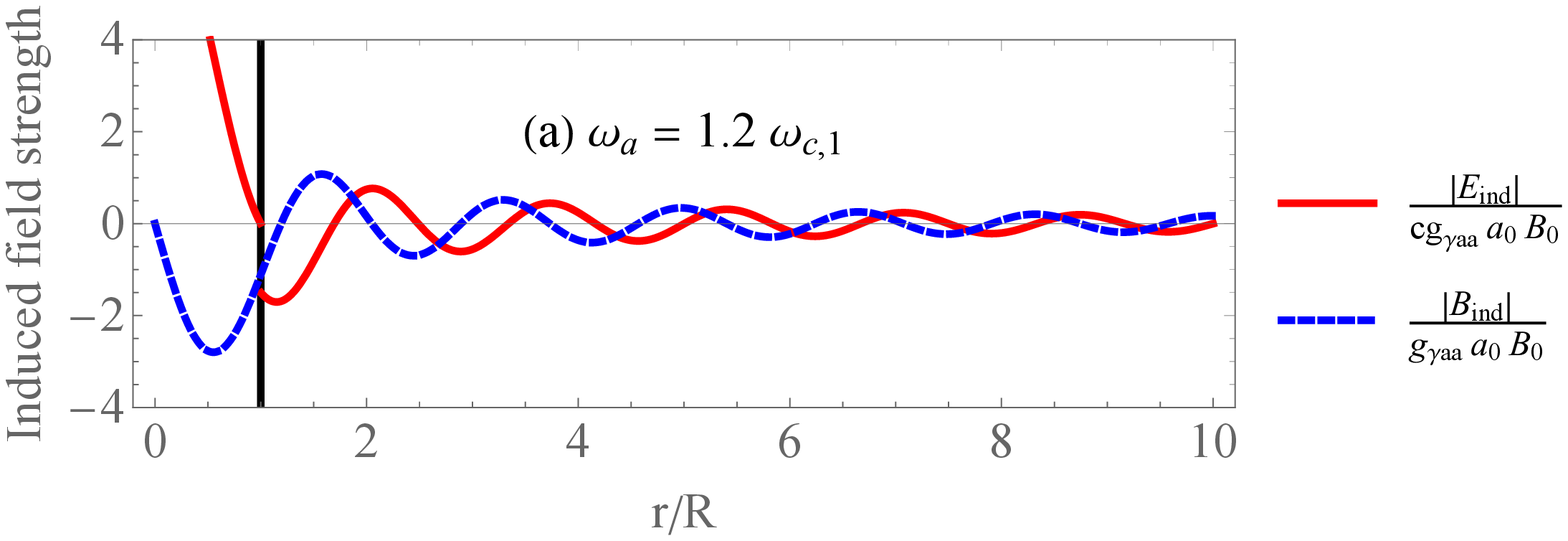}
    \includegraphics[width = 0.45 \textwidth]{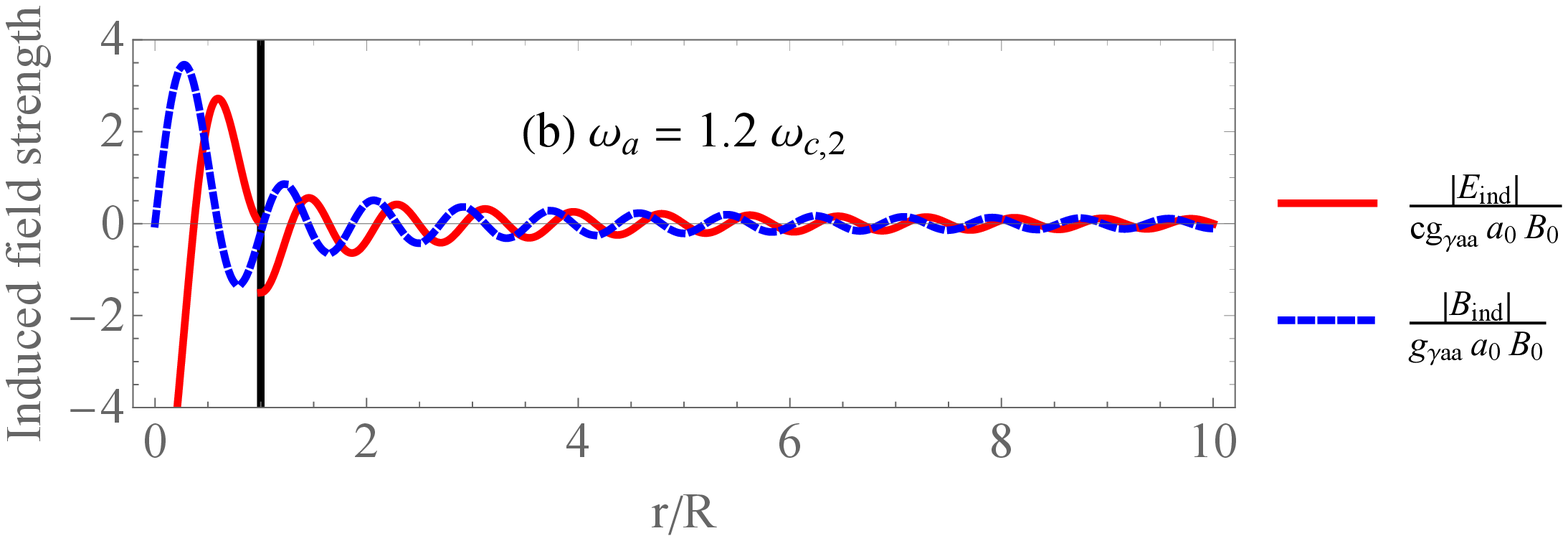}
    \includegraphics[width = 0.45 \textwidth]{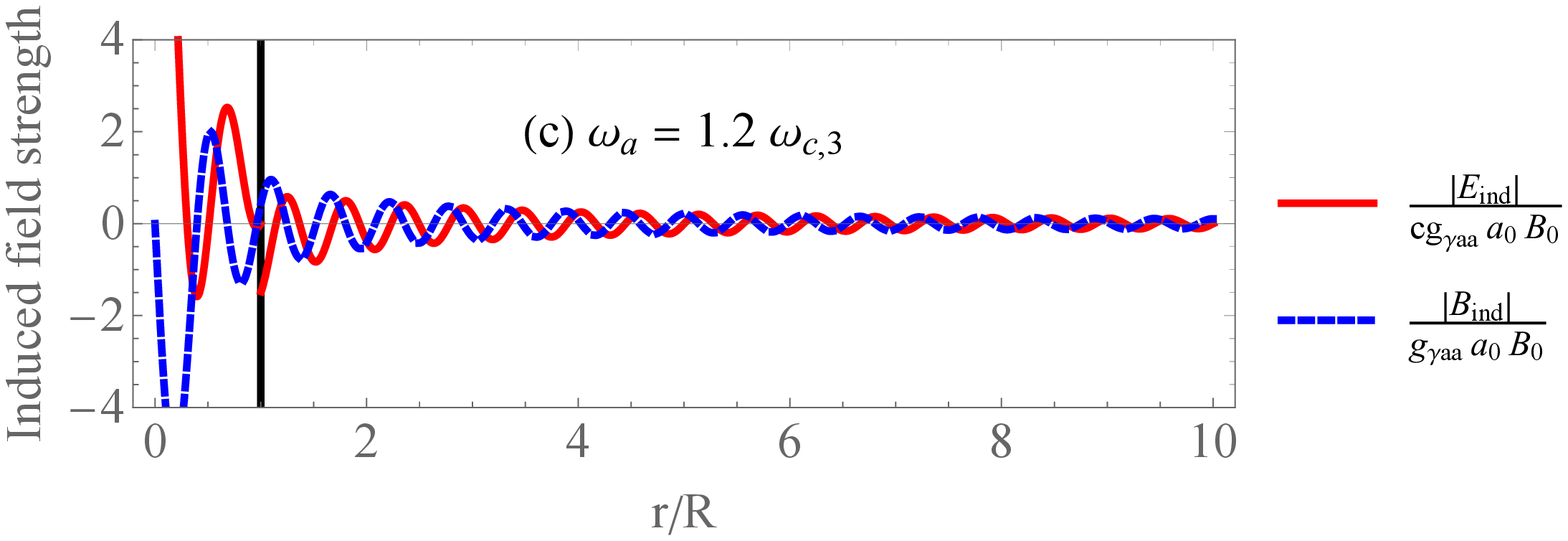}
    \caption{(a,b,c) The axion-induced electromagnetic fields strength $E_{ind}, B_{ind}$ near first three resonant modes $\omega _a = 1.2 \omega _{c,1}, 1.2 \omega _{c,2}, 1.2 \omega _{c,3}$ vary in space respectively.}
    \label{fig:case4EM}
\end{figure}

Figure \ref{fig:case4EM} shows axion-induced electromagnetic fields strength $E_{ind}, B_{ind}$ near first three resonant modes varying in space while Figure \ref{fig:case4EMstore} indicates the distribution of the axion-induced electromagnetic stored energy density at the equator of the spherical cavity.

\subsection{\label{sec:4e}Infinite two-parallel-sheet cavity}
The infinite two-parallel-sheet cavity consists two infinite plates which are placed parallel to each other and distances by $2R$. In this cavity, the surface DC-current and its applied magnetic field are given in Cartesian coordinates $(x,y,z)$ as
\begin{eqnarray}
\left\{
\begin{array}{l}
\mathbf{J}_e (\mathbf{r}) =  \dfrac{2 B_0}{\mu _0} \left[ \delta ( x - R ) - \delta ( x + R ) \right] \hat{\mathbf{y}}, \\
\mathbf{B}_0 (\mathbf{r})  =  B_0 \left[ \theta \left( x -R \right) - \theta \left( x + R \right) \right] \hat{\mathbf{z}}.
\end{array}
\right. \label{eqn:case5} .
\end{eqnarray} 
Replacing $\mathbf{B}_0$ of Equation \eqref{eqn:case5} into Equation \eqref{eqn:frakB} followed by solving this boundary value problem in Cartesian coordinates $(x, y, z)$ leads us to the exact expression of axion-induced electric and magnetic fields inside the cavity $|x| < R$
\begin{subequations} \begin{eqnarray}
\pmb{\mathfrak{B}} (\rho, \omega) && = B_0 \dfrac{\cos (x \omega / c) }{\cos (R \omega / c)} \hat{\mathbf{z}} , \label{eqn:frakB-case5} \\
\mathbf{E}_{ind} (\rho,t) && = c g_{a\gamma\gamma} a_0 B_0 \left[ 1 - \dfrac{\cos (x \omega / c) }{\cos (R \omega / c)} \right] e^{ - \imath \omega _a t} \hat{\mathbf{z}} . \label{eqn:E1-res-case5}\\
\mathbf{B}_{ind} (\rho, t) && =  \imath g_{a\gamma\gamma} a_0 B_0  \dfrac{\sin (x \omega / c) }{\cos (R \omega / c)}  e^{ - \imath \omega _a t} \hat{\mathbf{y}}, \label{eqn:B1-res-case5}
\end{eqnarray} \end{subequations} 
As long as the explicit expression of axion-induced electromagnetic field is known, both total/difference EM form factor $C_{\pm}$ and their ratio $\mathfrak{R}$ are known 
\begin{subequations} \begin{eqnarray}
C_{+} && = 1 - \dfrac{\tan \left( R \omega _a / c \right) }{R \omega _a / c} + \dfrac{\tan ^2 \left( R \omega _a / c \right)}{2} \label{eqn:total-case5}\\
C_{-} && = \dfrac{1}{2} \left[ 1 - \dfrac{\tan \left( R \omega _a / c \right) }{R \omega _a / c} \right] \label{eqn:diff-case5}\\
\mathfrak{R} && = \dfrac{1}{2} \left| \dfrac{ 1 - \dfrac{\tan \left( R \omega _a / c \right) }{R \omega _a / c}  }{ 1 - \dfrac{\tan \left( R \omega _a / c \right) }{R \omega _a / c} + \dfrac{\tan ^2 \left( R \omega _a / c \right)}{2} } \right|. \label{eqn:R-case5}
\end{eqnarray} \end{subequations} 
When $\omega _a \to \omega _{c,n} = (n-1/2) \pi c / R \; (n = 1,2, \ldots) $, the form factor $C_{+}$ goes to infinity and the ratio $\mathfrak{R}$ vanishes in PEC cavity. Therefore, the axion-to-photon conversion power for imperfect electric conductive cavity with the same shape is resonant at $\omega _{c,n} = (n-1/2) \pi c / R$.

\begin{figure}[H]
    \centering
    \includegraphics[width = 0.45 \textwidth]{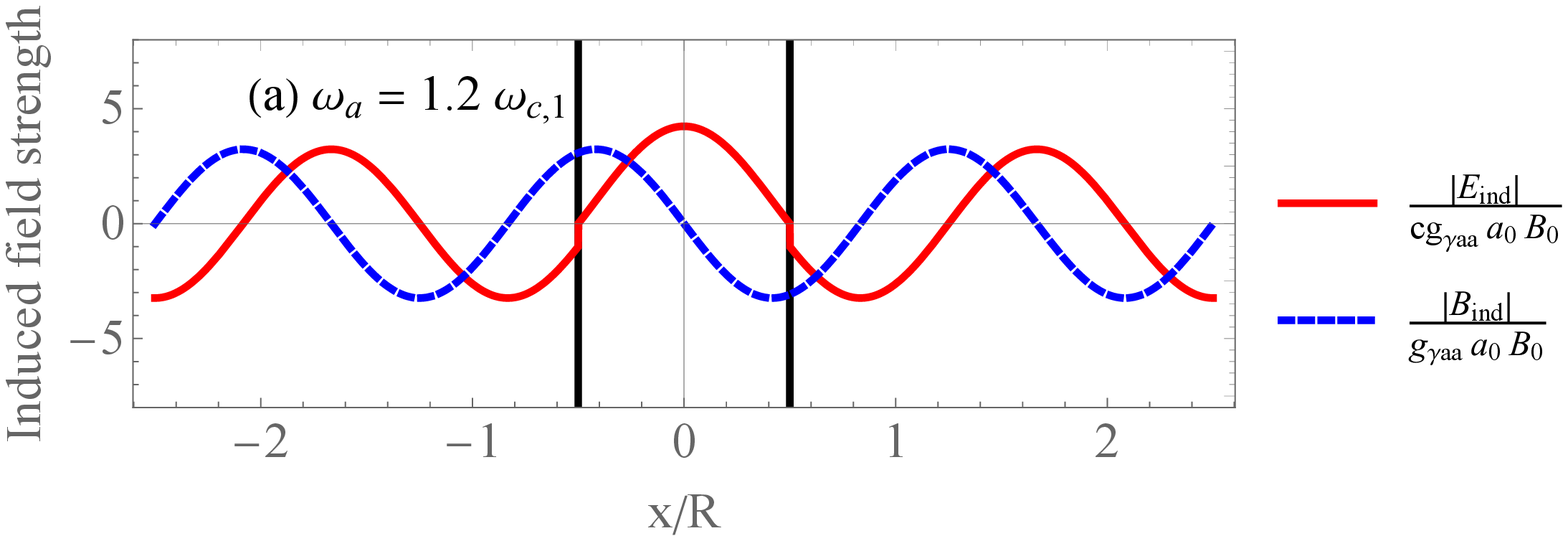}
    \includegraphics[width = 0.45 \textwidth]{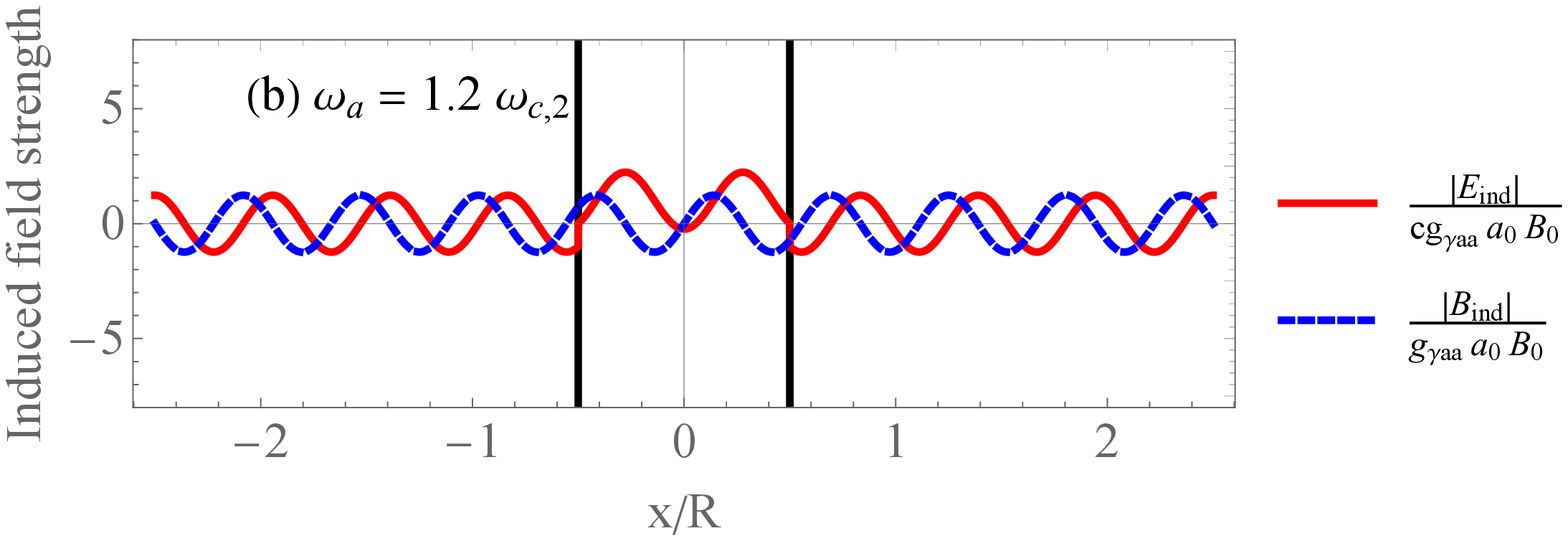}
    \includegraphics[width = 0.45 \textwidth]{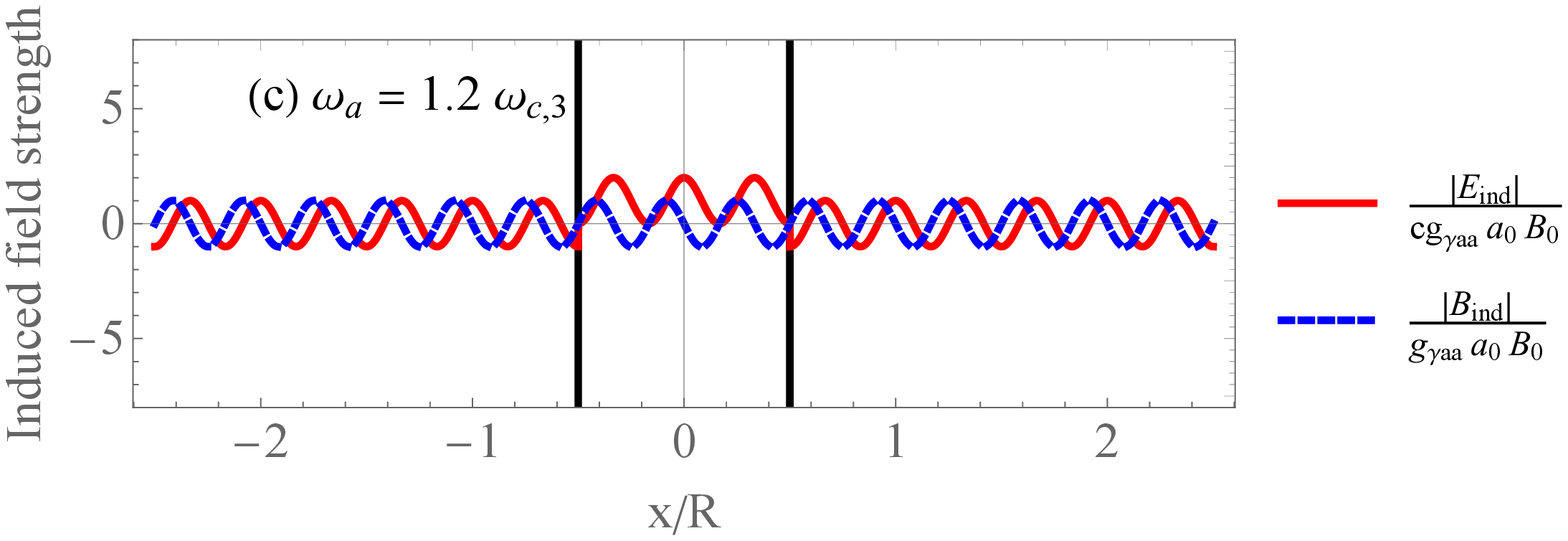}
    \caption{(a,b,c) The axion-induced electromagnetic fields strength $E_{ind}, B_{ind}$ near first three resonant modes $\omega _a = 1.2 \omega _{c,1}, 1.2 \omega _{c,2}, 1.2 \omega _{c,3}$ vary in space respectively.}
    \label{fig:case5EM}
\end{figure}

\begin{figure}[H]
    \centering
    \includegraphics[width = 0.45 \textwidth]{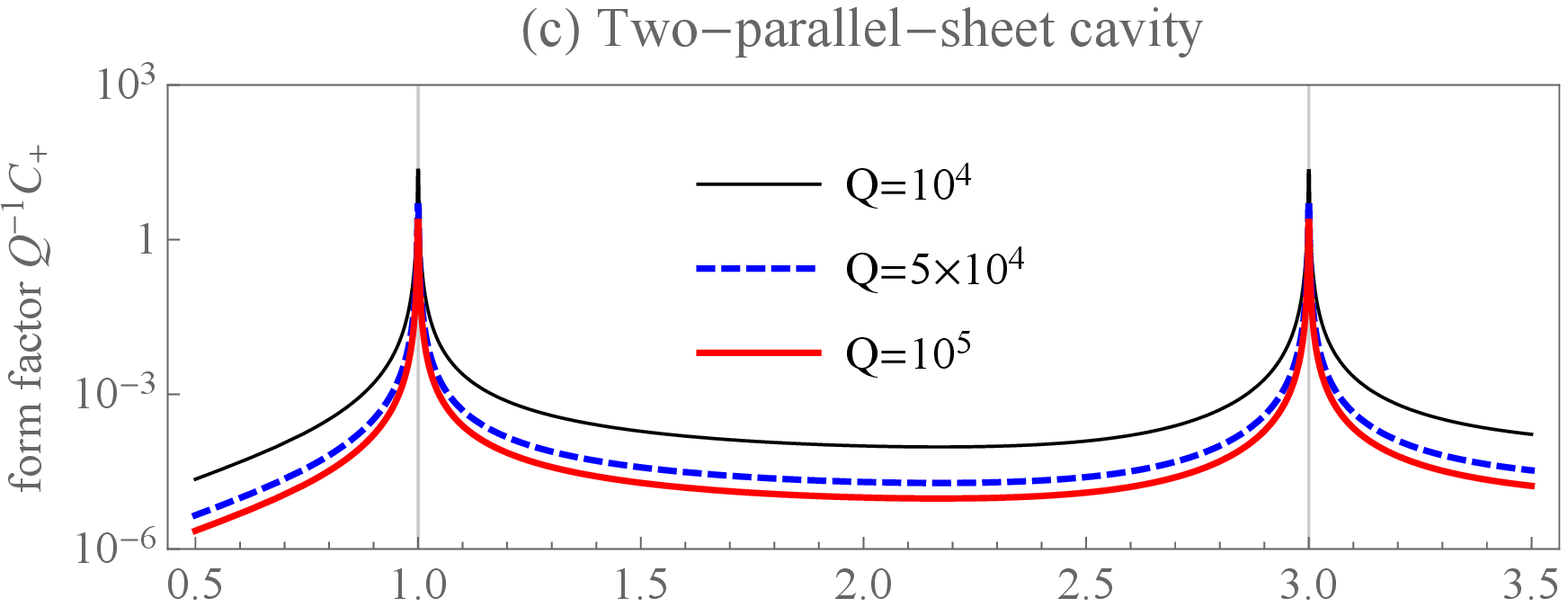}
    \includegraphics[width = 0.45 \textwidth]{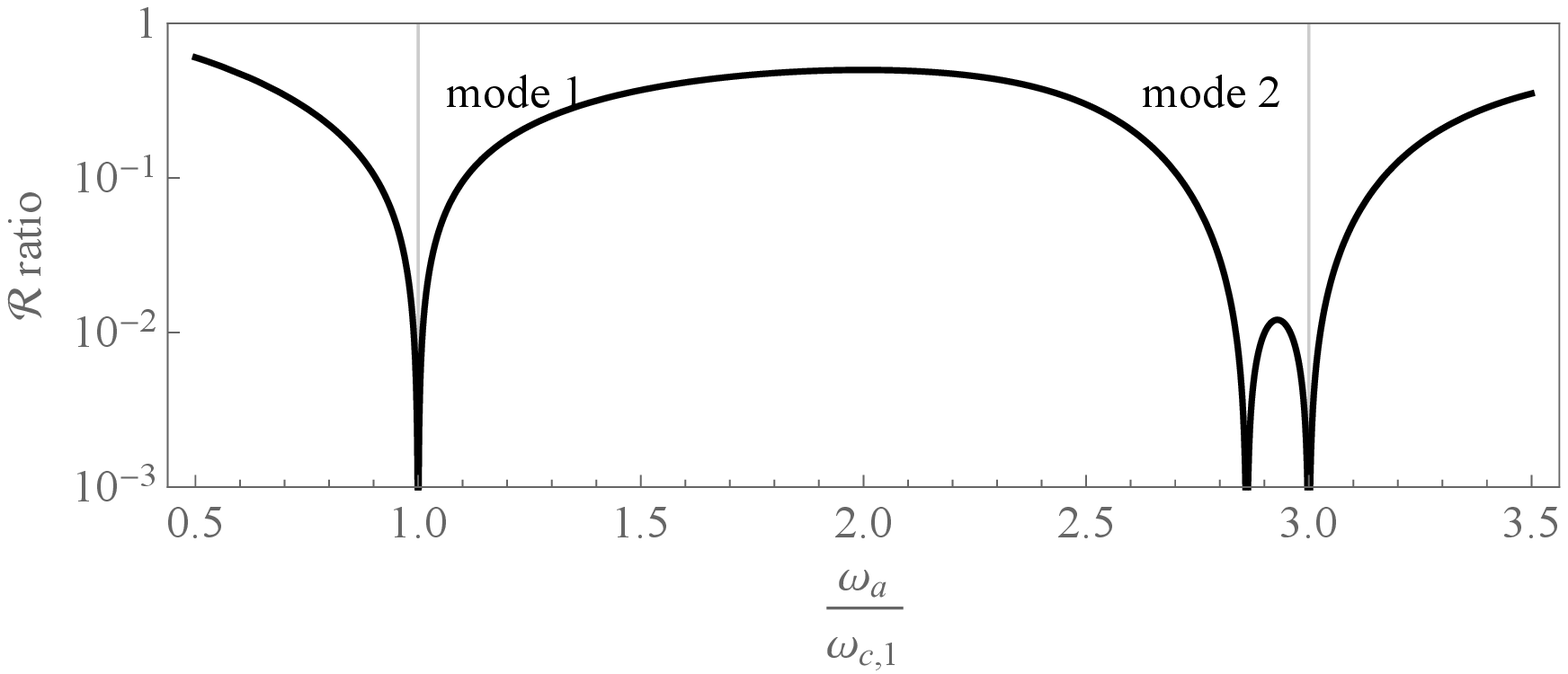}
    \caption{Profile of the modified form factor of imperfect electric conductive cavity $\left| Q^{-1} C_{+} \left(\omega_a \left(1 + \imath 2^{-1} Q^{-1} \right) \right) \right|$ as well as the ratio $\mathfrak{R} (\omega _a)$ versus frequency ratio $\omega_a / \omega _{c,1}$ for two-parallel-sheet cavity. Black, blue and red curves are for effective $Q$-factor are $10^4$, $5\times 10^4$, $10^5$ respectively.}
    \label{fig:case5}
\end{figure}

The gap between the sheets is set at $R = 3.141595 \text{ cm}$ so that the axion signal is resonant at $m_a \sim 10 \mu eV$. Figure \ref{fig:case5EM} shows axion-induced electromagnetic fields strength $E_{ind}, B_{ind}$ near first three resonant modes varying in space while Figure \ref{fig:case5EMstore} indicates the distribution of the axion-induced electromagnetic stored energy density. Figure \ref{fig:case5} shows the dependence of the modified form factor of imperfect electric conductive cavity $\left| Q^{-1} C_{+} \left(\omega_a \left(1 + \imath 2^{-1} Q^{-1} \right) \right) \right|$ and the ratio $\mathfrak{R} (\omega _a)$ on $\omega_a / \omega _{c,1}$.

\subsection{\label{sec:4b}Toroidal solenoid with rectangular cross section}
Considering the rectangular cross-section toroidal cavity with half width $R$, height $h$ and average radius $D$, the cavity surfaces in cylindrical coordinate are $\rho = D \pm R$ and $z = \pm h/2$. The applied magnetic field exists only inside the cavity
\begin{equation}\label{eqn:case2}
\mathbf{B}_0 \left( \mathbf{r} \right) = \dfrac{B_0 D}{\rho} \left[ \hat{\mathbf{y}} \cos \phi - \hat{\mathbf{x}} \sin \phi \right]
\end{equation} 
when $|\rho - D| \leq R$ and $|z| \leq h/2$. Solving the boundary value problem \eqref{eqn:frakB} inside the cavity gives
\begin{widetext}
\begin{subequations} \begin{eqnarray}
\pmb{\mathfrak{B}}(\rho, \omega) && = B_0 \left[\alpha J_1 \left( \rho \omega / c \right) + \beta Y_1 \left( \rho \omega  / c \right) \right] \left[ \hat{\mathbf{y}} \cos \phi - \hat{\mathbf{x}} \sin \phi \right], \label{eqn:frakB-case2} \\
\mathbf{E}_{ind} (\rho, \omega) && = c g_{a\gamma\gamma} a_0 B_0 \left[ \dfrac{D}{\rho} - \alpha J_1 (\rho \omega /c) - \beta Y_1 (\rho \omega /c) \right] e^{-\imath \omega t} \left[ \hat{\mathbf{y}} \cos \phi - \hat{\mathbf{x}} \sin \phi \right] \label{eqn:E1-res-case2}\\
\mathbf{B}_{ind} (\rho, \omega) && = \imath g_{a\gamma\gamma} a_0 B_0 \left[ \alpha J_0 (\rho \omega /c) + \beta Y_0 (\rho \omega /c)\right] e^{-\imath \omega t} \hat{\mathbf{z}} \label{eqn:B1-res-case2}
\end{eqnarray} \end{subequations} 
in which
\begin{eqnarray*}
\begin{pmatrix}
\alpha \\
\beta
\end{pmatrix} && =\dfrac{D/(D^2 - R^2)}{ J_1 ((D-R)\omega / c) Y_1 ((D+R)\omega / c) - J_1 ((D+R)\omega / c) Y_1 ((D-R)\omega / c) } \times \\
&& \begin{pmatrix}
(D+R) Y_1 ((D+R)\omega / c) - (D-R) Y_1 ((D-R)\omega / c)  \\
(D-R) J_1 ((D-R)\omega / c) - (D+R) J_1 ((D+R)\omega / c)
\end{pmatrix}, \quad \quad Y_n (x) \text{ is Bessel Y function.}
\end{eqnarray*}
This result coincides to the one in Reference \cite{Kim2019}. The form factors $C_{\pm}$ as well as their ratio $\mathfrak{R}$ are acquired using some integrals of Bessel function provided in \cite{gradshteyn2014table}
\begin{subequations} \begin{eqnarray}
&& C_{\pm} = 1 + \left[ \frac{2 \omega _a^2 D^2}{c^2} \ln \left( \frac{D+R}{D-R} \right) \right]^{-1}  \biggl\{ \frac{\alpha ^2 z^2}{2} \left( J_1^2(z) - J_2(z) J_0(z) \pm (J_1^2(z) + J_0^2(z))  \right) \nonumber\\
&& \quad \quad \quad \quad \quad \quad \biggl. + \frac{\beta ^2 z^2}{2} \left( Y_1^2(z) - Y_2(z) Y_0(z) \pm (Y_1^2(z) + Y_0^2(z))  \right) - \frac{2 \omega _a D}{c} \left( \alpha J_0(z) + \beta Y_0 (z) \right)  \nonumber\\
&& \quad \quad \quad \quad \quad \quad \biggl.  + \alpha \beta z^2 \left( J_1(z) Y_1(z) - \frac{J_0(z)Y_2(z) + J_2(z) Y_0(z)}{2} \pm ( J_0(z) Y_0(z) + J_1(z) Y_1(z) ) \right)  \biggr\}_{\omega_a (D-R)/c}^{\omega_a (D+R)/c} , \\
&& \mathfrak{R} = \left| \dfrac{C_{-}}{C_{+}} \right|.
\end{eqnarray} \end{subequations} 
\end{widetext}

\begin{figure}[H]
    \centering
    \includegraphics[width = 0.45 \textwidth]{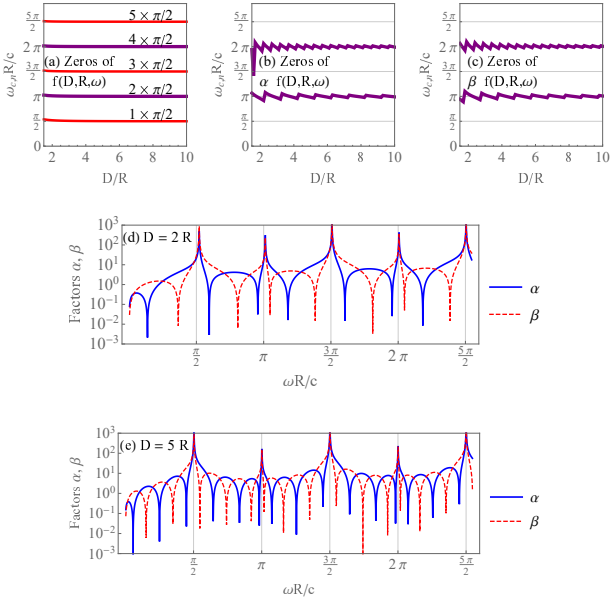}
    \caption{Zeros of (a) denominator $f (D,R, \omega)$ and (b,c) numerators $g_{\alpha, \beta} (D,R, \omega)$ versus $D/R$ when $D \geq2 R$. Since the zeros of denominator $f (D,R, \omega)$ and (b,c) numerators $g_{\alpha, \beta} (D,R, \omega)$ coincide near $\omega \approx n \pi c / R$, the possible resonant frequencies are $(n - 1/2)\pi c / R$. (d,e) Profile of coefficients $\alpha$ (blue, thick), $\beta$ (red, dashed) versus $\omega R/c$ when $D=2R$ and $D=5R$ respectively. $\alpha$ and $\beta$ go to very large value when $\omega$ approach $\omega _{c,n} \approx (n - 1/2)\pi c / R$.}
    \label{fig2a}
\end{figure}

About resonant angular frequencies $\omega _{c,n}$, we look on the nodes of the denominator of $\alpha$ and $\beta$ coefficient i.e
\begin{equation*}
\begin{array}{ll}
f (D,R, \omega) = &  J_1 ((D-R)\omega / c) Y_1 ((D+R)\omega / c) \\
& - J_1 ((D+R)\omega / c) Y_1 ((D-R)\omega / c),
\end{array}
\end{equation*}
and the nodes of their numerators $g_{\alpha} (D,R, \omega) = \alpha f (D,R, \omega), g_{\beta} (D,R, \omega) = \beta f (D,R, \omega)$. When the average radius of the toroidal solenoid is large $D \to \infty$, its cylinders becomes flat sheets distanced by $2R$. This suggests the resonance of the toroidal solenoid should occurs near the resonance of two-parallel-sheet cavity gapped by $2R$. Meanwhile, $\lim\limits_{D/R \to \infty} \omega _{c,n} = (n - 1/2)\pi c / R$ and the approximation $\omega _{c,n} \approx (n - 1/2)\pi c / R$ is meaningful when the average radius of the rectangular-torus cavity is larger than a certain value. Observation on the nodes of $f (D,R, \omega)$ and $g_{\alpha, \beta} (D,R, \omega)$, which is illustrated in Figure \ref{fig2a}, indicates that zeros of $f (D,R, \omega)$ tend to $(n - 1/2)\pi c / R$ or $n \pi c / R$ for $D > 2 R$ while zeros of $g_{\alpha, \beta} (D,R, \omega)$ tend to $n \pi c / R$ for $D > 2 R$. The resonant angular frequencies $\omega _{c,n}$ must be the frequencies at which the denominator $f (D,R, \omega)$ vanishes while the numerators $g_{\alpha, \beta} (D,R, \omega)$ do not. So the approximation $\omega _{c,n} \approx (n - 1/2)\pi c / R$ is hold for $D > 2 R$.

\begin{figure}[H]
    \centering
    \includegraphics[width = 0.45 \textwidth]{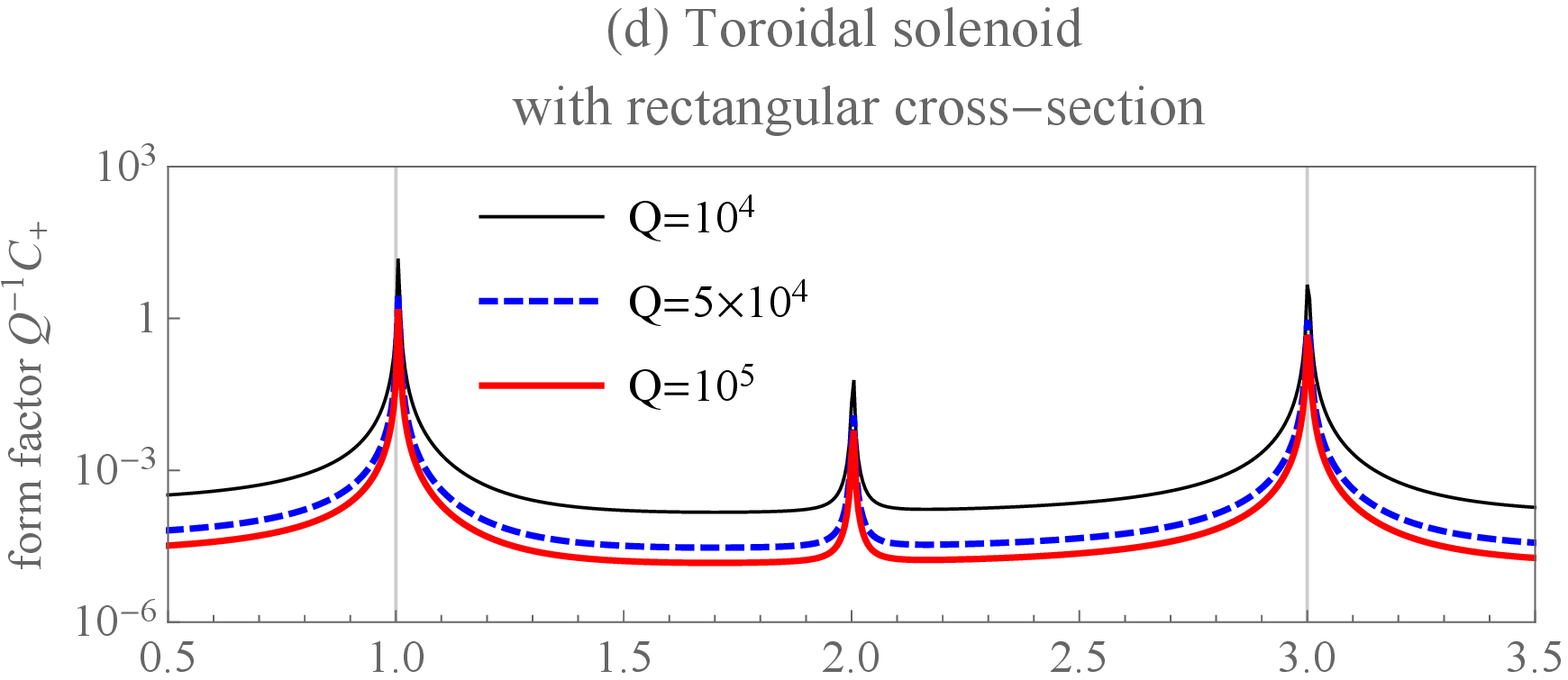}
    \includegraphics[width = 0.45 \textwidth]{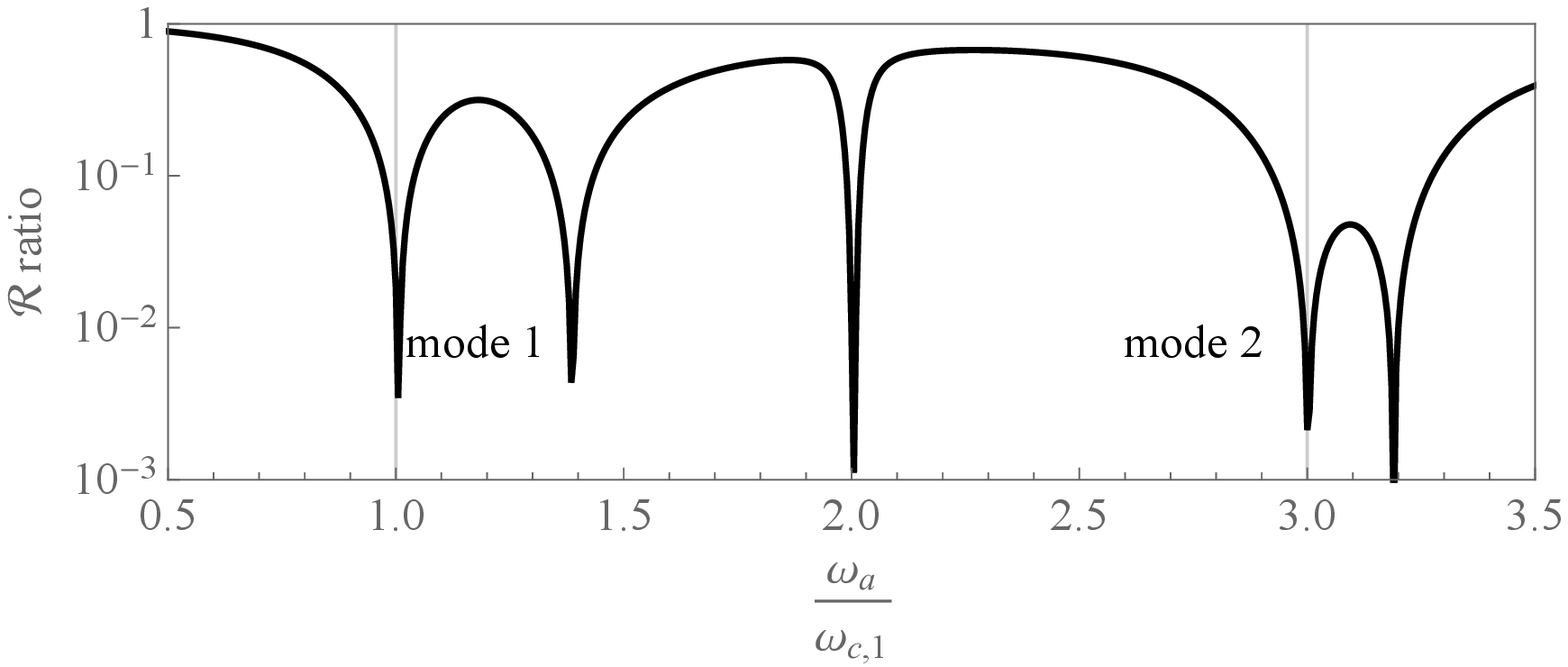}
    \caption{Profile of the modified form factor of imperfect electric conductive cavity $\left| Q^{-1} C_{+} \left(\omega_a \left(1 + \imath 2^{-1} Q^{-1} \right) \right) \right|$ as well as the ratio $\mathfrak{R} (\omega _a)$ versus frequency ratio $\omega_a / \omega _{c,1}$ for toroidal solenoid with a rectangular cross-section in case $D = 5R$. Black, blue and red curves are for effective $Q$-factor are $10^4$, $5\times 10^4$, $10^5$ respectively.}
    \label{fig:case2}
\end{figure}

Similar to the previous cases, we choose the half distance between two cylinders of the rectangular-torus cavity $R = 3.141595 \text{ cm}$ so that the axion signal is approximately resonant at $m_a \sim 10 \mu eV$. The average radius is chosen as $D = 5R$ i.e the outer cylinder is $1.5$ times bigger than the inner one. Figure \ref{fig:case2} shows the dependence of the modified form factor of imperfect electric conductive cavity $\left| Q^{-1} C_{+} \left(\omega_a \left(1 + \imath 2^{-1} Q^{-1} \right) \right) \right|$ and the ratio $\mathfrak{R} (\omega _a)$ on $\omega_a / \omega _{c,1}$.

\begin{figure}[H]
    \centering
    \includegraphics[width = 0.45 \textwidth]{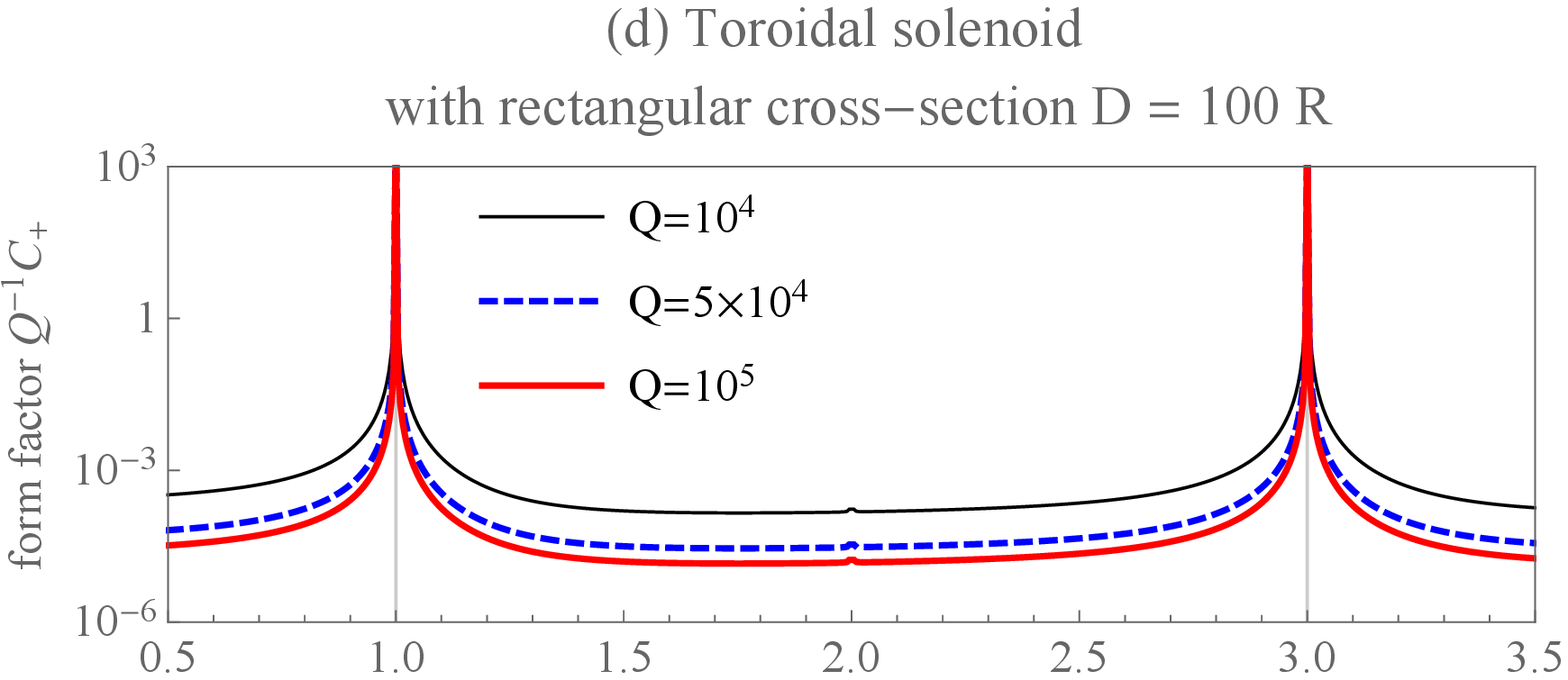}
    \includegraphics[width = 0.45 \textwidth]{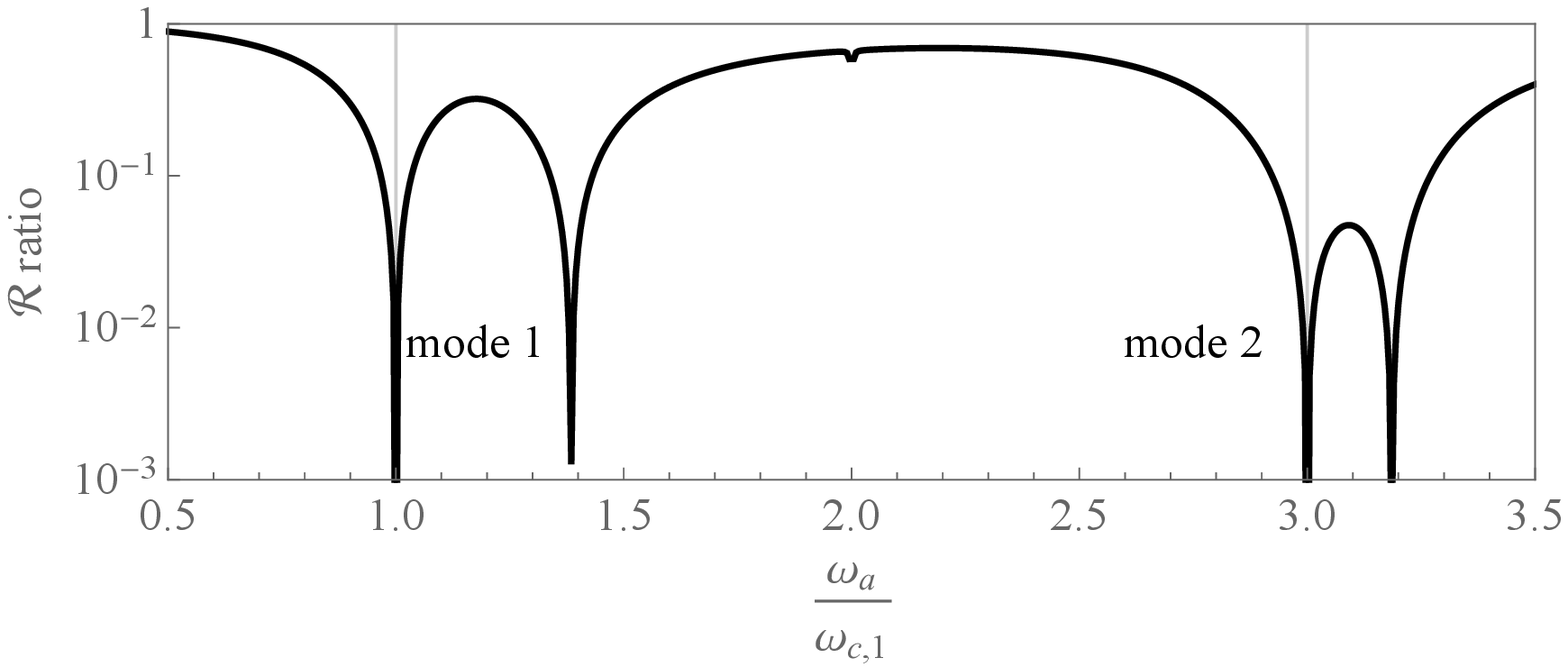}
    \caption{Profile of the modified form factor of imperfect electric conductive cavity $\left| Q^{-1} C_{+} \left(\omega_a \left(1 + \imath 2^{-1} Q^{-1} \right) \right) \right|$ as well as the ratio $\mathfrak{R} (\omega _a)$ versus frequency ratio $\omega_a / \omega _{c,1}$ for toroidal solenoid with a rectangular cross-section in case $D = 100 R$. Black, blue and red curves are for effective $Q$-factor are $10^4$, $5\times 10^4$, $10^5$ respectively.}
    \label{fig:case22}
\end{figure}

When looking at the Figure \ref{fig:case2} for the first moment, one may think that there is resonant at $\omega _a = 2 \omega _{c,1}$. However, this is the false resonant arising from numerical error of the estimation $\omega _{c,n} \approx (n - 1/2)\pi c / R$. When increasing the ratio $D/R$ the signal at this point goes down significantly and the profile will look like the one of two-parallel-sheet cavity when $D$ is large enough. For example, Figure \ref{fig:case22} shows the profile when $D = 100 R$.

\begin{figure}[H]
    \centering
    \includegraphics[width = 0.45 \textwidth]{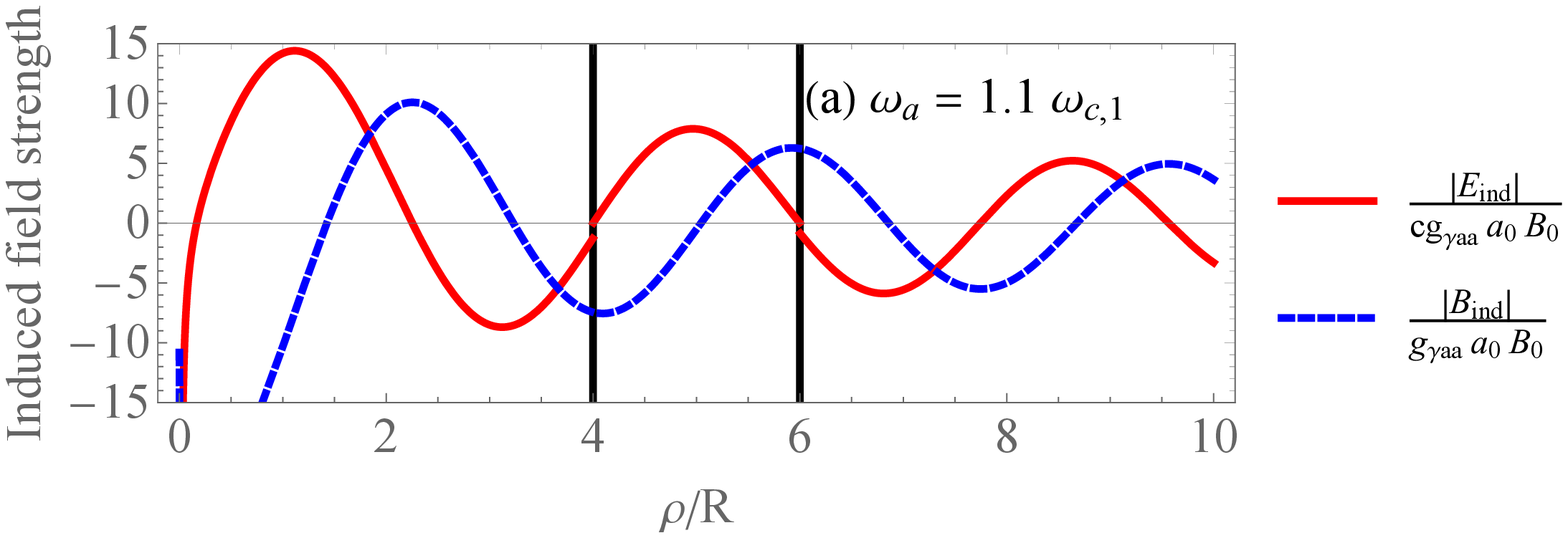}
    \includegraphics[width = 0.45 \textwidth]{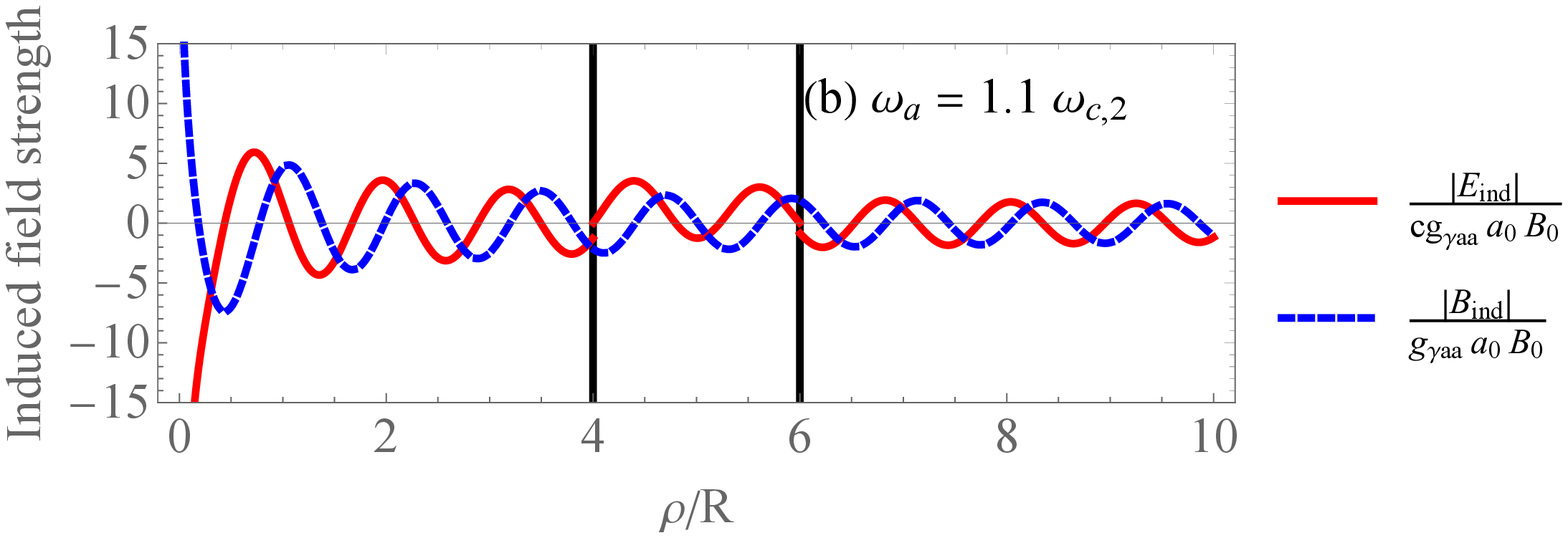}
    \includegraphics[width = 0.45 \textwidth]{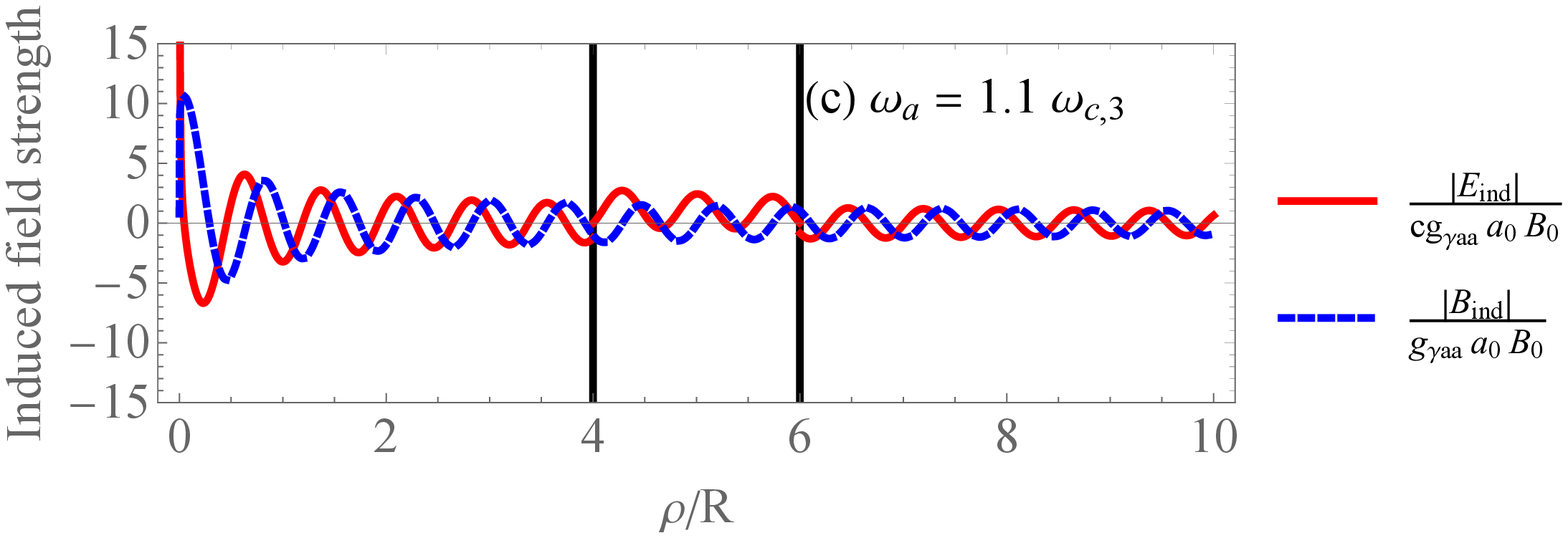}
    \caption{(a,b,c) The axion-induced electromagnetic fields strength $E_{ind}, B_{ind}$ near first three resonant modes $\omega _a = 1.1 \omega _{c,1}, 1.1 \omega _{c,2}, 1.1 \omega _{c,3}$ vary in space respectively.}
    \label{fig:case2EM}
\end{figure}

Figure \ref{fig:case2EM} shows axion-induced electromagnetic fields strength $E_{ind}, B_{ind}$ near first three resonant modes varying in space while Figure \ref{fig:case2EMstore} indicates the distribution of the axion-induced electromagnetic stored energy density.

\subsection{\label{sec:4c}Toroidal solenoid with circular cross section}

One of the most attention of theoretical investigation on cold axion searches is about using circular cross-section toroidal solenoid as a resonant cavity because this cavity is more practical than infinite cylindrical solenoid and its magnetic field is nearly uniform. Different from previous cases, the set of orthogonal coordinates namely toroidal coordinates $(\xi, \eta, \varphi)$ \cite{toroid}
\begin{equation*}
\left\{ \begin{array}{l}
x = \dfrac{d \sinh \xi \cos \varphi}{\cosh \xi - \cos \eta}, \\
y = \dfrac{d \sinh \xi \sin \varphi}{\cosh \xi - \cos \eta}, \\
z = \dfrac{d \sin \eta}{\cosh \xi - \cos \eta}.
\end{array}
\right. 
\end{equation*}
In this paper, we use a different coordinate system from other works such as in \cite{Hoang2017, Kim2019} because the one used here is orthogonal and Helmholtz equation has been solved in this kind of coordinate system \cite{Weston1958}. The surface of a torus with major radius $D$ and minor radius $R$ is simply described as a surface $\xi = \ln \left[\left( D + d \right) / 2 R \right]$ with $d = \sqrt{D^2 - R^2}$. As Laplace equation is separably solvable in this coordinate system, the applied magnetic field inside this toroidal cavity can be analytically expressed by a well-know form
\begin{subequations} \begin{eqnarray}\label{eqn:case3}
\mathbf{B}_0 (\mathbf{r}) && = \dfrac{B_0 D}{\sqrt{x^2 + y^2}} \left[ \hat{\mathbf{y}} \cos \varphi - \hat{\mathbf{x}} \sin \varphi \right] \nonumber\\
&& = \dfrac{B_0 D \left( \cosh \xi - \cos \eta \right) }{d \sinh \xi} \left[ \hat{\mathbf{y}} \cos \varphi - \hat{\mathbf{x}} \sin \varphi \right]
\end{eqnarray} \end{subequations} 
Regretfully, the Helmholtz equation \eqref{eqn:frakB} is inseparable in toroidal coordinetes so that the boundary value problem \eqref{eqn:frakB} must be solved by recurrence method \cite{Weston1958}. Right at the surface inside the cavity $\xi \to \xi _0 = \ln \left[\left( D + d \right) / 2 R \right] ^{-}$, three components of applied magnetic field are $\left( B_{0,x}, B_{0,y}, B_{0,z} \right) = B_0 D d^{-2} \left( D - R \cos \eta \right) \times \left( - \sin \varphi, \cos \varphi, 0 \right)$. Therefore, the solution of Equation \eqref{eqn:frakB} should take the form 
\begin{equation*}
\pmb{\mathfrak{B}}(\mathbf{r}, \omega) = B_0 F \left(\xi, \eta, \omega \right) \left[ \hat{\mathbf{y}} \cos \varphi - \hat{\mathbf{x}} \sin \varphi \right] , 
\end{equation*}
where $F \left(\xi, \eta, \omega \right) = \sum_{n} c_n W_n (\xi, \eta, \omega)$ is linear combination of non-separated eigenfunctions $W_n (\xi, \eta, \omega)$ with the coefficients $c_n$ satisfying $F \left(\xi _0, \eta, \omega \right) = D d^{-2} \left( D - R \cos \eta \right)$. Reference \cite{Weston1958} showed that non-separated eigenfunctions $W_n (\xi, \eta, \omega)$ had to take the following forms
\begin{eqnarray*}
W_n (\xi, \eta, \omega) && = \sum _{r = T}^{\infty} A_{r} (\xi) \left( \dfrac{\cosh \xi - \cos \eta}{d \sinh \xi} \right)^{-r}, \\
W_0 (\xi, \eta, \omega) && = \sin \eta \sum _{r = T^{\prime}}^{\infty} B_{r} ( \xi) \left( \dfrac{\cosh \xi - \cos \eta}{d \sinh \xi} \right)^{-r},
\end{eqnarray*}
whereas $A _r (s)$ and $B_r (s)$ are solutions of a set of recurrence differential equations. Comparing with the boundary condition of $F \left(\xi , \eta, \omega \right)$, its general solution should be
\begin{eqnarray*}
F (\xi, \eta, \omega) && = D f_{-1} ( \xi, \omega) \left( \dfrac{\cosh \xi - \cos \eta}{d \sinh \xi} \right) + \nonumber\\
&&  \sum _{r = 0}^{\infty} D^{-r} f_{r} ( \xi, \omega) \left( \dfrac{\cosh \xi - \cos \eta}{d \sinh \xi} \right)^{-r}.
\end{eqnarray*} 
To avoid cumbersome arising from full solution of $F \left(\xi , \eta, \omega \right)$, we perform the approximate solution by two following assumption: (i) the major radius $D$ is much larger than minor radius $R$, (ii) axion frequency $\omega$ is quite small enough. The former assumption comes from the asymptotic behaviour that the circular toroidal cavity coincides the circular cylindrical solenoid when  $D / R \to \infty$. The second assumption is based on the other asymptotic behaviour that $F \left(\xi , \eta, \omega \right)$ becomes $|\mathbf{B}_0 (\xi , \eta)| / B_0$ when $\omega \to 0$. As the results, we neglect all of higher order terms $r \geq 0$ i.e $f_r (\xi, \omega) \approx 0$. The suitable approximate solution is given as
\begin{equation*}
F (\xi, \eta, \omega) \approx D f (\xi, \omega) \left( \dfrac{\cosh \xi - \cos \eta}{d \sinh \xi} \right).
\end{equation*} 
Introducing new variable $z = \omega d  c^{-1} \cosh ^{-1} \xi \ll 1$ and neglecting $\eta$ terms as well as higher order terms of $z$, we show that $f(z)$ obeys the approximate ordinary differential equation
\begin{equation*}
 z^2 f^{''} (z) + z f^{'} (z) + z^2 f (s) \approx 0,
\end{equation*}
whose solution is the Bessel $J$ function: $f (z) \sim J_0 (z).
$ \cite{gradshteyn2014table}.
\begin{widetext}
Finally, we obtain \footnote{It is noticed that if $\rho $ is distance from a point to the circle $r = D$ in $Oxy$ plane and the major radius $D$ is much larger than minor radius $R$, then $\rho \approx d/ \cosh \xi$ and hence, above results coincides to the known one in References \cite{Kim2019, Hoang2017} $\pmb{\mathfrak{B}} (\mathbf{r}, \omega) \approx \mathbf{B}_0 \dfrac{J_0 (\rho \omega / c) }{J_0 (R \omega / c)}$. }
\begin{eqnarray}\label{eqn:frakB-case3}
\pmb{\mathfrak{B}}(\mathbf{r}, \omega) \approx \mathbf{B}_0 (\mathbf{r}) \dfrac{J_0\left( \dfrac{\omega \sqrt{D^2-R^2} }{c} \dfrac{1}{\cosh \xi} \right)}{ J_0\left( \dfrac{\omega R}{c} \dfrac{\sqrt{D^2-R^2}}{D} \right)  } ,
\end{eqnarray}
from which the axion-induced fields, the total and difference EM form factors as well as their ratio are calculated
\begin{subequations} \begin{eqnarray}
&& \mathbf{E}_{ind} (\mathbf{r}, \omega)  \approx c g _{a \gamma \gamma} a_0 B_0 e^{- \imath \omega t} \left[ \hat{\mathbf{y}} \cos \varphi - \hat{\mathbf{x}} \sin \varphi \right] \dfrac{D (\cosh \xi - \cos \eta)}{d \sinh \xi} \left[ 1 - \dfrac{J_0\left( \frac{\omega \sqrt{D^2-R^2} }{c} \frac{1}{\cosh \xi} \right)}{ J_0\left( \frac{\omega R}{c} \frac{\sqrt{D^2-R^2}}{D} \right)  } \right]. \\
&& \mathbf{B}_{ind} (\mathbf{r}, \omega)  \approx \imath g_{a \gamma \gamma} a_0 B_0 e^{-\imath \omega t} \frac{D \left( \cosh \xi - \cos \eta \right)^2}{d \cosh ^2 \xi} \dfrac{J_1 \left( \frac{\omega \sqrt{D^2-R^2} }{c} \frac{1}{\cosh \xi} \right)}{ J_0\left( \frac{\omega R}{c} \frac{\sqrt{D^2-R^2}}{D} \right)  } \nonumber\\
&& \times \left[ \frac{\cosh \xi \sin \eta}{\cosh \xi - \cos \eta} \left( \hat{\mathbf{x}} \cos \varphi + \hat{\mathbf{y}} \sin \varphi \right) + \frac{1 - \cosh \xi \cos \eta}{\cosh \xi - \cos \eta} \hat{\mathbf{z}} \right], 
\end{eqnarray}
\begin{eqnarray}
&& C_{\pm} \approx 1 - \dfrac{2}{\frac{\omega R}{c} \frac{\sqrt{D^2-R^2}}{D}} \dfrac{J_1 \left( \frac{\omega R}{c} \frac{\sqrt{D^2-R^2}}{D} \right)}{ J_0\left( \frac{\omega R}{c} \frac{\sqrt{D^2-R^2}}{D} \right)} + \frac{1}{2} \dfrac{J_1^2 \left( \frac{\omega R}{c} \frac{\sqrt{D^2-R^2}}{D} \right)}{ J_0^2 \left( \frac{\omega R}{c} \frac{\sqrt{D^2-R^2}}{D} \right)} \pm \frac{1}{2} \dfrac{J_1^2 \left( \frac{\omega R}{c} \frac{\sqrt{D^2-R^2}}{D} \right)}{ J_0^2\left( \frac{\omega R}{c} \frac{\sqrt{D^2-R^2}}{D} \right)} \mp \frac{1}{2} \dfrac{J_2 \left( \frac{\omega R}{c} \frac{\sqrt{D^2-R^2}}{D} \right)}{ J_0\left( \frac{\omega R}{c} \frac{\sqrt{D^2-R^2}}{D} \right)}   \\
&& \mathfrak{R} = \dfrac{1}{\left| 3 - \frac{2 J_1^2 \left( \frac{\omega R}{c} \frac{\sqrt{D^2-R^2}}{D} \right)}{J_0 \left( \frac{\omega R}{c} \frac{\sqrt{D^2-R^2}}{D} \right) J_2 \left( \frac{\omega R}{c} \frac{\sqrt{D^2-R^2}}{D} \right)} \right|} .
\end{eqnarray} \end{subequations} 
\end{widetext}
As can be seen from above expression, the results in circular toroidal cavity is similar to the one in infinite cylindrical cavity whereas the frequency is scaled by a factor $\sqrt{D^2-R^2}/D$ which is closed to $1$. Therefore, these results obviously take infinite cylindrical cavity as its limit. In the same manner as infinite cylindrical cavity, the resonant frequency of circular toroidal cavity is given as $\omega _{c,n} = \chi _n c D / R \sqrt{D^2 - R^2}$ in which $\chi _n$ is $n^{th}$-zero of $J_0(\chi)$ function.

\begin{figure}[H]
    \centering
    \includegraphics[width = 0.45 \textwidth]{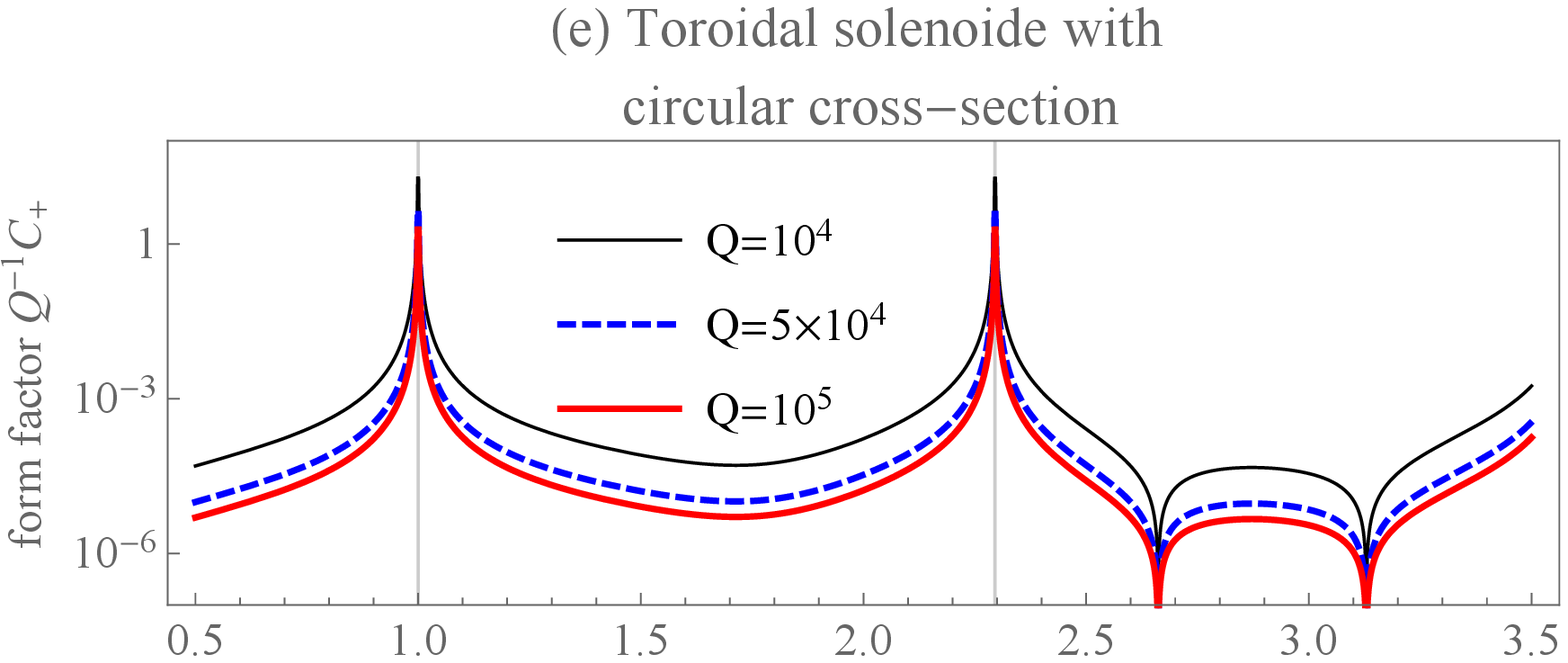}
    \includegraphics[width = 0.45 \textwidth]{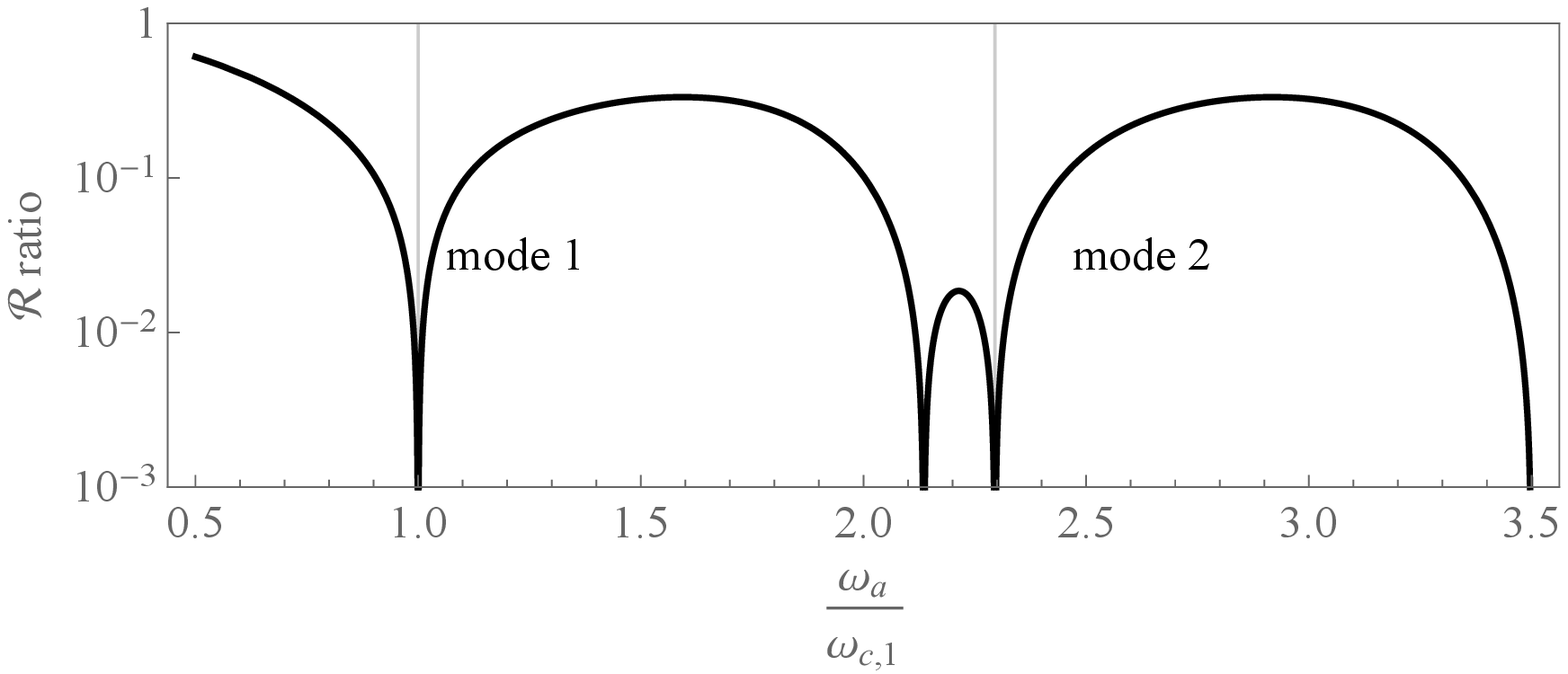}
    \caption{Profile of the modified form factor of imperfect electric conductive cavity $\left| Q^{-1} C_{+} \left(\omega_a \left(1 + \imath 2^{-1} Q^{-1} \right) \right) \right|$ as well as the ratio $\mathfrak{R} (\omega _a)$ versus frequency ratio $\omega_a / \omega _{c,1}$ for circular toroidal solenoid cavity. Black, blue and red curves are for effective $Q$-factor are $10^4$, $5\times 10^4$, $10^5$ respectively. The higher quality factor is, the sharper resonant peaks are.}
    \label{fig:case3}
\end{figure}

Similar to infinite cylindrical cavity, we choose the minor radius $R = 4.90883 \text{ cm}$ while the major radius is 5 times of that $D = 5R$. As the result, the axion signal is resonant at $m_a \sim 10 \mu eV$. Figure \ref{fig:case3} illustrates the modified form factor of imperfect electric conductive cavity $\left| Q^{-1} C_{+} \left(\omega_a \left(1 + \imath 2^{-1} Q^{-1} \right) \right) \right|$ as well as the ratio $\mathfrak{R} (\omega _a)$ versus $\omega_a / \omega _{c,1}$.

\begin{figure}[H]
    \centering
    \includegraphics[width = 0.45 \textwidth]{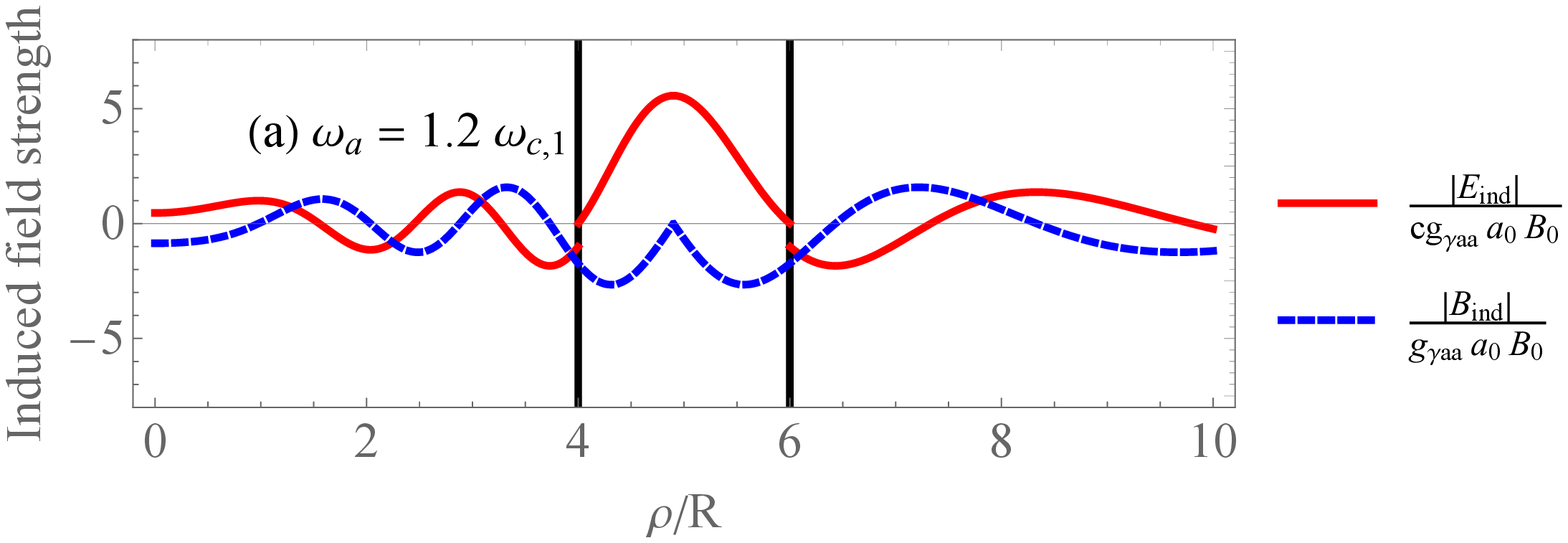}
    \includegraphics[width = 0.45 \textwidth]{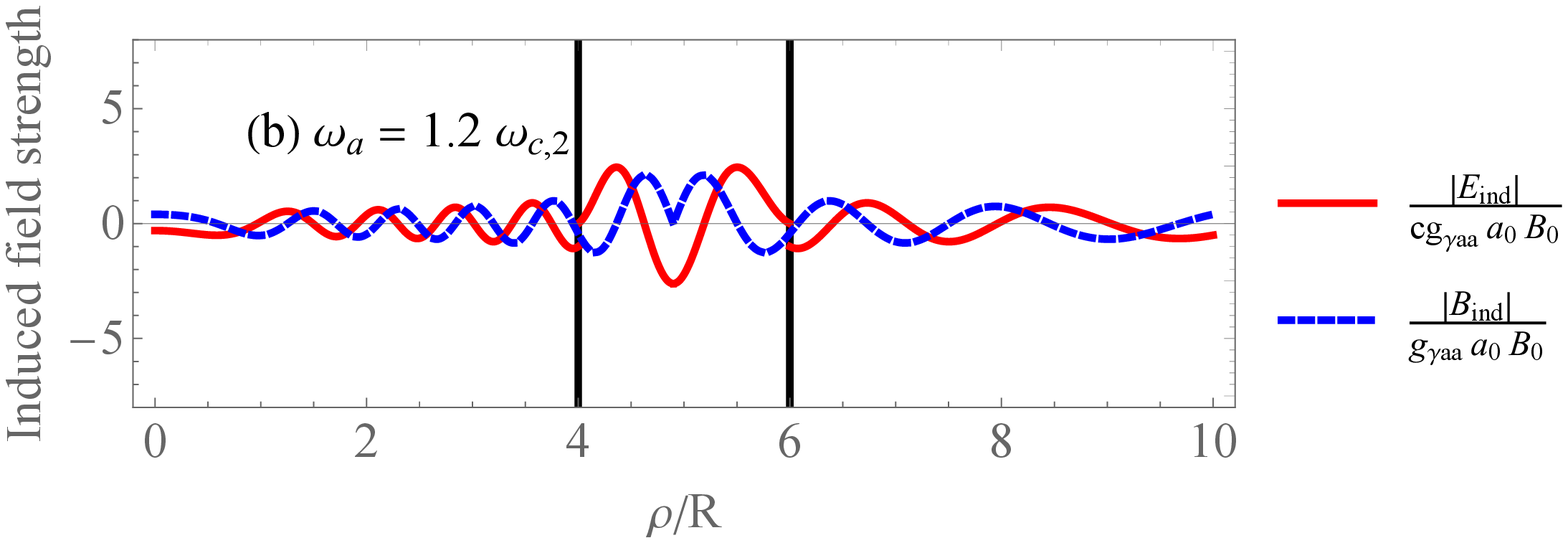}
    \includegraphics[width = 0.45 \textwidth]{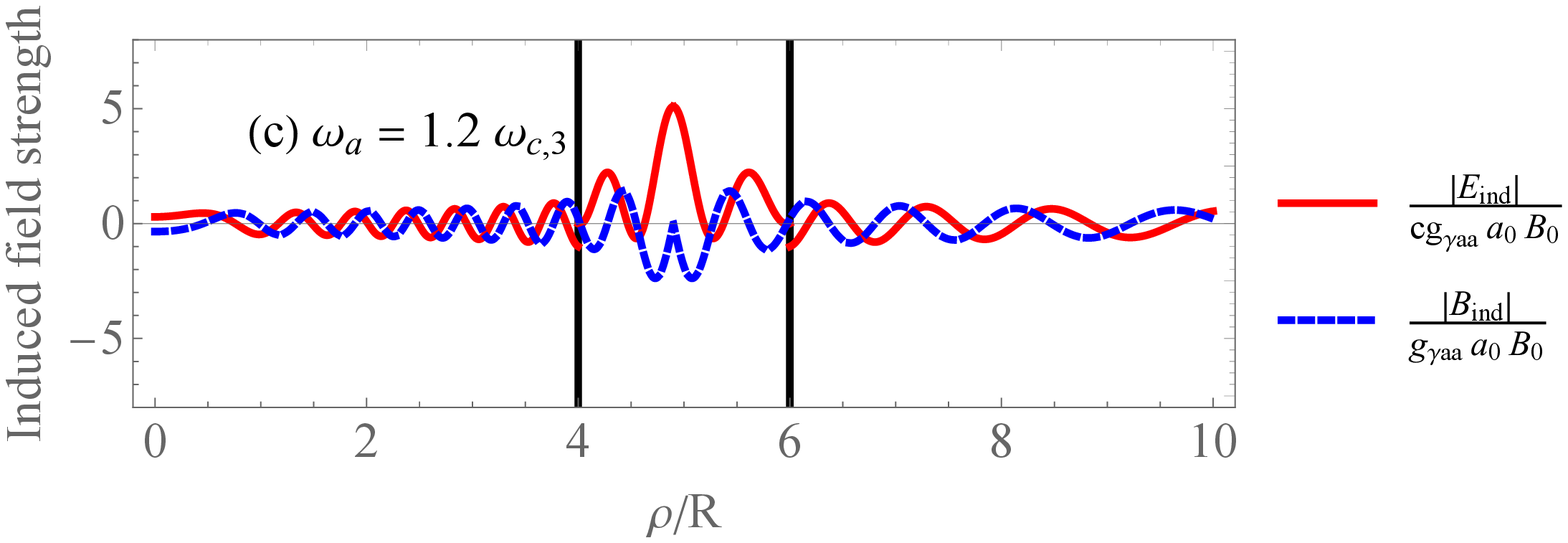}
    \caption{(a,b,c) The axion-induced electromagnetic fields strength $E_{ind}, B_{ind}$ near first three resonant modes $\omega _a = 1.2 \omega _{c,1}, 1.2 \omega _{c,2}, 1.2 \omega _{c,3}$ vary in space respectively.}
    \label{fig:case3EM}
\end{figure}

Figure \ref{fig:case3EM} shows axion-induced electromagnetic fields strength $E_{ind}, B_{ind}$ near first three resonant modes varying in space while Figure \ref{fig:case3EMstore} indicates the distribution of the axion-induced electromagnetic stored energy density.

\section{\label{sec:5}Miscellaneous discussion between conventional and dual symmetric point of view}
To compare conventional and dual symmetric theories of axion electrodynamics, we derive the second order D'Alembert equation governing the axion induced electromagnetic field $\mathbf{E}_{ind}$, $\mathbf{B}_{ind}$. In the literature, the second partial differential equations for induced fields are more practically used than general solution; therefore, it is reasonable to write them down in dual symmetric version of axion electrodynamics equations. First, it is obvious that the applied and radiation fields must be solutions of
\begin{subequations}\label{eqn:d-eb-0}
\begin{eqnarray}
\square \mathbf{E}_0 & = & - \dfrac{\pmb{\nabla} \rho_e}{\varepsilon _0} - \mu _0 \dfrac{\partial \mathbf{J}_e}{\partial t}, \label{eqn:d-e-0}\\
\square \mathbf{B}_0 & = & \mu _0 \pmb{\nabla} \times \mathbf{J}_e , \label{eqn:d-b-0}
\end{eqnarray} \end{subequations} 
{Now, we aim to derive the second order D'Alembert equation governing the axion induced electromagnetic field $\mathbf{E}_{ind}$, $\mathbf{B}_{ind}$ in haloscope experiment for both dual symmetric and conventional  scenarios.}

\subsection{\label{sec:5a}In dual symmetric scenario of haloscope experiment}

{Taking D'Alembert operator $\square$ act on first-order fields $\mathbf{E}_{ind}$ and $\mathbf{B}_{ind}$ given in Equation \eqref{eqn:1-order} leads to the following equations
\begin{subequations}\label{eqn:d-eb-1}
\begin{eqnarray}\label{eqn:d-e-1}
\square \left[ \mathbf{E}_{ind}^{dual} - c g_{a\gamma\gamma} a \mathbf{B}_0 \right] && \approx \square \mathbf{E}_{rad} = 0,
\end{eqnarray}
\begin{eqnarray}\label{eqn:d-b-1}
\square \left[ \mathbf{B}_{ind}^{dual} + \dfrac{g_{a\gamma\gamma} a}{c}\mathbf{E}_0 \right] && \approx \square \mathbf{B}_{rad} = 0.
\end{eqnarray} \end{subequations} 
Equation \eqref{eqn:axion} and Equation \eqref{eqn:d-eb-0} are used and second order terms of coupling constant $g_{a\gamma\gamma}$ are neglected. In the static $DC$ haloscope experiments, applied fields $\mathbf{E}_0$ and $\mathbf{B}_0$ are static i.e time-independent. They are induced from surface electric current $\mathbf{J}_e$ and surface electric charge $\rho _e$ on cavity 's surface. Therefore, on the free space whereas $\rho _e = 0$ and $\mathbf{J}_e = \mathbf{0}$, Equations \eqref{eqn:axion}, \eqref{eqn:d-eb-0}, \eqref{eqn:d-eb-1} in first order approximation of dual symmetric scenario becomes
\begin{eqnarray}
&& \left( \square + \dfrac{\omega _a^2}{c^2} \right) a \approx  0 \label{eqn:axion-approx},
\end{eqnarray}
\begin{subequations} \label{eqn:d-eb-0-approx} \begin{eqnarray}
&&\pmb{\nabla}^2 \mathbf{E}_0  =  0 \label{eqn:d-e-0-approx},\\
&&\pmb{\nabla}^2 \mathbf{B}_0 = 0 \label{eqn:d-b-0-approx},\\
&&\square \left[ \mathbf{E}_{ind}^{dual} - c g_{a\gamma\gamma} a \mathbf{B}_0 \right] \approx 0 \label{eqn:d-e-1-approx},\\
&&\square \left[ \mathbf{B}_{ind}^{dual} + \dfrac{g_{a\gamma\gamma} a}{c}\mathbf{E}_0 \right] \approx 0 \label{eqn:d-b-1-approx}.
\end{eqnarray} \end{subequations} }

\subsection{\label{sec:5b}In conventional scenario of haloscope experiment}
{
As we have discussed in Section \ref{sec:intro}, in our point of view, the conventional scenario of axion electrodynamics is equivalent to the dual symmetric axion electrodynamics with magnetic monopole
\begin{subequations}\label{eqn:cond5}
\begin{eqnarray}
\rho _m^{conv} && = c g_{a\gamma\gamma} a \rho _e + \dfrac{g_{a\gamma\gamma}}{\mu _0 c} \left( \mathbf{E} + c g_{a \gamma \gamma} a \mathbf{B} \right) \cdot \pmb{\nabla} a \label{eqn:cond5-rho-m} \\
\mathbf{J}_m^{conv} && = c g_{a\gamma\gamma} a \mathbf{J}_e - \dfrac{g_{a\gamma\gamma}}{c \mu _0} \dfrac{\partial a}{\partial t} \left( \mathbf{E} + c g_{a \gamma \gamma} a \mathbf{B} \right) \nonumber\\
&& \quad \quad \quad \quad \quad - \dfrac{g_{a\gamma\gamma}}{\mu _0} \left(c \mathbf{B} - g_{a\gamma\gamma} a \mathbf{E} \right) \times \pmb{\nabla} a \label{eqn:cond5-J-m}.
\end{eqnarray}
\end{subequations}
On the free space i.e out of the cavity 's surface on haloscope experiment, there are no electric charge $\rho _e = 0$ or current $\mathbf{J}_e = \mathbf{0}$, then the conventional magnetic monopole sources in the first order approximation are given as follows:
\begin{subequations}\label{eqn:cond5b}
\begin{eqnarray}
\rho _m^{conv} && \approx \dfrac{g_{a\gamma\gamma}}{\mu _0 c} \mathbf{E}_0 \cdot \pmb{\nabla} a \label{eqn:cond5b-rho-m} \\
\mathbf{J}_m^{conv} && \approx - \dfrac{g_{a\gamma\gamma} c}{\mu _0} \left[ \dfrac{\partial a}{\partial t} \dfrac{\mathbf{E}_0}{c^2} + \mathbf{B}_0 \times \pmb{\nabla} a \right] \label{eqn:cond5b-J-m}.
\end{eqnarray}
\end{subequations}}

{
To obtain the D'Alembert equations for axion-induced electromagnetic field in this conventional scenario $\mathbf{E}_{ind}^{conv}$ and $\mathbf{B}_{ind}^{conv}$, we act the curl operator $\pmb{\nabla}\times$ on Equations \eqref{eqn:rot-e-general-intro} and \eqref{eqn:rot-b-general-intro} followed by using the identity $\pmb{\nabla}\times\pmb{\nabla}\times\mathbf{A} = \pmb{\nabla} \left( \pmb{\nabla}\cdot \mathbf{A} \right) - \pmb{\nabla}^2 \mathbf{A}$ as well as Equations \eqref{eqn:div-e-general-intro} and \eqref{eqn:div-b-general-intro}. Finally the D'Alembert equations for axion-induced electromagnetic field in this conventional scenario are
\begin{subequations}\label{eqn:d-eb-1-conv}
\begin{eqnarray}\label{eqn:d-e-1-conv}
\square \left[ \mathbf{E}_{ind}^{conv} - c g_{a\gamma\gamma} a \mathbf{B}_0 \right] && \approx - \mu _0 \pmb{\nabla}\times \mathbf{J}_m^{conv},
\end{eqnarray}
\begin{eqnarray}\label{eqn:d-b-1-conv}
\square \left[ \mathbf{B}_{ind}^{conv} + \dfrac{g_{a\gamma\gamma} a}{c}\mathbf{E}_0 \right] && \approx - \mu _0 \left[\pmb{\nabla} \rho _m^{conv} + \dfrac{\partial \mathbf{J}_m^{conv}}{c^2 \partial t} \right].
\end{eqnarray} \end{subequations} 
Equation \eqref{eqn:d-eb-1-conv} actually comes from the generalization of Equation \eqref{eqn:d-hat} with non-zero magnetic monopole.
}

{
For free axion, the propagation of axion field is the plane wave $a(\mathbf{r},t) = a_0 (\mathbf{k},\omega) \exp \left[ \imath \left( \mathbf{k}\cdot \mathbf{r} - \omega t \right) \right]$. If we write the wave vector $\mathbf{k}$ as follow $\mathbf{k} = \dfrac{\omega}{c} \vec{n}$ then the module of $\vec{n}$ is $|\vec{n}| = v_a /c$. $\vec{n}$ is the dimensionless quantity whose module relates to the relative speed of free axion particles. For cold axion or cosmology axion, there velocity is often small $v_a/c \sim 10^{-3}$ \cite{Irastorza2012, Jaeckel2016,Millar2017, Knirck2018}. Then the resonant mainly occurs near the ground state of axion $\omega \approx \omega _a$. The gradient and time derivative of axion field are approximately
\begin{equation}
\pmb{\nabla} a\approx \imath \omega _a a c^{-1} \mathbf{n}, \quad \partial  a / \partial t = - \imath \omega_a a.   
\end{equation}
Consequently the conventional magnetic monopole sources in the first order approximation becomes:
\begin{subequations}\label{eqn:cond5c}
\begin{eqnarray}
\rho _m^{conv} && \approx \dfrac{\imath \omega _a g_{a\gamma\gamma} a}{\mu _0 c^2} \mathbf{E}_0 \cdot \mathbf{n} \label{eqn:cond5c-rho-m} \\
\mathbf{J}_m^{conv} && \approx \dfrac{\imath \omega _a g_{a\gamma\gamma} a}{\mu _0} \left[ \mathbf{n} \times \mathbf{B}_0 + \dfrac{\mathbf{E}_0}{c} \right] \label{eqn:cond5c-J-m}.
\end{eqnarray}
\end{subequations}
Substituting Equations \eqref{eqn:cond5c} into Equations \eqref{eqn:d-eb-1-conv}, keeping the first order of axion field or axion velocity and noticing that $\square a = - \omega _a^2 c^{-2} a$, we finally obtain the D'Alembert equations governs free propagation of axion-induced electromagnetic fields in the conventional scenario of axion haloscope experiments
\begin{subequations}\label{eqn:d-eb-2-conv}
\begin{eqnarray}\label{eqn:d-e-2-conv}
\square \left[ \mathbf{E}_{ind}^{conv} - c g_{a\gamma\gamma} a \mathbf{B}_0 + g_{a\gamma\gamma} a \mathbf{n} \times \mathbf{E}_0 \right] && \approx 0,
\end{eqnarray}
\begin{eqnarray}\label{eqn:d-b-2-conv}
\square \left[ \mathbf{B}_{ind}^{conv} - g_{a\gamma\gamma} a \mathbf{n} \times \mathbf{B}_0 \right] && \approx 0.
\end{eqnarray} \end{subequations}
}
\subsection{\label{sec:5c}Comparision between conventional and dual symmetric point of view in axion haloscope experiments}
{As can be seen from Equations \eqref{eqn:d-eb-1} and \eqref{eqn:d-eb-2-conv}, the axion-induced electromagnetic in haloscope experiments in both theories are sightly different. While the axion-induced fields in dual symmetric scenario are simply
\begin{subequations}\label{eqn:d-eb-1-sol}
\begin{eqnarray}\label{eqn:d-e-1-sol}
\mathbf{E}_{ind}^{dual} \approx c g_{a\gamma\gamma} a \mathbf{B}_0 + \mathbf{E}_{rad}^{dual},
\end{eqnarray}
\begin{eqnarray}\label{eqn:d-b-1-sol}
\mathbf{B}_{ind}^{dual} \approx - \dfrac{g_{a\gamma\gamma} a}{c}\mathbf{E}_0 + \mathbf{B}_{rad}^{dual},
\end{eqnarray} \end{subequations}
the one in conventional scenario are more complicated
\begin{subequations}\label{eqn:d-eb-2-sol}
\begin{eqnarray}\label{eqn:d-e-2-sol}
\mathbf{E}_{ind}^{conv} \approx c g_{a\gamma\gamma} a \mathbf{B}_0 - g_{a\gamma\gamma} a \mathbf{n} \times \mathbf{E}_0 + \mathbf{E}_{rad}^{conv},
\end{eqnarray}
\begin{eqnarray}\label{eqn:d-b-2-sol}
\mathbf{B}_{ind}^{conv} \approx g_{a\gamma\gamma} a \mathbf{n} \times \mathbf{B}_0 + \mathbf{B}_{rad}^{conv}.
\end{eqnarray} \end{subequations}
where $\mathbf{E}_{rad}, \mathbf{B}_{rad}$ is electromagnetic radiation arising from boundary effect of haloscope cavity. For simplicity, in further investigation, we only consider PEC cavity in which the induced electric field vanishes at the surface $C$ of the cavity: $\mathbf{E}_{ind} (\mathbf{r} \in C) = 0$. Then the differences between dual symmetric and conventional theories are
\begin{subequations}\label{eqn:d-eb-diff}
\begin{eqnarray}\label{eqn:d-e-diff}
\delta \mathbf{E}_{ind} && = \mathbf{E}_{ind}^{conv} - \mathbf{E}_{ind}^{dual} \nonumber\\
&& \approx - g_{a\gamma\gamma} a \mathbf{n} \times \mathbf{E}_0 + \delta \mathbf{E}_{rad},
\end{eqnarray}
\begin{eqnarray}\label{eqn:d-b-diff}
\delta \mathbf{B}_{ind} && = \mathbf{B}_{ind}^{conv} - \mathbf{B}_{ind}^{dual} \nonumber\\
&& \approx \dfrac{g_{a\gamma\gamma} a}{c}\mathbf{E}_0 + g_{a\gamma\gamma} a \mathbf{n} \times \mathbf{B}_0 + \delta \mathbf{B}_{rad}.
\end{eqnarray} \end{subequations}
From boundary condition of PEC cavity, we can estimate the order of different radiation terms $\delta \mathbf{E}_{rad} = \mathbf{E}_{rad}^{conv}-\mathbf{E}_{rad}^{dual}$ and $\delta \mathbf{B}_{rad} = \mathbf{B}_{rad}^{conv}-\mathbf{B}_{rad}^{dual}$ are $\delta E_{rad} \sim g_{a\gamma\gamma} a_0 v_a c^{-1} E_0$ and $\delta B_{rad} \sim c^{-1} \delta E_{rad}$ respectively. Now we start to examine the difference between two theories some specific cases of axion haloscope experiments.
} 

{
First of all, we examine the traditional axion haloscope experiments using applied DC magnetic field $\mathbf{B}_0 \neq 0$ on a resonant cavity without any electric field $\mathbf{E}_0 = 0$. Then $\delta E_{rad}, \delta B_{rad} \to 0$ and hence,  the differences between dual symmetric and conventional theories   only appears in the axion-induced magnetic field and its magnitude is proportional to axion velocity $v_a$: 
\begin{subequations}\label{eqn:d-eb-diff-case1}
\begin{eqnarray}\label{eqn:d-e-diff-case1}
\delta \mathbf{E}_{ind} = 0,
\end{eqnarray}
\begin{eqnarray}\label{eqn:d-b-diff-case1}
\delta \mathbf{B}_{ind} = g_{a\gamma\gamma} a \mathbf{n} \times \mathbf{B}_0 \sim g_{a\gamma\gamma} a v_a c^{-1} B_0.
\end{eqnarray} \end{subequations}
Under the LWA region, $\pmb{\nabla}a \to 0$ or $v_a \to 0$, both theories coincides to each other: $\delta \mathbf{E}_{ind} = 0$, $\delta \mathbf{B}_{ind} = 0$. This is confirmed since the solution of conventional axion Maxwell equations under LWA approxiation in previous works \cite{Ouellet2019, Kim2019} is the same as ours in Equations \eqref{eqn:eb-3-sol}. The equivalence between two theories in LWA region is understandable since the required monopole sources in free space for conventional theory are proportional to gradient of axion field  $\rho _m^{conv} = \mu _0 ^{-1} g_{a\gamma\gamma} \mathbf{E}_0 \cdot \pmb{\nabla} a \to 0$, $\mathbf{J} _m^{conv} = \mu _0 ^{-1} g_{a\gamma\gamma} c \mathbf{B}_0 \times \pmb{\nabla} a \to 0$ and hence, become zero in LWA region $\pmb{\nabla}a \to 0$.
Another case in which the difference between two theories vanishes is when the axion source is homogeneous and the cavity is symmetry. In this case, the contribution of axion particles in one direction $\mathbf{n}$ will be cancelled by the contribution of axion particles in opposite direction $-\mathbf{n}$. This could happen when using resonant cavity such as cylindrical solenoid, spherical solenoid, two-parallel-sheet cavity, toroidal solenoid with a rectangular cross-section, or with a circular cross-section.
}

{Next, from above observations, we consider two other possibilities of detecting the differences between dual symmetric and conventional scenarios of axion electrodynamics. The first possible experiments is using directional axion detection \cite{Irastorza2012, Jaeckel2016,Millar2017, Knirck2018}. Enhancement of the directional sensitivity by using asymmetric cavity such as long tube or spheroidal cavities, we can keep the different axion-induced magnetic field non zero $\delta \mathbf{B}_{ind} \sim g_{a\gamma\gamma} a v_a c^{-1} B_0$. Another possible set up is applying an DC electric field $\mathbf{E}_0 \neq 0$ on axion haloscope cavity. {To the best of our knowledge, there are no proposals about applying external DC electric field in haloscope experiment to distinct between dual symmetric and conventional scenarios of axion electrodynamics}. Even when searching axion particles in LWA region or using symmetric cavity, axion-induce fields calculated from two theories are still sightly different
\begin{subequations}\label{eqn:d-eb-diff-case2}
\begin{eqnarray}\label{eqn:d-e-diff-case2}
\delta \mathbf{E}_{ind} \approx 0,
\end{eqnarray}
\begin{eqnarray}\label{eqn:d-b-diff-case2}
\delta \mathbf{B}_{ind} \approx \dfrac{g_{a\gamma\gamma} a}{c}\mathbf{E}_0 \sim g_{a\gamma\gamma} a_0 c^{-1} E_0.
\end{eqnarray} \end{subequations}
If we using high enough DC electric field $E_0 > v_a B_0 > 10^{-3} c B_0$, the distinguish between two theories is much more sensitive than directional axion detection. 
}

{Before ending our discussion, we have to emphasize again that the compact solution of axion-induced fields in dual symmetric theory \eqref{eqn:d-eb-1} are not only correct for static axion haloscope experiments but also for the AC ones. Thus, dual symmetric theory of axion electrodynamics provided potential benchmark for various types of axion haloscope experiments.}
\section{\label{sec:conl}Conclusion}
In this paper, we have considered the axion electrodynamics in haloscope experiments under the perspective of dual symmetry. First, the dual symmetric revision of the axion Maxwell equation allows us to obtain an exact analytical solution in haloscopes without LWA approximation. Moreover, we pointed out that our solution from the dual symmetric theory is identical to the one from conventional theory when the wavelength of the axion is in the long-wavelength regime. It can be explained by the fact that required monopole sources in free space for conventional theory are proportional to gradient of axion field i.e from Equation \eqref{eqn:cond-intro}: $\rho _m^{conv} = \mu _0 ^{-1} g_{a\gamma\gamma} \mathbf{E}_0 \cdot \pmb{\nabla} a \to 0$, $\mathbf{J} _m^{conv} = \mu _0 ^{-1} g_{a\gamma\gamma} c \mathbf{B}_0 \times \pmb{\nabla} a \to 0$, then the magnetic monopole sources are such small that can be ignored. Hence, the dual symmetry approach examined in our work provides the same results as the conventional approach in Reference \cite{Sudbery1986} without using LWA approximation. Particularly, we apply to (im)perfect electric conductive resonant cavity in various shapes such as cylinder, two parallel sheets, sphere, circular torus, and rectangular torus. The characteristic quantities of these geometries like form factors related to summation/difference of induced electric and magnetic stored energy and their ratio have also been examined. The resonance of the axion-to-photon conversion power $P_{a\to \gamma\gamma}$ in imperfect electric conductive and the zeros of the ratio $\mathfrak{R}$ are also observed. This research may support scientists and technicians in the field regarding the search for axions or axion-like particles and provide more precise analytical frame to characterize axion dynamics in haloscope experiments with different geometries of resonant cavities. Finally, we also discussed about the difference bewteen dual symmetric and conventional theories of axion haloscope experiments. Our discussion could suggest some possibilities of distinguish these two theories by using known directional axion detection or {external DC electric field} on haloscopes.

\section*{Acknowledgement}

One of the author, Dai-Nam Le, thanks to Dr. Eibun Senaha (Science and Technology Advanced Institute, Van Lang University, Vietnam) for his helpful discussion when finalizing this work. Dai-Nam Le was funded by Vingroup Joint Stock Company and supported by the Domestic Master/ PhD Scholarship Programme of Vingroup Innovation Foundation (VINIF), Vingroup Big Data Institute (VINBIGDATA), code VINIF.2020.TS.03. The authors also thank the referees for their constructive comments.

\bibliographystyle{apsrev4-2}
\bibliography{axion_EM}

\begin{widetext}
\appendix
\section{\label{sec:A}The spatial distribution of axion-induced electromagnetic stored energy density}

\subsubsection{Infinite cylindrical solenoid}

\begin{figure}[H]
    \centering
    \includegraphics[width = 0.3 \textwidth]{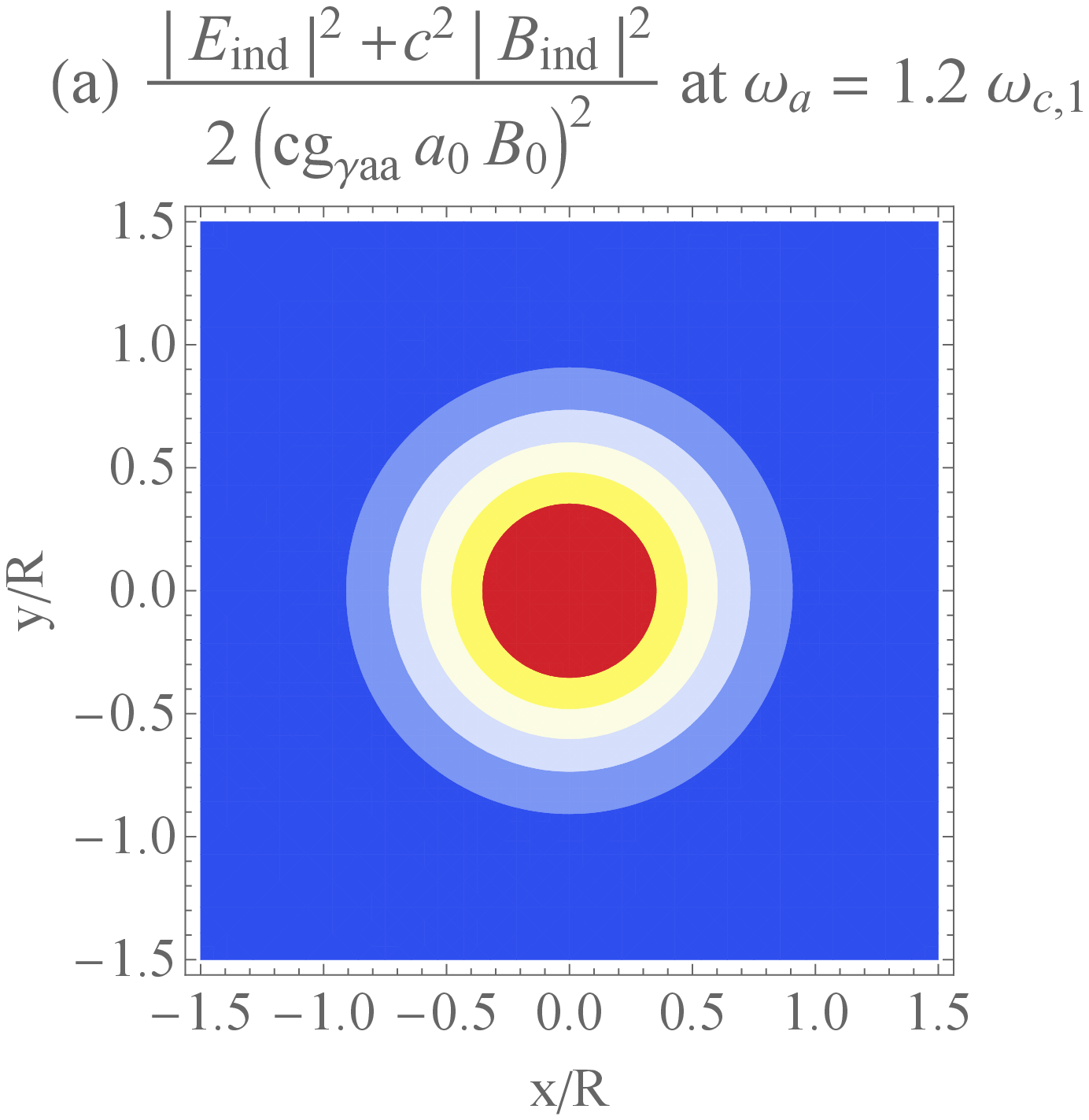}
    \includegraphics[width = 0.3 \textwidth]{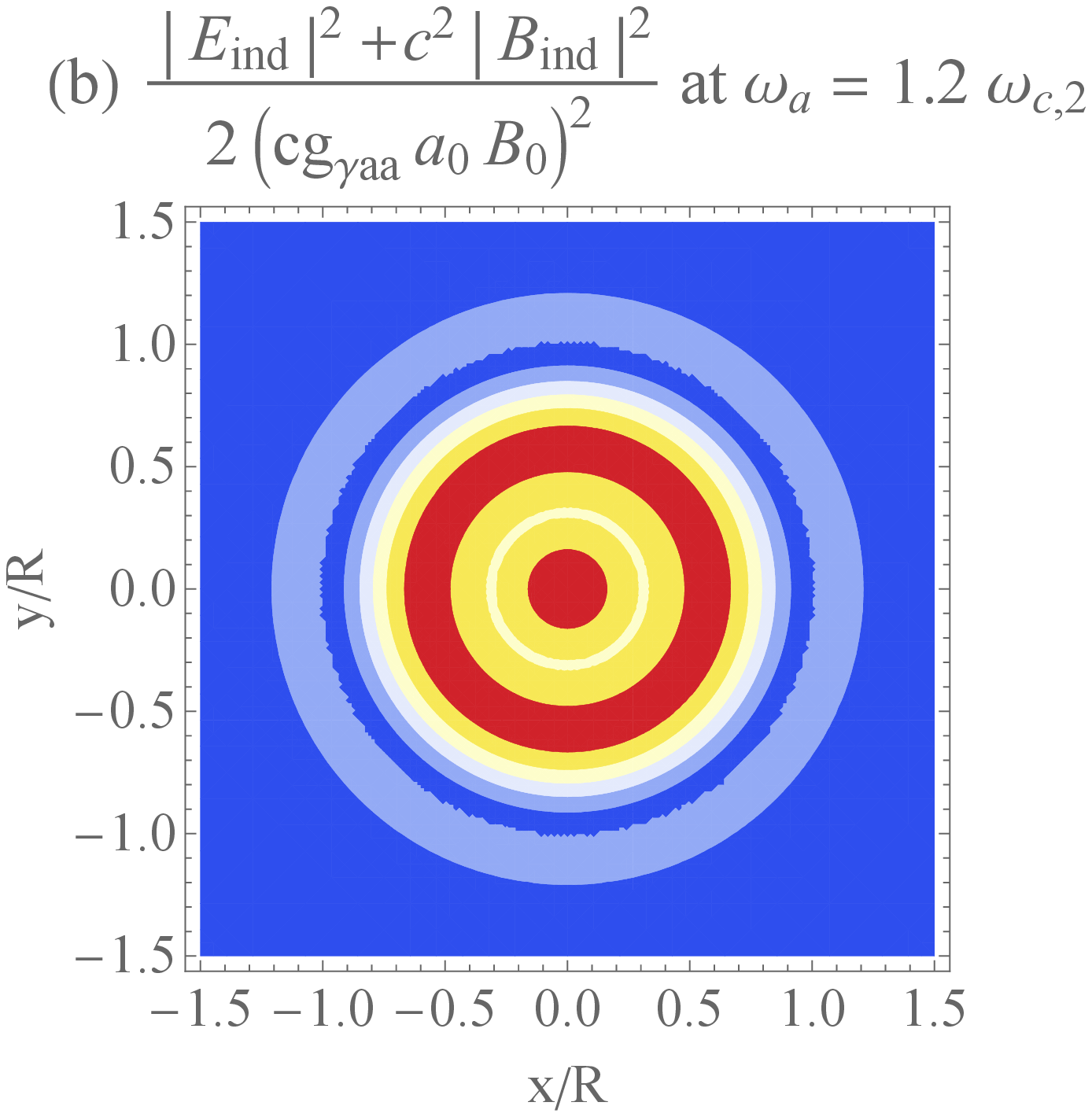}
    \includegraphics[width = 0.3 \textwidth]{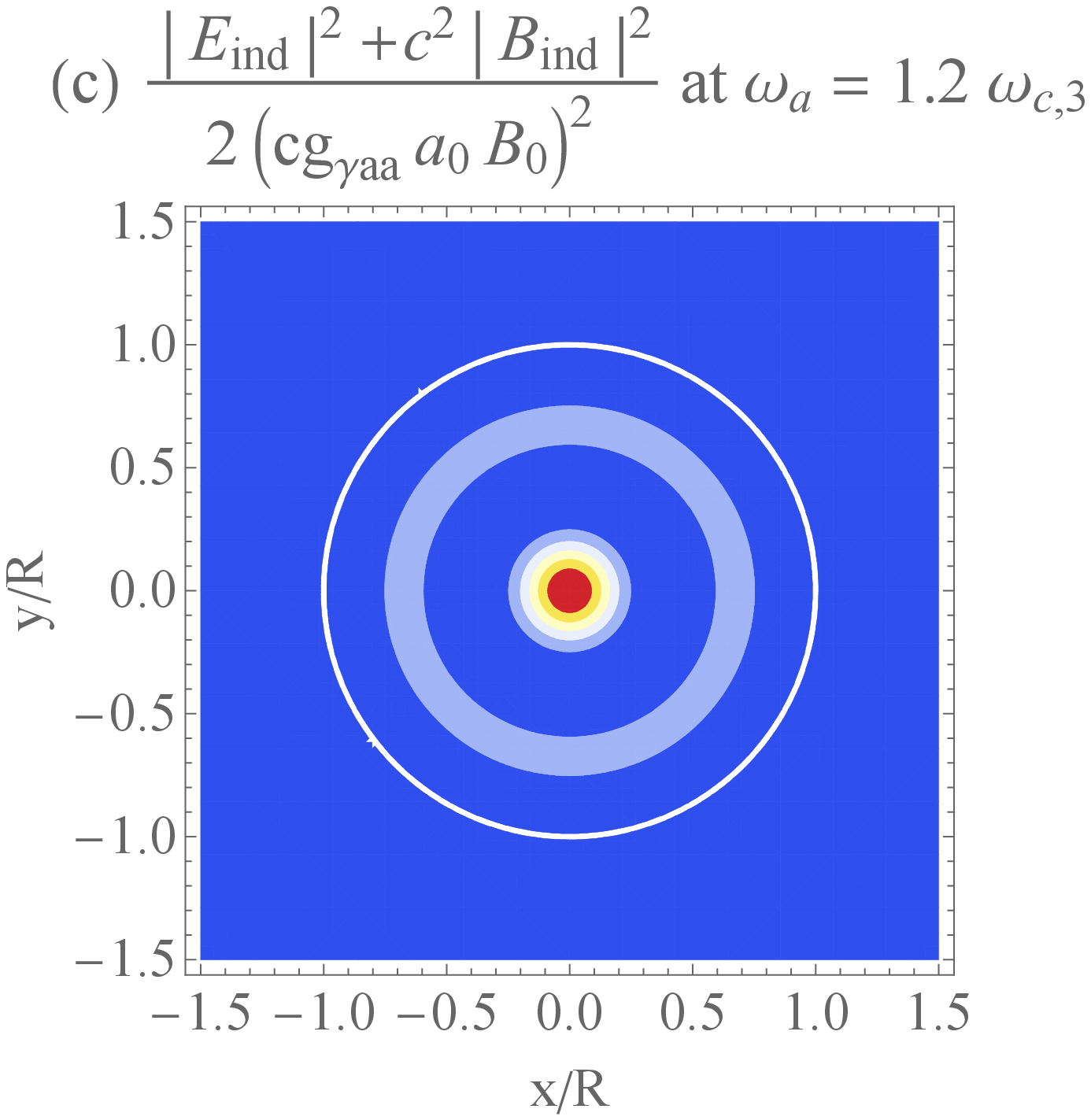}
    \caption{(a,b,c) The spatial distribution of axion-induced electromagnetic stored energy density inside and outside the cavity when $\omega _a = 1.2 \omega _{c,1}, 1.2 \omega _{c,2}, 1.2 \omega _{c,3}$.}
    \label{fig:case1EMstore}
\end{figure}

\subsubsection{Spherical solenoid}

\begin{figure}[H]
    \centering
    \includegraphics[width = 0.3 \textwidth]{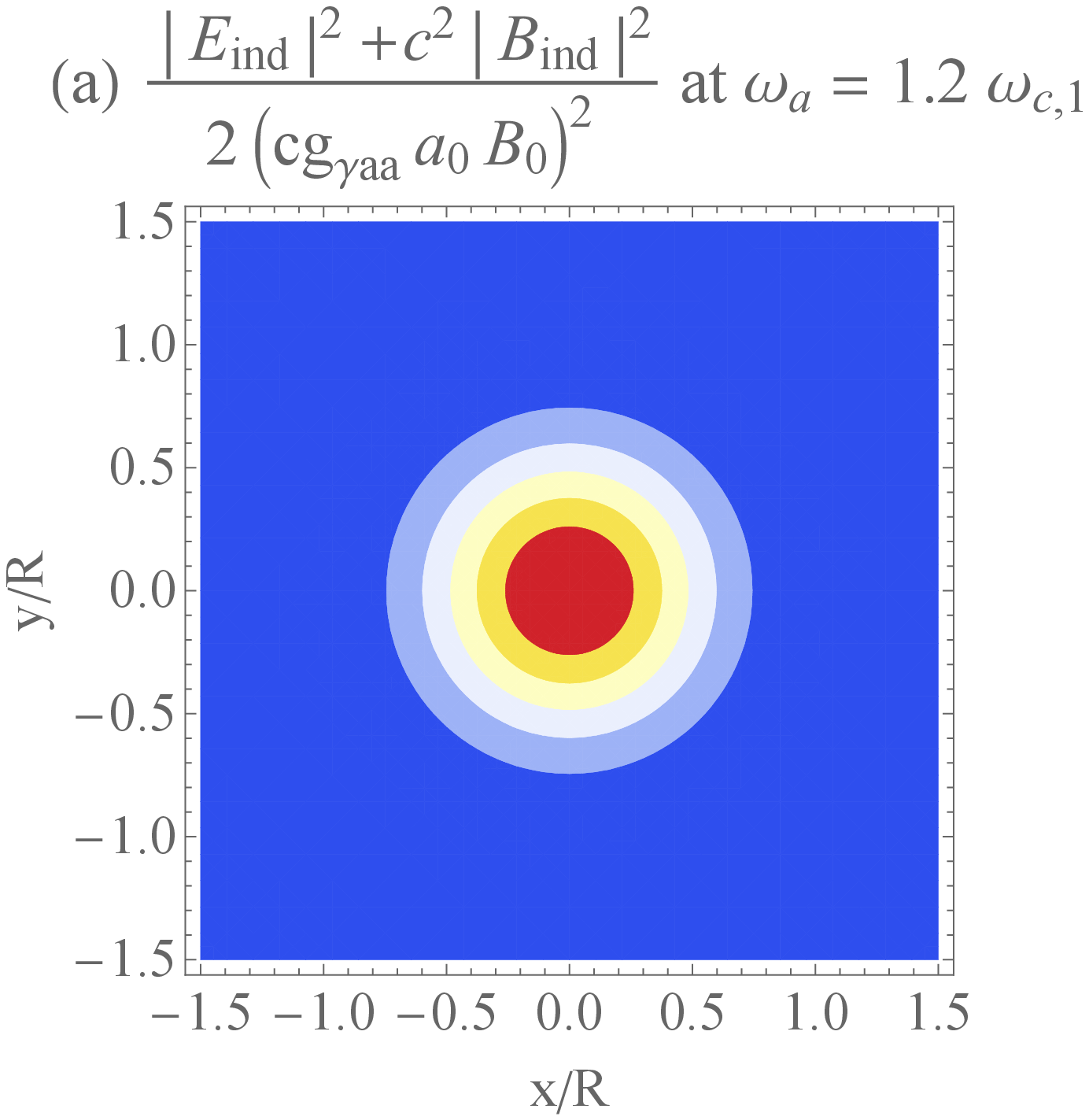}
    \includegraphics[width = 0.3 \textwidth]{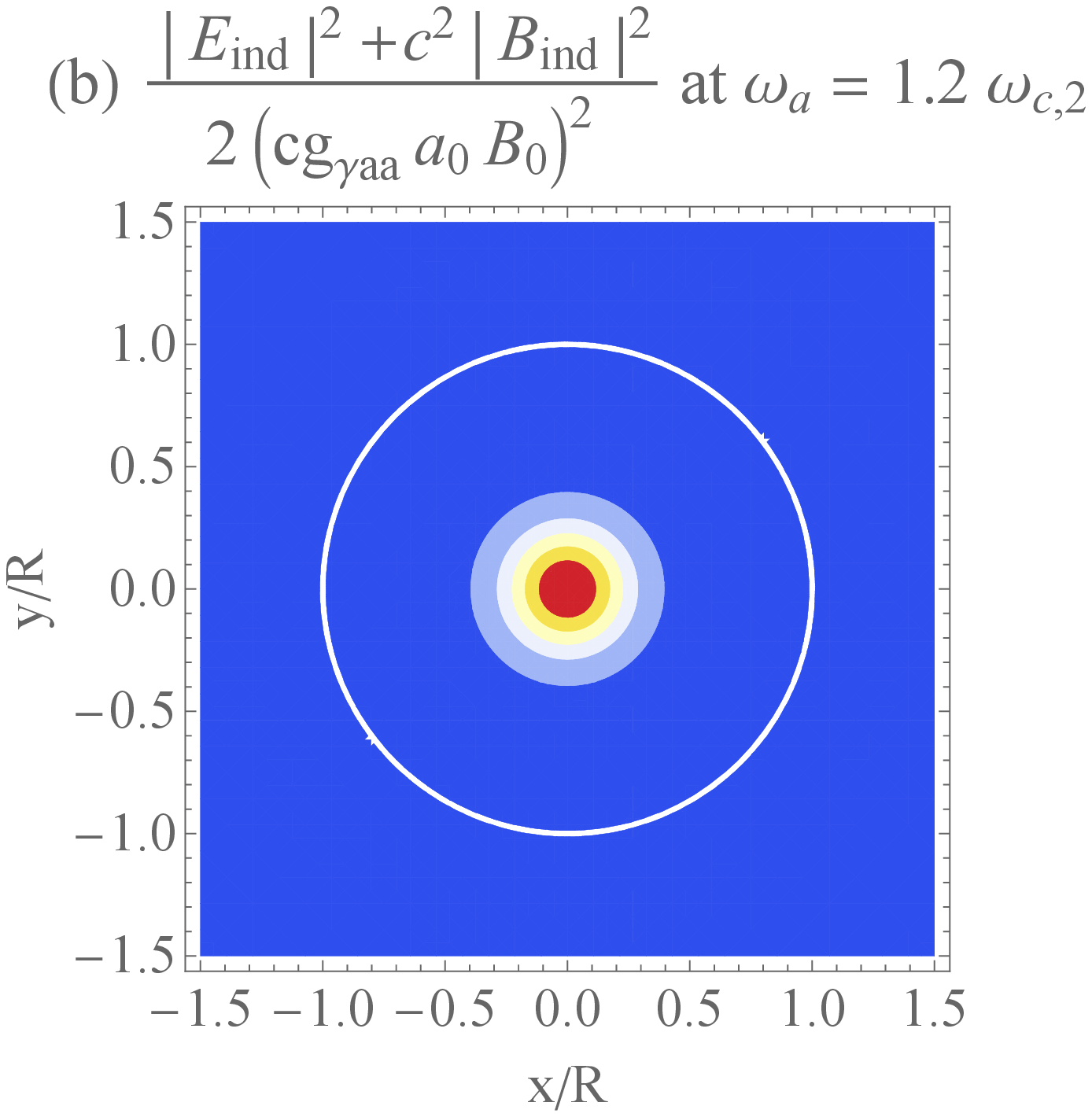}
    \includegraphics[width = 0.3 \textwidth]{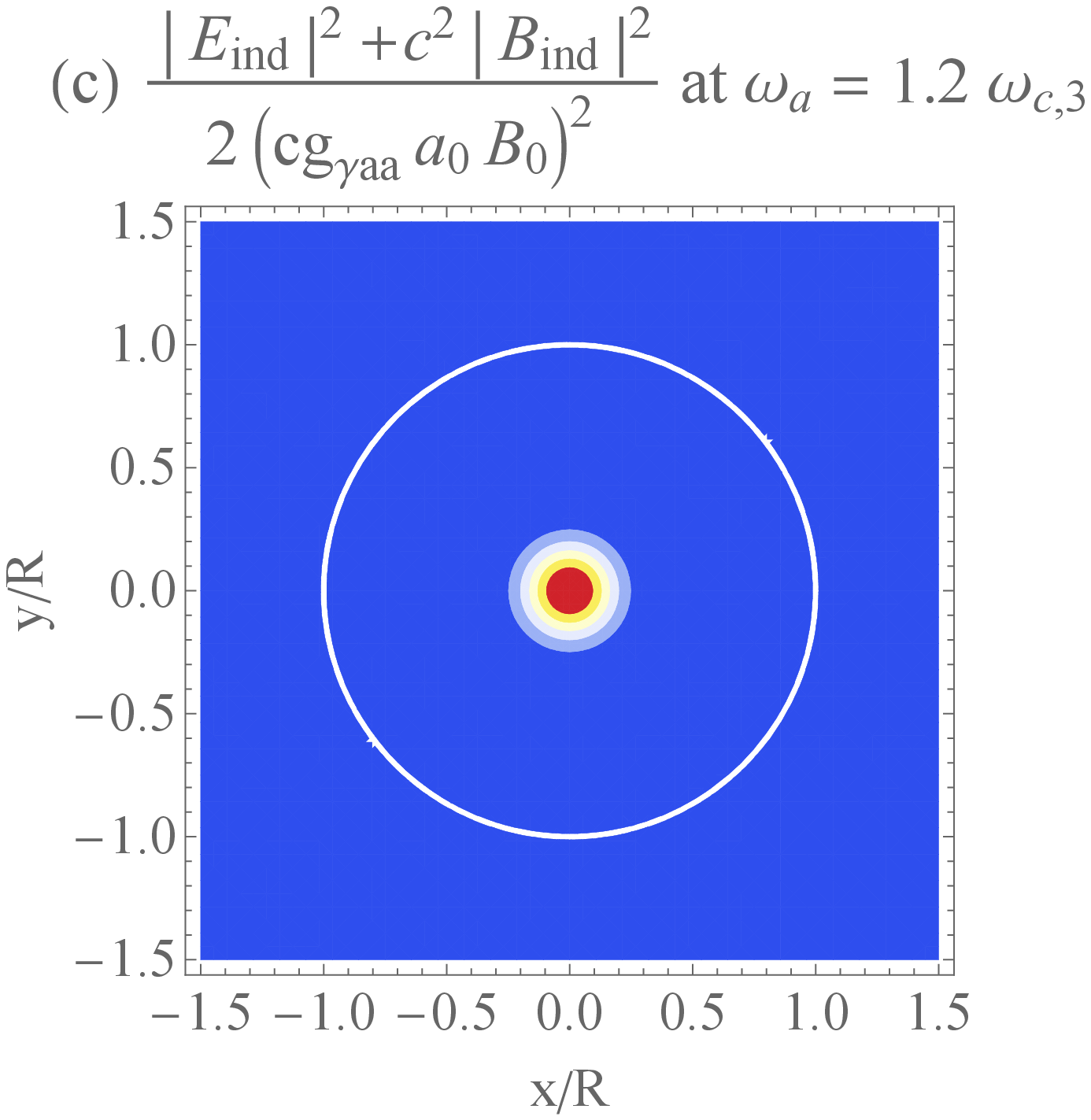}
    \caption{(a,b,c) The spatial distribution of axion-induced electromagnetic stored energy density inside and outside the cavity when $\omega _a = 1.2 \omega _{c,1}, 1.2 \omega _{c,2}, 1.2 \omega _{c,3}$ at equator of spherical cavity $\theta = 0$.}
    \label{fig:case4EMstore}
\end{figure}

\subsubsection{Infinite two-parallel-sheet cavity}
\begin{figure}[H]
    \centering
    \includegraphics[width = 0.3 \textwidth]{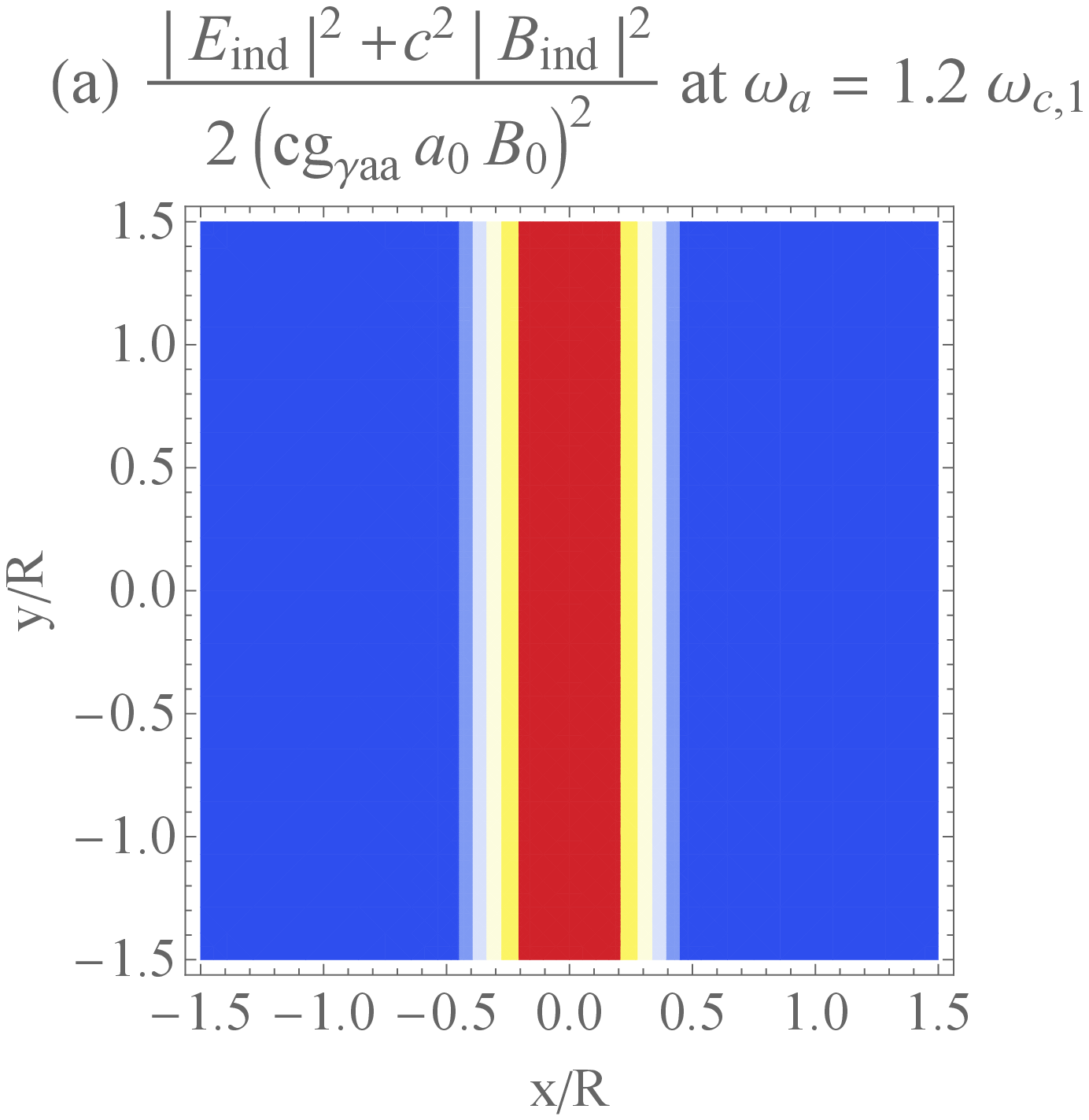}
    \includegraphics[width = 0.3 \textwidth]{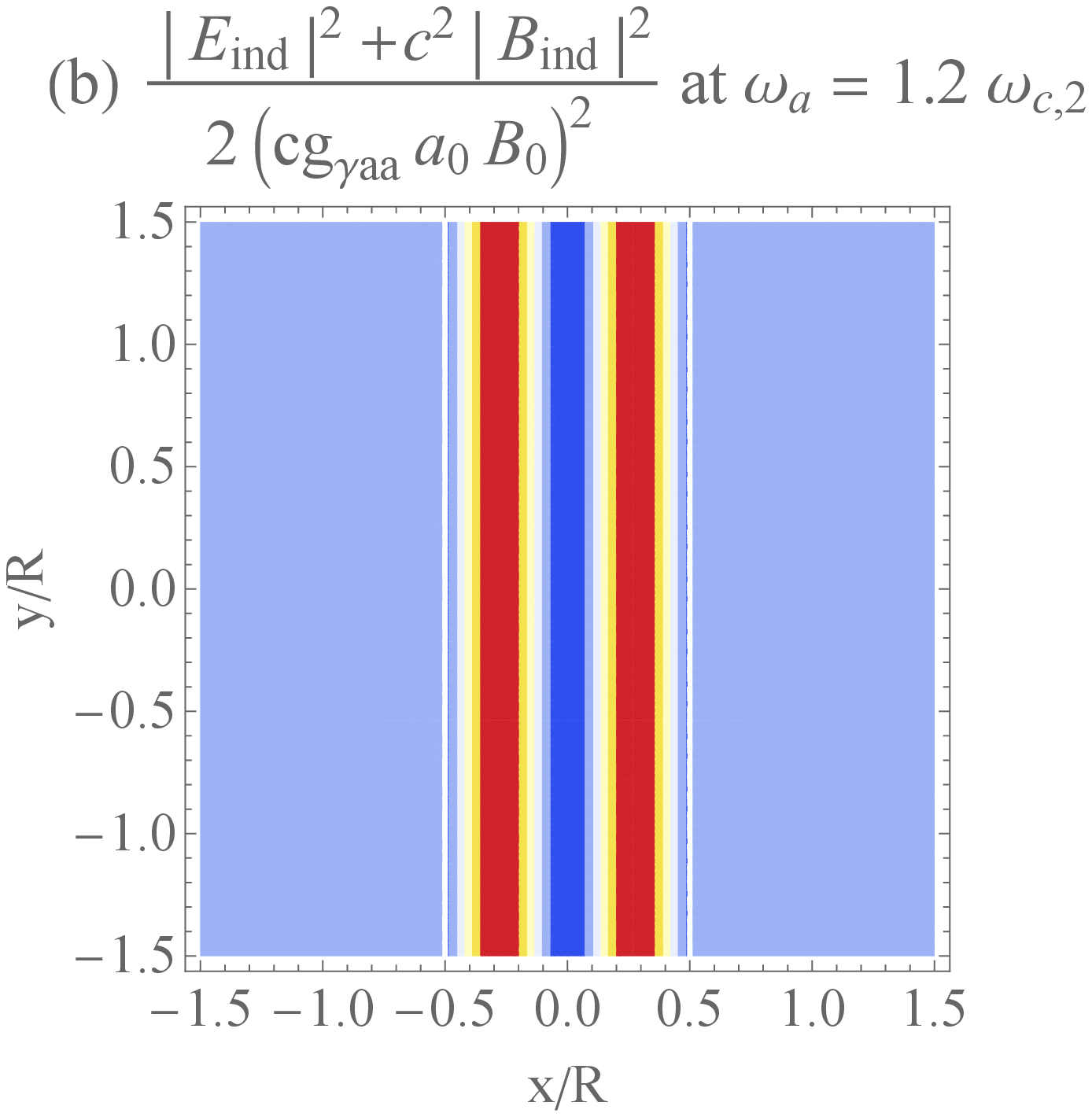}
    \includegraphics[width = 0.3 \textwidth]{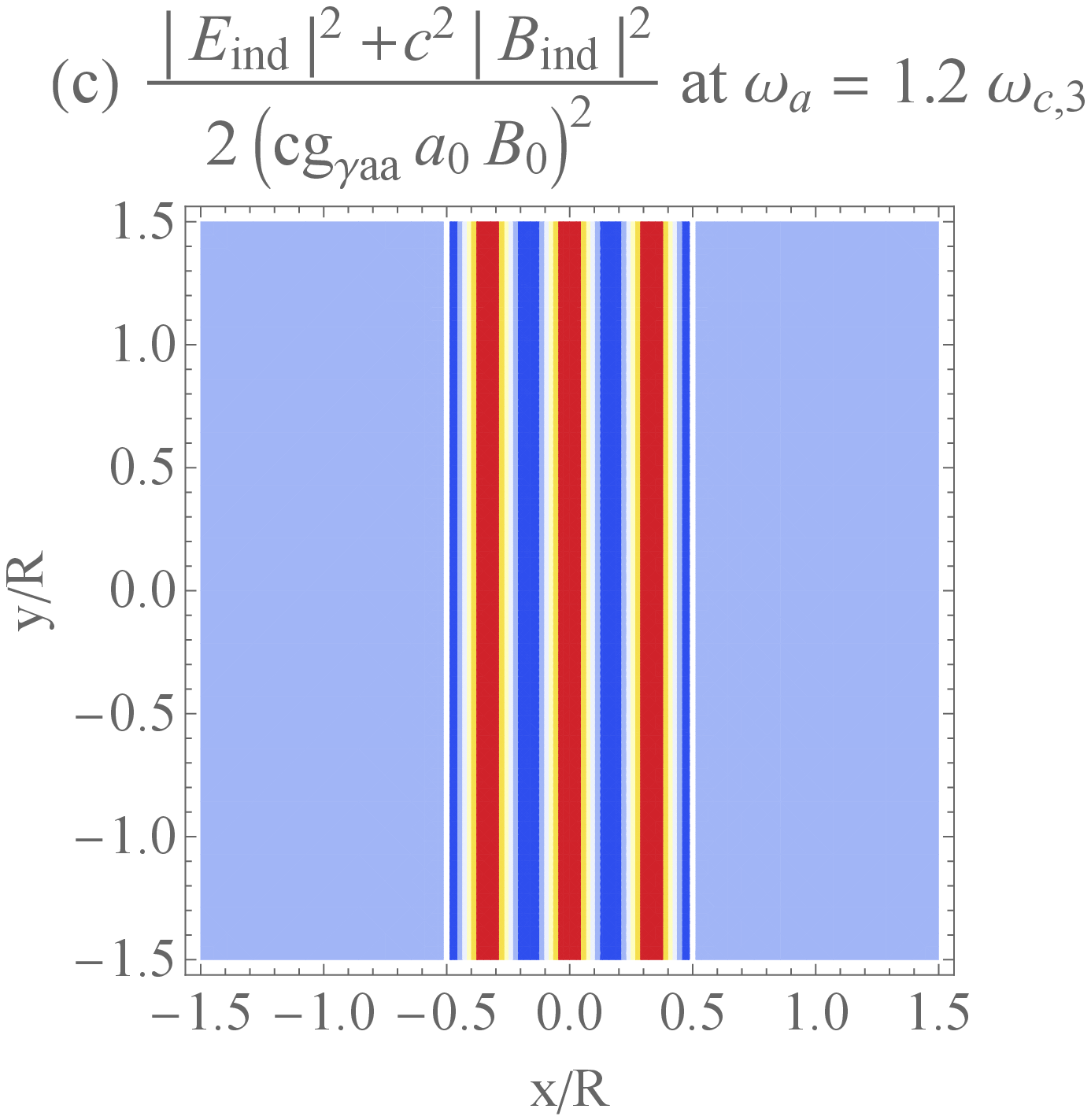}
    \caption{(a,b,c) The spatial distribution of axion-induced electromagnetic stored energy density inside and outside the cavity when $\omega _a = 1.2 \omega _{c,1}, 1.2 \omega _{c,2}, 1.2 \omega _{c,3}$.}
    \label{fig:case5EMstore}
\end{figure}
\subsubsection{Toroidal solenoid with rectangular cross section}
\begin{figure}[H]
    \centering
    \includegraphics[width = 0.3 \textwidth]{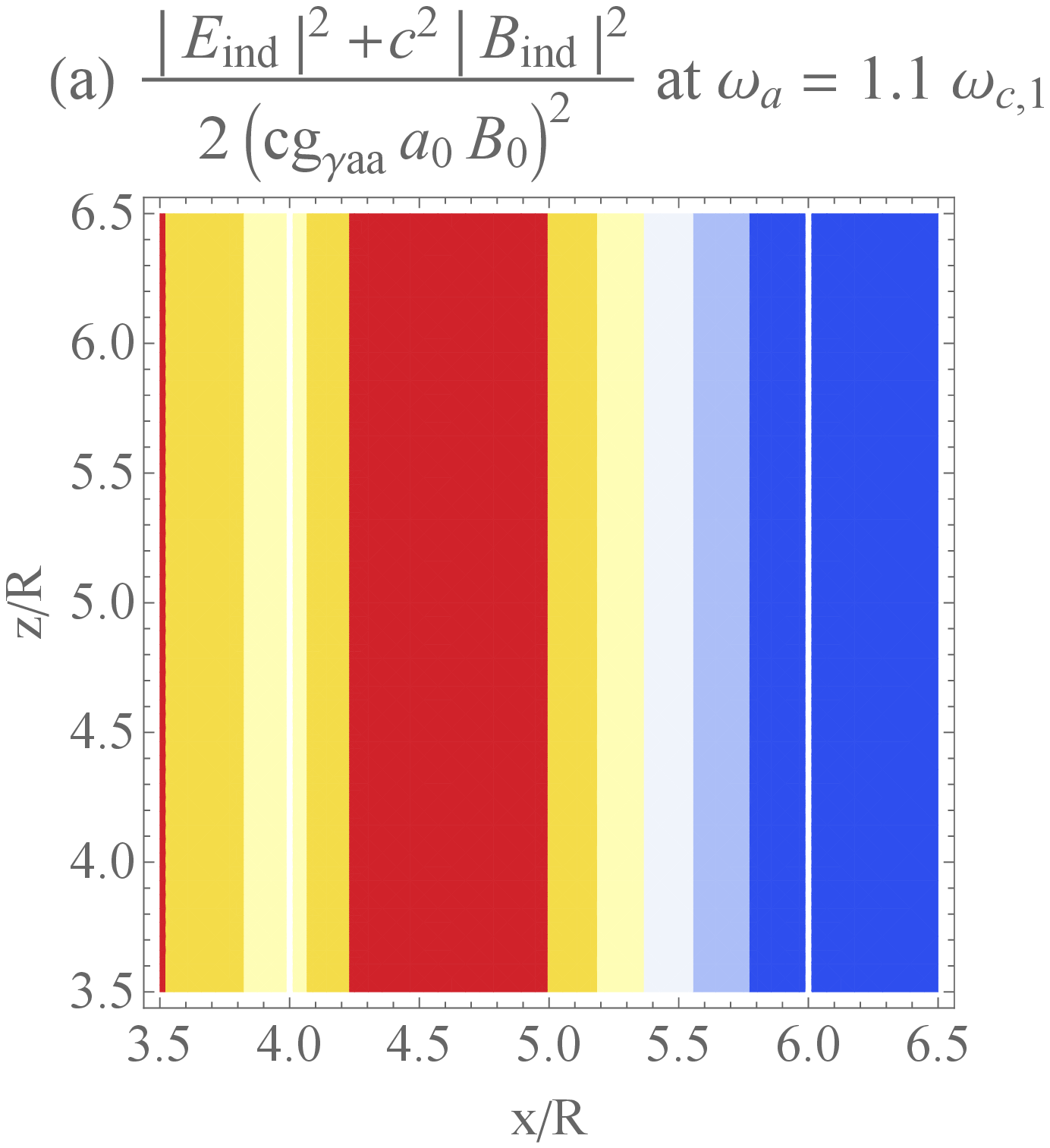}
    \includegraphics[width = 0.3 \textwidth]{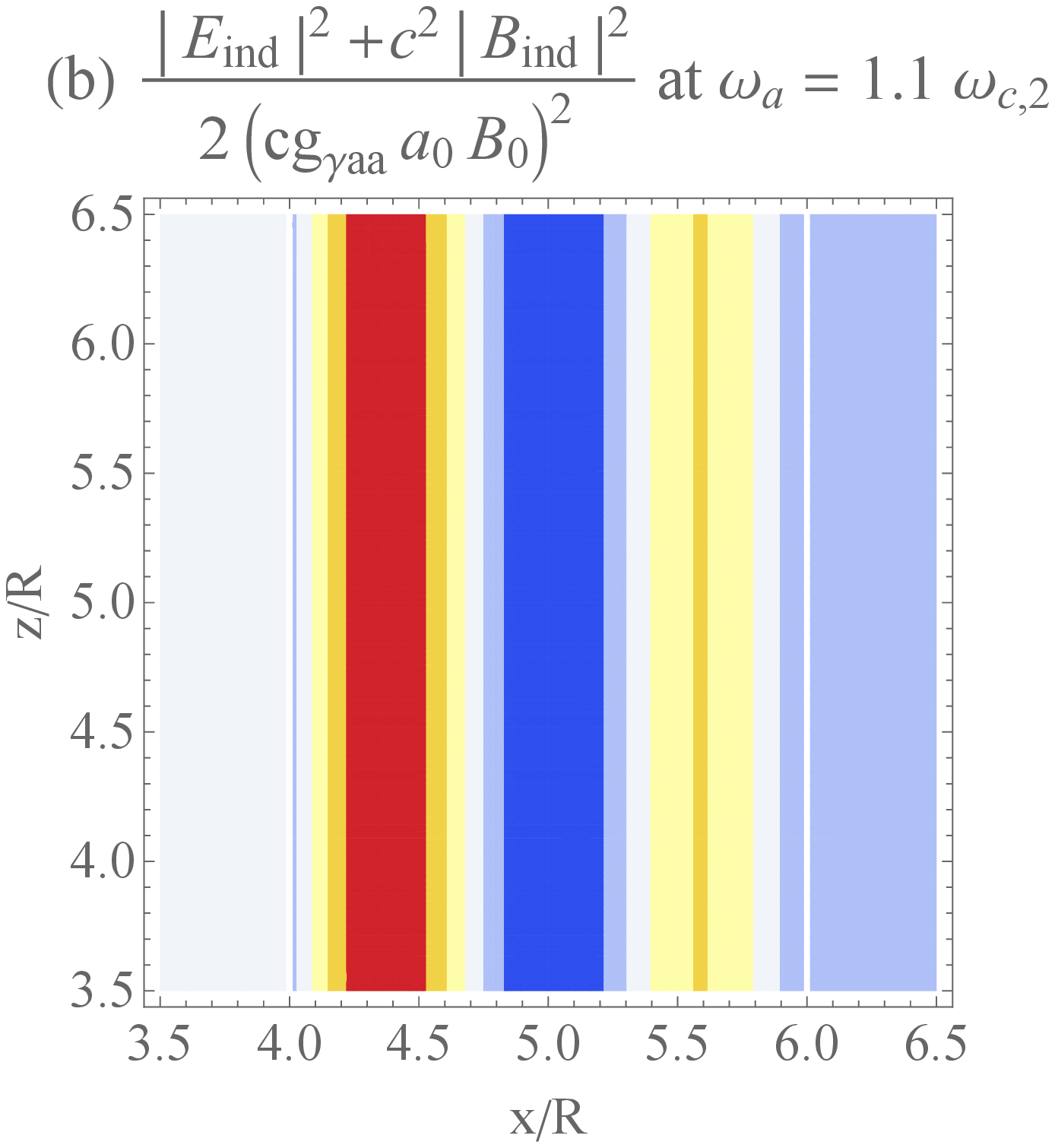}
    \includegraphics[width = 0.3 \textwidth]{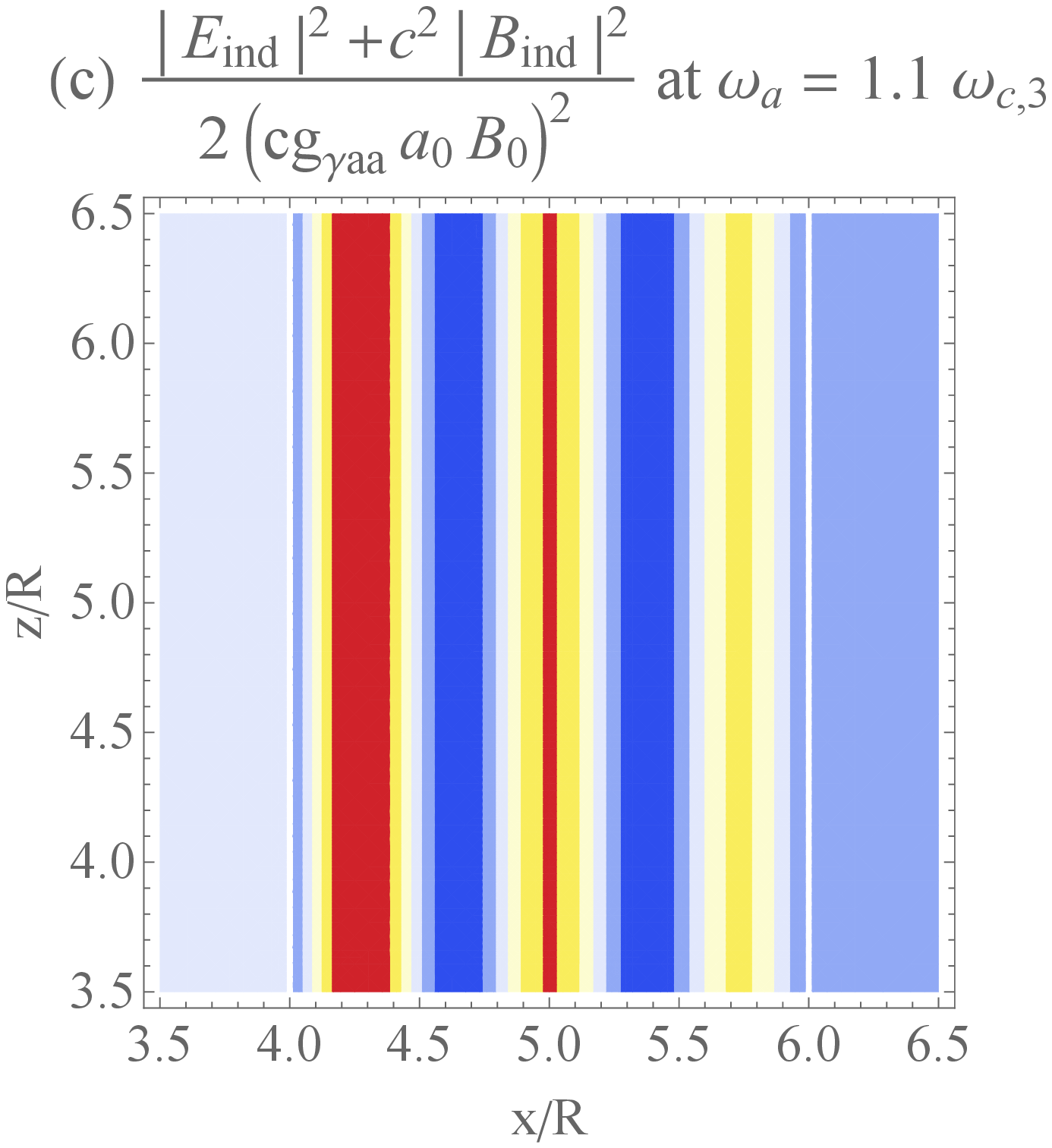}
    \caption{(a,b,c) The spatial distribution of axion-induced electromagnetic stored energy density inside and outside the cavity when $\omega _a = 1.1 \omega _{c,1}, 1.1 \omega _{c,2}, 1.1 \omega _{c,3}$.}
    \label{fig:case2EMstore}
\end{figure}

\subsubsection{Toroidal solenoid with circular cross section}

\begin{figure}[H]
    \centering
    \includegraphics[width = 0.3 \textwidth]{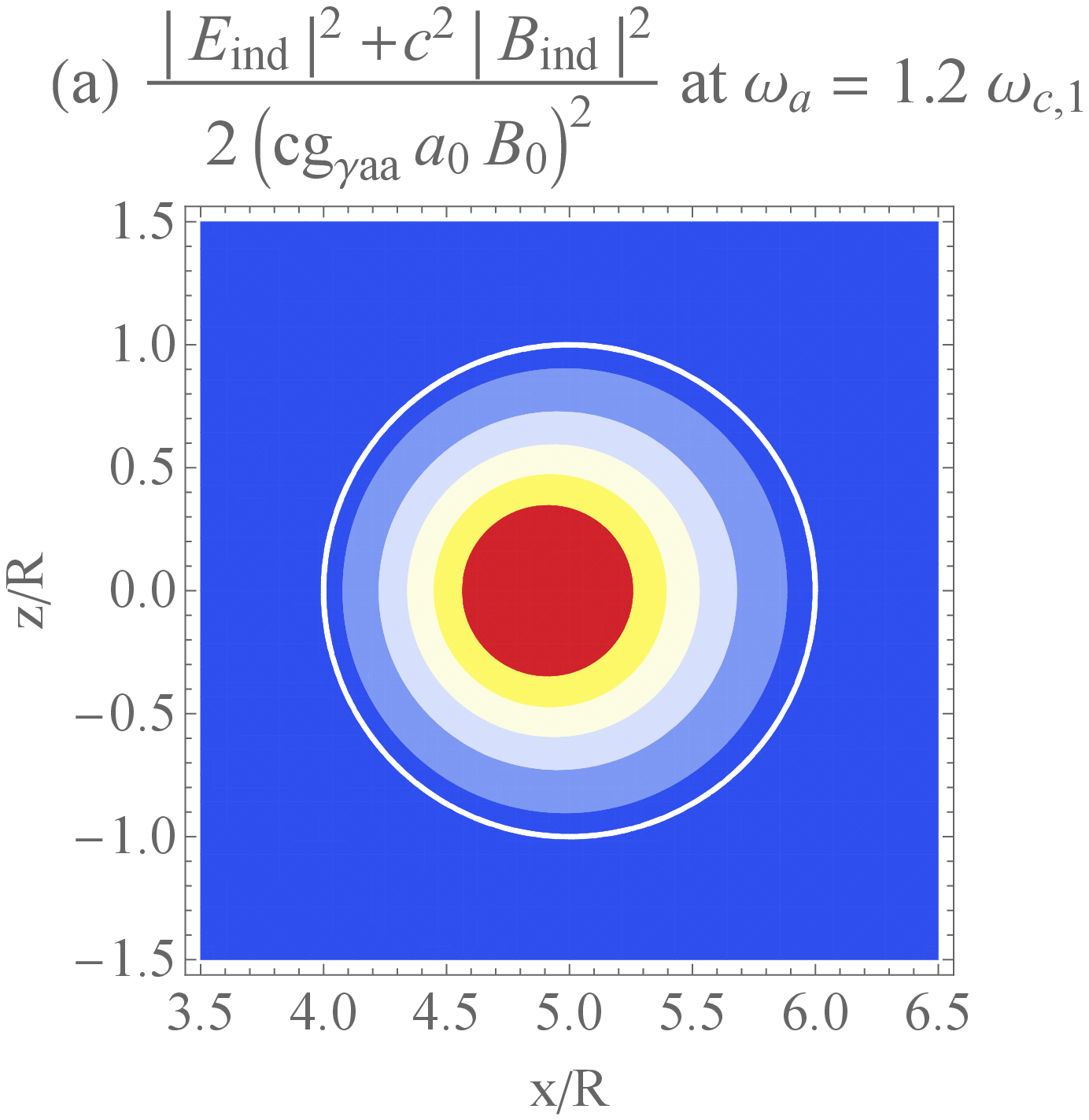}
    \includegraphics[width = 0.3 \textwidth]{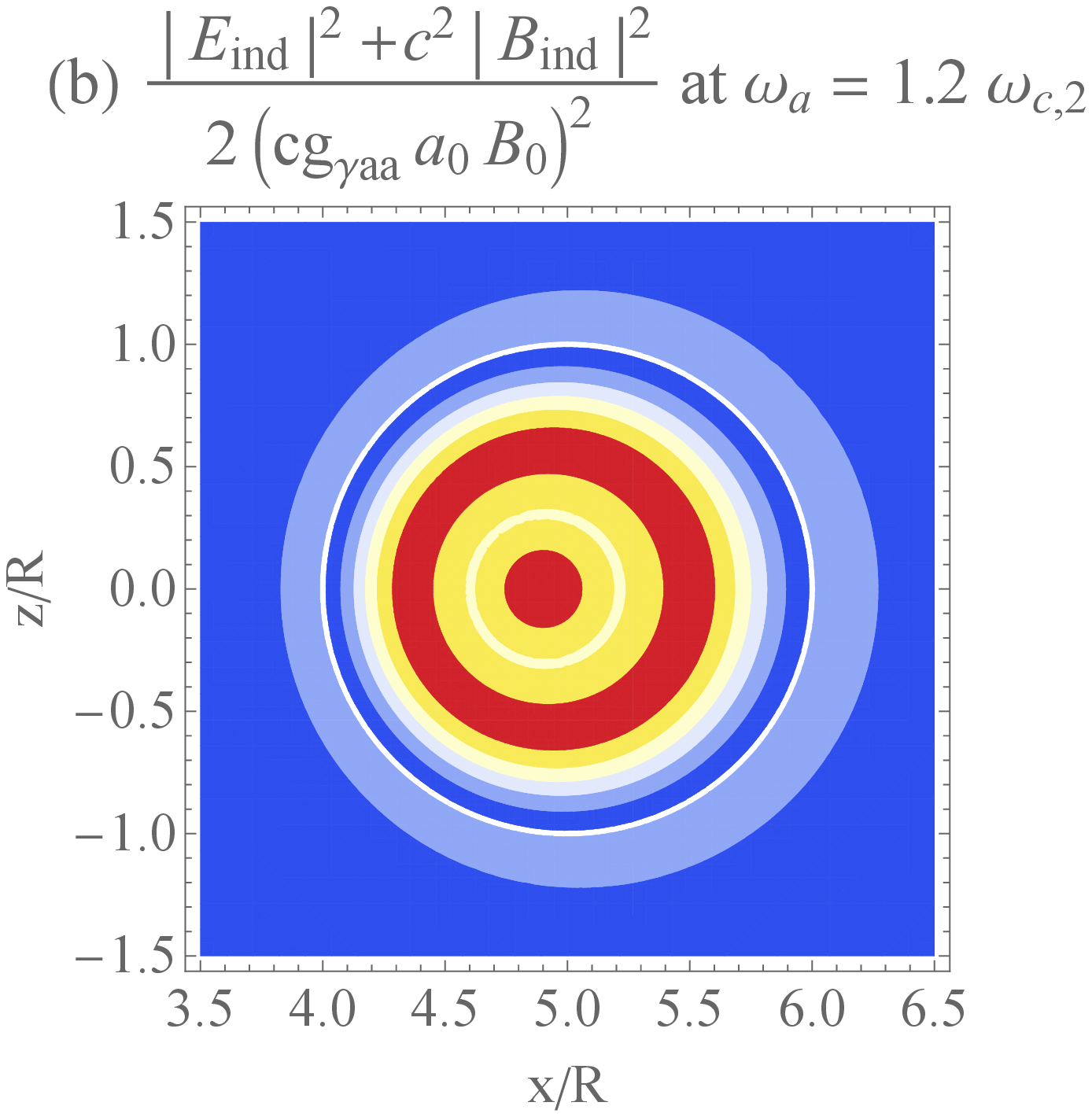}
    \includegraphics[width = 0.3 \textwidth]{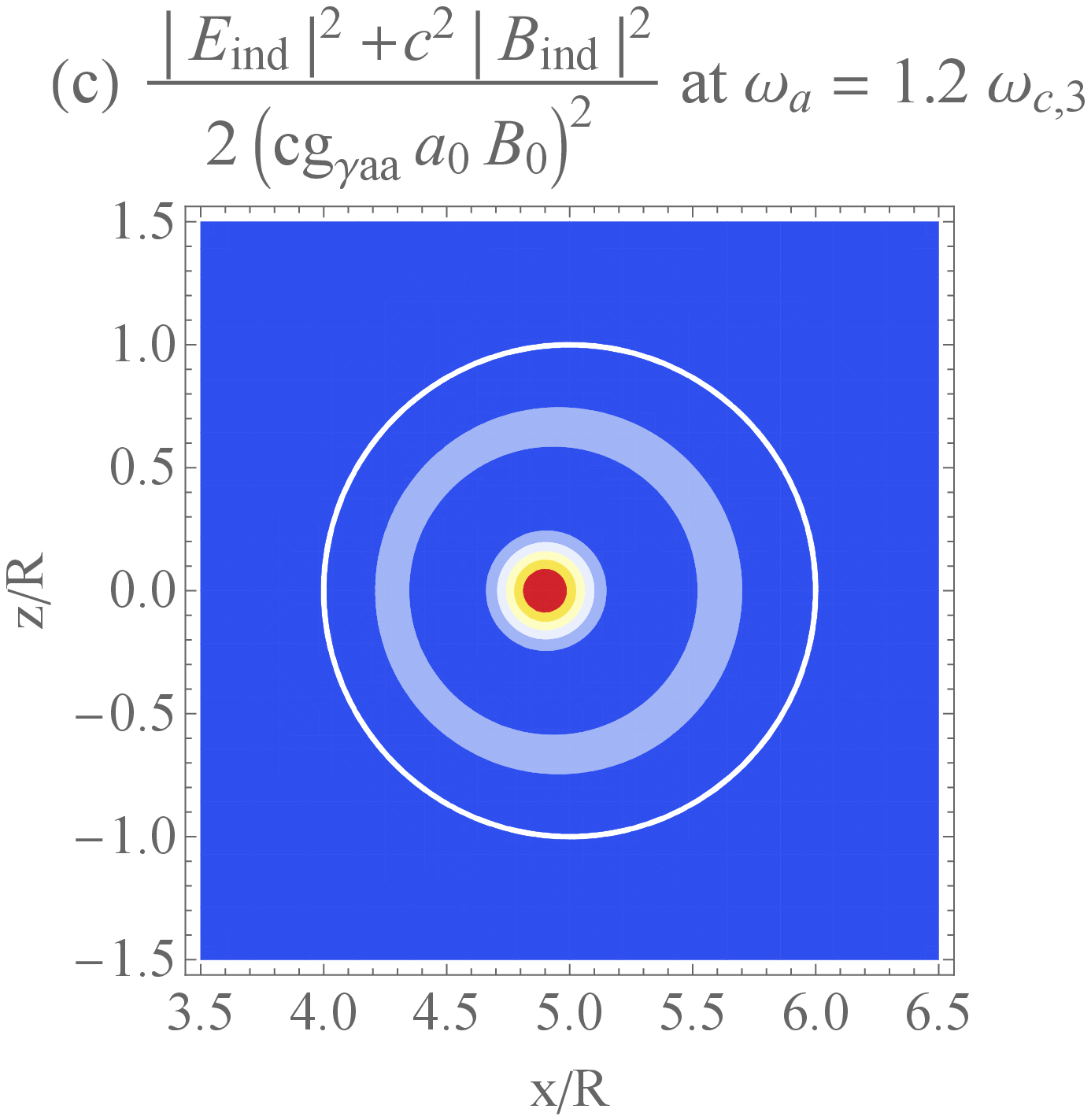}
    \caption{(a,b,c) The spatial distribution of axion-induced electromagnetic stored energy density inside and outside the cavity when $\omega _a = 1.2 \omega _{c,1}, 1.2 \omega _{c,2}, 1.2 \omega _{c,3}$.}
    \label{fig:case3EMstore}
\end{figure}

\end{widetext}
\end{document}